\documentclass[letterpaper, 12pt, oneside]{book}
\pdfoutput=1
    
    \usepackage[left=108pt, right=74pt, top=108pt, bottom=81pt, dvips, pdftex]{geometry}
    \usepackage{verbatim}
    \usepackage{subfigure}
    \usepackage{fancyhdr} 
    \renewcommand{\bibname}{References}

    \usepackage{amsmath,amsthm, amsfonts,amssymb} 
    \usepackage{setspace} 
    \usepackage[linktocpage,bookmarksopen,bookmarksnumbered,
                pdftitle={UC Davis Ph.D. Doctoral Thesis},
                pdfauthor={Department of Mathematics},%
                pdfsubject={UC Davis Ph.D. Doctoral Thesis},%
                pdfkeywords={UC Davis Ph.D. Doctoral Thesis}]{hyperref}
    
     \usepackage{graphicx} 

    \numberwithin{equation}{section}
    \numberwithin{figure}{section}

    \newcommand{\n}{\vspace{12pt}} 
      
    \newcommand{\newchapter}[3] 
	{                           
        \chapter[#2]{#3}

 \fancypagestyle{plain}{
  \fancyhf{}
  \fancyhead[R]{\hfill \thepage}
}

	}

     \newcommand{\mytitlea}{Spatial and Temporal Heterogeneity }
     \newcommand{\mytitleb}{of Host-Parasitoid Interactions }
     \newcommand{\mytitlec}{in Lupine Habitat }

\begin{document}
    

    \pagenumbering{roman}
    \pagestyle{plain}

    %
    %

    \singlespacing

    ~\vspace{-0.75in} 
    \begin{center}
\pdfbookmark[0]{Title Page}{title}
        \begin{huge}
           \mytitlea 
           \mytitleb
           \mytitlec
        \end{huge}\\\n
        By\\\n
        {\sc Roy Werner Wright}\\
        B.S. (University of California, Irvine) 2004\\\n\n
        DISSERTATION\\\n
        Submitted in partial satisfaction of the requirements for the degree of\\\n
        DOCTOR OF PHILOSOPHY\\\n
        in\\\n
        APPLIED MATHEMATICS\\\n
        in the\\\n
        OFFICE OF GRADUATE STUDIES\\\n
        of the\\\n
        UNIVERSITY OF CALIFORNIA\\\n
        DAVIS\\\n\n
        
        Approved:\\\n\n
        
        \rule{4in}{1pt}\\
        ~Alan Hastings (Chair)\\\n\n
        
        \rule{4in}{1pt}\\
        ~John Hunter\\\n\n
        
        \rule{4in}{1pt}\\
        ~Alex Mogilner\\
        
        \vfill
        
        Committee in Charge\\
        ~2008

    \end{center}

    \newpage

    \doublespacing

    %
    %
    
    \pdfbookmark[0]{Contents}{contents}
    \tableofcontents
    
    \newpage

    %
    %
    
    %

    ~\vspace{-1in} 
    \begin{flushright}
        \singlespacing
        Roy Werner Wright\\
        ~September 2008\\
        Applied Mathematics
    \end{flushright}

    \centerline{\large \mytitlea \mytitleb}
    \centerline{\large \mytitlec}
    
    \centerline{\textbf{\underline{Abstract}}}
    

\pdfbookmark[0]{Abstract}{abstract}

The inhabitants of the bush lupine in coastal California have been the subject of scientific scrutiny in recent years. Observations of a host-parasitoid interaction in the shrub's foliage, in which victims are significantly less motile than their exploiters, record stable spatial patterns in a fairly homogeneous environment. Though such pattern formation has been found in reaction-diffusion models, the correspondence of these models to continuous-time predator-prey interactions does not reflect the reality of the system being studied. Near the root of the lupine, another host-parasitoid interaction is also of considerable interest. In some cases this interaction, which promotes the health of the lupine, has been observed to be much more persistent than suggested by mathematical models. 

In this work a discrete-time spatial model of the first host-parasitoid system is introduced. We analyze the model, describing its transient behavior and finding the conditions under which spatial patterns occur, as well as an estimate of outbreak size under those conditions. We consider the feasibility of the necessary conditions in the natural system by modeling the mechanisms responsible for them, and discuss the effects of variable habitat on pattern formation. We also explore one possible explanation for the persistence of the second host-parasitoid system -- the existence of an alternate host. Under certain surprising conditions and by means quite different from previous models of similar situations, an alternate host can greatly enhance persistence of the nematode parasitoid.

    \newpage

    %
    %

    \chapter*{\vspace{-1.5in}Acknowledgments}
    \pdfbookmark[0]{Acknowledgments}{acks}



I would like to thank my teachers from throughout the years, and a few of these deserve particular mention. Mr. Frank Ferencz taught me two years of high school mathematics with care and creativity. Dr. Hong-Kai Zhao introduced me to mathematical modeling and strongly encouraged me to come to graduate school. And Dr. Alan Hastings, of course, gave me a thorough introduction to theoretical ecology and the support I needed to begin contributing to this field. 

Countless people, many of whom I've never met, contributed to my preparation and this dissertation's completion. But I would also like to thank my wife Ashley and our children Rae and Blaise, who contributed little to the process, but gracefully put up with it and made the rest of life enjoyable.

This work was funded in part by the National Science Foundation through Grant DMS-0135345.

    \newpage

    %
    %

    \pagestyle{fancy}
    \pagenumbering{arabic}

%
%
%
%

   \newchapter{Introduction} 
   {Introduction: The Lupine Habitat} 
   {Introduction: The Lupine Habitat} 
   \label{The Lupine Habitat}

The ecological system motivating the present work centers on bush lupines on the coast of California.  Above ground, these shrubs are home to flightless tussock moths, which feed on the bush but have little or no effect on its year-to-year dynamics. In turn, the tussock moths are parasitized by a variety of much more motile enemies \cite{Harrison}. Below ground, the bush lupine's roots provide shelter and sustenance to destructive ghost moth larvae, which are in turn parasitized by entomopathogenic nematodes. In this way, the nematodes promote the health of the shrubs through a trophic cascade.

Let us briefly review some pertinent results, to which we will return in the following chapters. In \cite{Harrison2} it is shown experimentally that the pupae of the tussock moth are subject to mortality by a predatory ant with a saturating functional response. In \cite{Hastings2}, measurements of tussock moth density from the field are compared with results from a reaction-diffusion model of a victim and its more motile exploiter. The model gives a good qualitative prediction of the somewhat counterintuitive spatial distribution of the tussock moth. An integrodifference model for the tussock moth is analyzed in \cite{Wilson}, but it proves to be incapable of some of the spatial patterns seen in the field.

In \cite{Dugaw} and \cite{Dugaw2}, deterministic and stochastic models, respectively, show the tendency of soil-dwelling nematode populations to oscillate wildly from year to year between very high and dangerously low levels. Some experimental work in \cite{Dugaw2} suggests that model values for nematode mortality have been too severe in some cases, but it is also noted that many other possible explanations for the observed persistence of nematodes in some rhizospheres have been all but ruled out.

In \cite{Harrison}, the lupine-centered community is discussed in detail and a preliminary step is taken to link the above- and below-ground subsystems. This step consists of analyzing the effects of changing carrying capacity -- which represents the health of the bush lupine -- on the tussock moth. 

\begin{figure}[ht]
\begin{center}
\includegraphics[width=.85\textwidth]{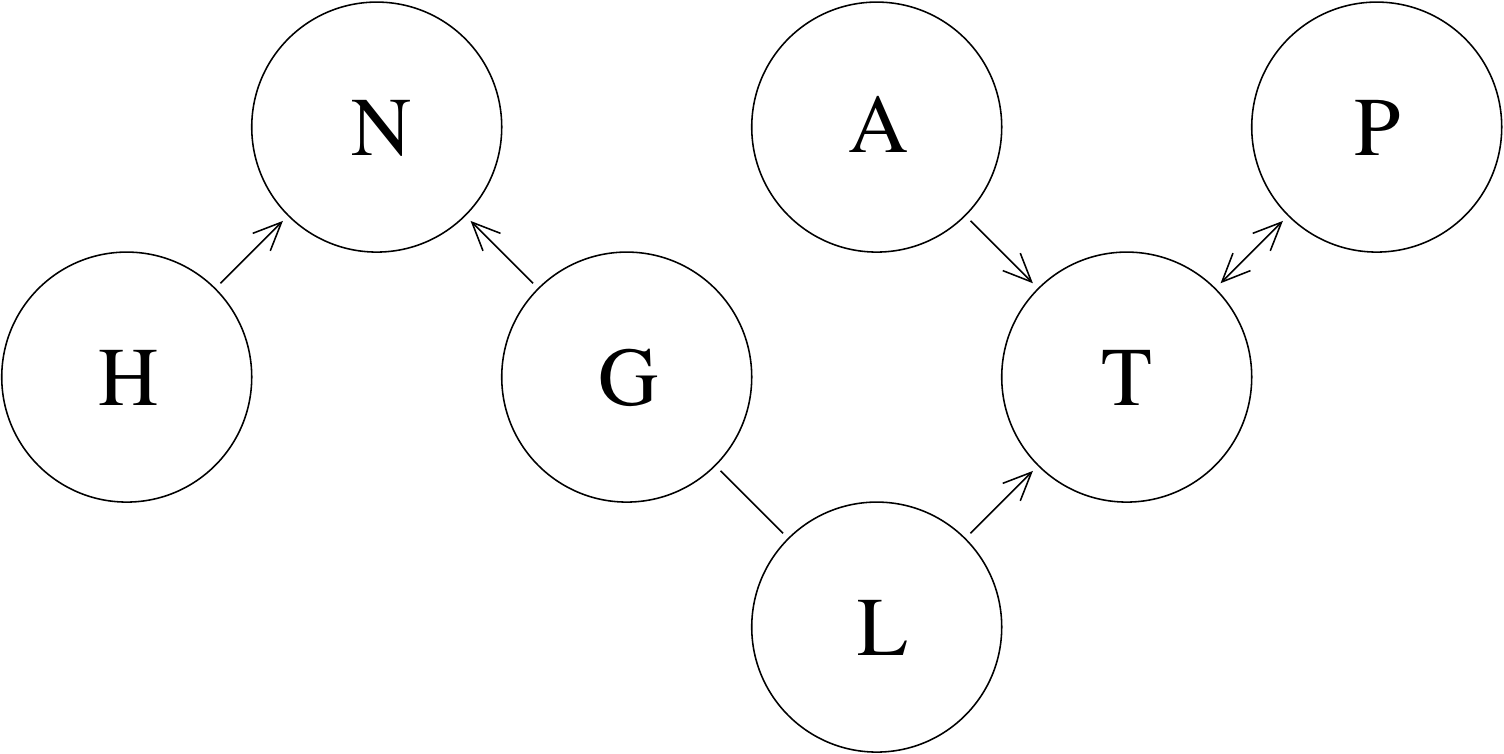}
\caption{Species of the lupine habitat: soil-dwelling {\bf N}ematodes with their larval {\bf G}host moth hosts and possible alternate {\bf H}osts, {\bf T}ussock moths with their motile {\bf P}arasitoids and predatory {\bf A}nts, and the bush {\bf L}upine. The meaning of arrows here is nonstandard (see text); trophic level is represented by vertical position.}
\label{community}
\end{center}
\end{figure}

A graphical overview of the community described above, as modeled in this dissertation, is given in Figure~\ref{community}. Arrows point in the direction of modeled ecological influences. For example, lupine (L) abundance affects the dynamics of the tussock moth (T), but defoliation by the tussock moth has no effect on the health of the lupine. The tussock moth and its parasitoids (P) are mutually interacting, while it is assumed that the generalist ant predator (A) is not limited by tussock moth abundance. We analyze an integrodifference model for the interaction of the tussock moth with its parasitoid enemies in Chapter~\ref{Spontaneous Patchiness}. In Chapter~\ref{The Effects of Habitat Quality}, we consider in greater detail the influence of predatory ants and the health of the lupine habitat.

The availability of the ghost moth host (G) is crucial to the nematode (N); however, while the ghost moth's in-year dynamics are modeled explicitly, they are reset at the beginning of each wet season, and the model becomes a single-species yearly return map for the nematode. This is also the case for the modeled alternate host~(H). Clearly, the ghost moth has a large and detrimental effect on the lupine, but this interaction is not modeled here. The persistence of the nematode is so precarious, and the health of the lupine is so dependent on the parasitoid's control of the ghost moth, that it suffices to consider the (generally transient) survival of the nematode. In Chapter~\ref{Nematode Chapter}, we search for a positive influence on the soil-dwelling nematode's persistence by examining the possible effects of a second host species.

%
%
%
%

    \newchapter{Spontaneous Patchiness} 
    {Spontaneous Patchiness in an Integrodifference Model} 
{Spontaneous Patchiness in a Host-Parasitoid Integrodifference Model} 
\label{Spontaneous Patchiness} 

\section{Introduction}
Host-parasitoid and other victim-exploiter interactions have long been a staple of mathematical investigations in ecology. An interesting subset of such systems are those in which the victim's and exploiter's dispersal behavior is crucial to the outcome of their interaction. In many cases, the exploiter has significantly greater motility than the victim (e.g.~\cite{Harrison}), which can cause the formation of intriguing and even counterintuitive spatial patterns~\cite{Comins}.

When the tussock moth and its parasitoid exploiters, described in Chapter~\ref{The Lupine Habitat}, are modeled by a pair of one-dimensional reaction-diffusion partial differential equations with negligible victim diffusion, a cursory singular perturbation analysis suggests the possibility, and predicts the shape, of a steady state solution~\cite{Conway}. The predicted solution consists of an outbreak region of nontrivial host density and a region in which only parasitoids exist. Surprisingly, the host density is highest near the edge of the outbreak, outside of which it falls rapidly to zero. This result is in qualitative agreement with data from the field~\cite{Hastings2}. More recent work has found that such a pair of reaction-diffusion equations produces traveling waves with fronts of the shape described above -- with highest host density near the front~\cite{Owen}. However, that work also places fairly strict limits on the types of victim growth behavior and boundary conditions that may produce spatially interesting steady state solutions. 

The main drawback to the partial differential equation formulation of the model is that it is continuous in time, while host-parasitoid systems often are not. The tussock moth is univoltine and it would therefore be more sensible to model the interaction in discrete time, as a pair of integrodifference equations. Integrodifference equations bring with them the additional benefit of greater flexibility in choosing the kernel describing the dispersal behavior~\cite{VanKirk}. A first step in this direction is taken in~\cite{Wilson}, where an integrodifference model is built upon the Nicholson-Bailey parasitoid model with density-independent host growth. After introducing and analyzing our model for the tussock moth system and demonstrating its spatial behavior, we will return to that earlier model for comparison.

We will now formulate our integrodifference model. Following this, we explore the dramatic spatial patterns possible for our model, and determine the requirements for their existence. First, using a singular perturbation approach we find that an Allee effect in the growth of the host is necessary for a patchy final state. We then visualize the process leading to the formation of a patchy spatial distribution through numerical simulation. With this process in mind, we form a more concrete understanding of the behavior of the system en route to its final state, first by relating that behavior to the local dynamics of our model, then by linear stability calculations, and finally by finding an estimate of the maximum spatial extent of a host outbreak.

\section{The Model}
\label{The Model}
Similar to~\cite{Wilson}, the basis for our model will be a Nicholson-Bailey host-parasitoid interaction:
\begin{equation}
  N_{t + 1}  = N_t e^{f(N_t ) - aP_t } ,
\label{no dispersal N}
\end{equation}
\begin{equation}
  P_{t + 1}  = N_t \left( {1 - e^{ - aP_t } } \right).
\label{no dispersal P}
\end{equation}
The function $f$ provides some form of density-dependent host growth. In this model, such density dependence is vital to the possibility of stable coexistence~\cite{Hastings}. As an example, simple density-dependent (Ricker) growth would be given by $f(N)=r\left(1-N\right)$~\cite{Beddington}. 

For any $x,y$ in our problem domain $\Omega$, let $k_d(x,y)$ be a dispersal kernel -- a density function for the probability that an organism at $y$ at time $t$ will be at $x$ at time $t+1$~\cite{Kot}. The parameter $d$ will describe the magnitude of dispersal, similarly to the diffusion coefficient in the continuous-time model. Then our spatial model is
  \begin{equation}
  N_{t + 1} (x) = \int\limits_\Omega  {k_\varepsilon  (x,y)N_t (y)e^{f(N_t (y)) - aP_t (y)} dy} ,
  \label{model N} 
  \end{equation}
  \begin{equation}
  P_{t + 1} (x) = \int\limits_\Omega  {k_D (x,y)N_t (y)\left( {1 - e^{ - aP_t (y)} } \right)dy},
  \label{model P} 
  \end{equation}
where $D\sim 1$ and $\varepsilon \ll D$, both positive, are the dispersal parameters for the parasitoid and host, respectively.

In much of the analysis to follow, we will use the Laplace kernel, 
\begin{equation}
k_d(x,y)  =   \frac{1}
{{2d}}e^{ - \left| x-y \right|/d}. 
\label{Laplace kernel} 
\end{equation}
This dispersal kernel is reasonable for many species~\cite{Kot} and has the additional benefit of mathematical tractability, as will be seen. Similar to previous models we will take our domain, for most of this paper, to be some interval $[0,L]\in\mathbb{R} $.

\section{Singular Perturbation Analysis of Steady State}
\subsection{Regular Perturbation}
\label{Regular Perturbation}
To obtain a first-order approximation to a time-invariant solution $(N,P)$, we formally take the limit as $\varepsilon\to 0$. Since $k_d(x,y)$ is an approximate identity in the sense of convolutions as $d\to0$, (\ref{model N}) becomes $N(x) = N(x)e^{f(N(x)) - aP(x)} $, so $N(x) \equiv 0$ or $e^{f(N(x)) - aP(x)}  = 1$. In the latter case, $f(N(x)) = aP(x)$, and we will say that $N(x) = f^{-1}(aP(x))$ in a sense to be explained below.

From~(\ref{model P}) we see immediately that choosing $N(x) \equiv 0$ can only lead to the trivial solution $(N,P) =(0,0)$. If we choose the other approximation, differentiating twice gives 
\begin{equation}
  { P'' = \frac{1}{{D^2 }}\left( {P - f^{ - 1} (aP)\left( {1 - e^{ - aP} } \right)} \right) }\;\rm{with}
  \label{ODE-P}
\end{equation} 
\[
   { P'(0) = \frac{1} 
{D}P(0),\; P'(L) =  - \frac{1}
{D}P(L), } 
\]
an equivalent problem to the integral equation for $P(x)$ as $\varepsilon\to 0$. Note that $P\equiv0$ is also a solution to this problem.

\subsection{Transition Layers}
\label{Allee result}
Following a derivation similar to that of~(\ref{ODE-P}), the integral equation for $\tilde N(x)$ with $\varepsilon>0$ is equivalent to the problem 
  \begin{equation}
  \varepsilon ^2 N'' = N\left( {1 - e^{f(N) - aP} } \right) \rm{ with}
  \label{ODE-N ODE} 
  \end{equation}
  \begin{equation}
  N'(0) = \frac{1}{\varepsilon}N(0),\;  N'(L) =  - \frac{1}{\varepsilon}N(L).
  \label{ODE-N BC} 
  \end{equation}

The previous regular approximation would lead us to believe that $N'(0)<0$ and $N'(L)>0$. This is because, as is evident from~(\ref{ODE-N ODE})-(\ref{ODE-N BC}), there is a layer at each of the domain endpoints in which $N$ is changing rapidly. These transition layers are not surprising given that~(\ref{model N})-(\ref{model P}) allows loss at the boundary and the dispersal of the host is small. The properties of the solution in and near these layers can be determined by singular perturbation analysis, which we omit here. Suffice it to report that they are not significantly different from previous results; the host density is highest at the edge before declining rapidly near the boundary.

Since we have found two possible regular approximations, we now turn our attention to the existence of internal transition layers. If such layers can exist, it would signal the possibility of striking spatial patterns, with stable patchiness -- i.e. separate coexistence and extinction subdomains -- despite the underlying spatial heterogeneity of the model. Suppose, then, that a transition layer occurs at some $x_0\in\Omega$. Because the dispersal of $P$ is much higher than that of $N$, the density $P$ should be approximately constant across the transition. So let $P\approx P_0$ there. Define the rescaled variable $\xi  = {{\left( {x - x_0 } \right)} \mathord{\left/
 {\vphantom {{\left( {x - x_0 } \right)} \varepsilon }} \right.
 \kern-\nulldelimiterspace} \varepsilon }$. Rewriting~(\ref{ODE-N ODE}) in terms of $\xi$, we have 
\begin{equation}
N'' = N\left( {1 - e^{f(N) - aP_0} } \right).
\label{ODE-int} 
\end{equation}
Suppose without loss of generality that the transition occurs between a subdomain where $N(x)\equiv0$ on the left and a subdomain where $N(x)=f^{-1}(aP(x))$ for $P(x)\not\equiv0$ on the right. Then the ``boundary" conditions for~(\ref{ODE-int}) are 
\begin{equation}
\mathop {\lim }\limits_{\xi  \to  - \infty } N(\xi ) = 0 \;\; {\rm{and }} \;\; \mathop {\lim }\limits_{\xi  \to  + \infty } N(\xi ) = f^{ - 1} \left( {aP_0 } \right).
\label{BC-int} 
\end{equation} 

It is instructive to consider~(\ref{ODE-int}) as a pair of first-order ordinary differential equations for $N$ and $N'$. That system has fixed points at $(N,N')=(0,0)$ and $\left(  f^{-1}(aP_0)     ,0\right)$. The conditions~(\ref{BC-int}) require these fixed points to be saddles with a heterocline connecting them. In order for them to be saddles, the determinant of the Jacobian must be negative at both points:
\[
  \det \mathbf J(0,0) = e^{f(0) - aP_0 }  - 1 < 0 \; {\rm{and}}
  \]\[
  \det \mathbf J\left( {f^{ - 1} \left( {aP_0 } \right),0} \right) = f^{ - 1} \left( {aP_0 } \right)f'\left( {f^{ - 1} \left( {aP_0 } \right)} \right) < 0.
\]
That is, $f(0) < aP_0 $ and $f'\left( {f^{ - 1} \left( {aP_0 } \right)} \right) < 0$. The first of these implies that, if $f$ is nonincreasing on its domain, $f(N) = aP_0$ cannot be solved for positive $N$. But the existence of transition layers requires such a solution. The second inequality implies that $f$ is decreasing at some point in its domain. Therefore interior transition layers are an impossibility if $f$ is monotonic. 

Suppose that $f$ satisfies the conditions for saddles at the fixed points in question. Then there remains the additional necessity of a heterocline connecting them. We determine how to satisfy this condition by manipulating~(\ref{ODE-int}) subject to~(\ref{BC-int}):\[
N''N' = N\left( {1 - e^{f(N) - aP_0 } } \right)N', \; {\rm{so}}
\]
\[
\frac{1}
{2}\int\limits_\mathbb{R}  {\left( {\left( {N'} \right)^2 } \right)'d\xi }  = \int\limits_\mathbb{R}  {N\left( {1 - e^{f(N) - aP_0 } } \right)N'd\xi } , \; {\rm{so}}
\]\[
\left. {\frac{1}
{2}\left( {N'} \right)^2 } \right|_{ - \infty }^\infty   = \int\limits_\mathbb{R}  {N\left( {1 - e^{f(N) - aP_0 } } \right)\frac{{dN}}
{{d\xi }}d\xi } , \; {\rm{so}}
\]
\begin{equation}
\int\limits_0^{f^{ - 1} \left( {aP_0 } \right)} {N\left( {1 - e^{f\left( N \right) - aP_0 } } \right)dN}  = 0.
\label{matcond} 
\end{equation}

The first fact evident from~(\ref{matcond}) is that there must be two values of $N$ at which $f(N)=aP_0$. If there were only one such value, it would be the only possible definition of $N=f^{ - 1} \left( {aP_0 } \right)$, and either $f(N)<aP_0$, or $f(N)>aP_0$, for all $N\in\left(0,f^{ - 1} \left( {aP_0 } \right)\right)$. If this were the case, the integral could not vanish. Therefore $f^{ - 1} \left( {aP_0 } \right)$ must be defined as the larger value of $N$ at which $f(N)=aP_0$ (here we dismiss as biologically unlikely a growth function $f$ that crosses $aP_0$ at three points). For continuous $f$, since $f$ must be decreasing at $f^{ - 1} \left( {aP_0 } \right)$, it must be increasing at the other solution of $f(N)=aP_0$. So $f$ has the unimodal form typical of an Allee effect. 

\subsection{Summary of Results}
We have analyzed our host-parasitoid model~(\ref{model N})-(\ref{model P}), probing the possibility of a patchy steady state -- a long-term spatial distribution with both extinction and coexistence subdomains, sharply segregated. The conclusion of our rigorous analysis is that such spatial patterns may only occur if the growth of the host has an Allee effect. That is, at small enough host densities, per-capita growth must decrease as density decreases.

\begin{figure}[htbp]
\begin{center}
\includegraphics[width=.48\textwidth]{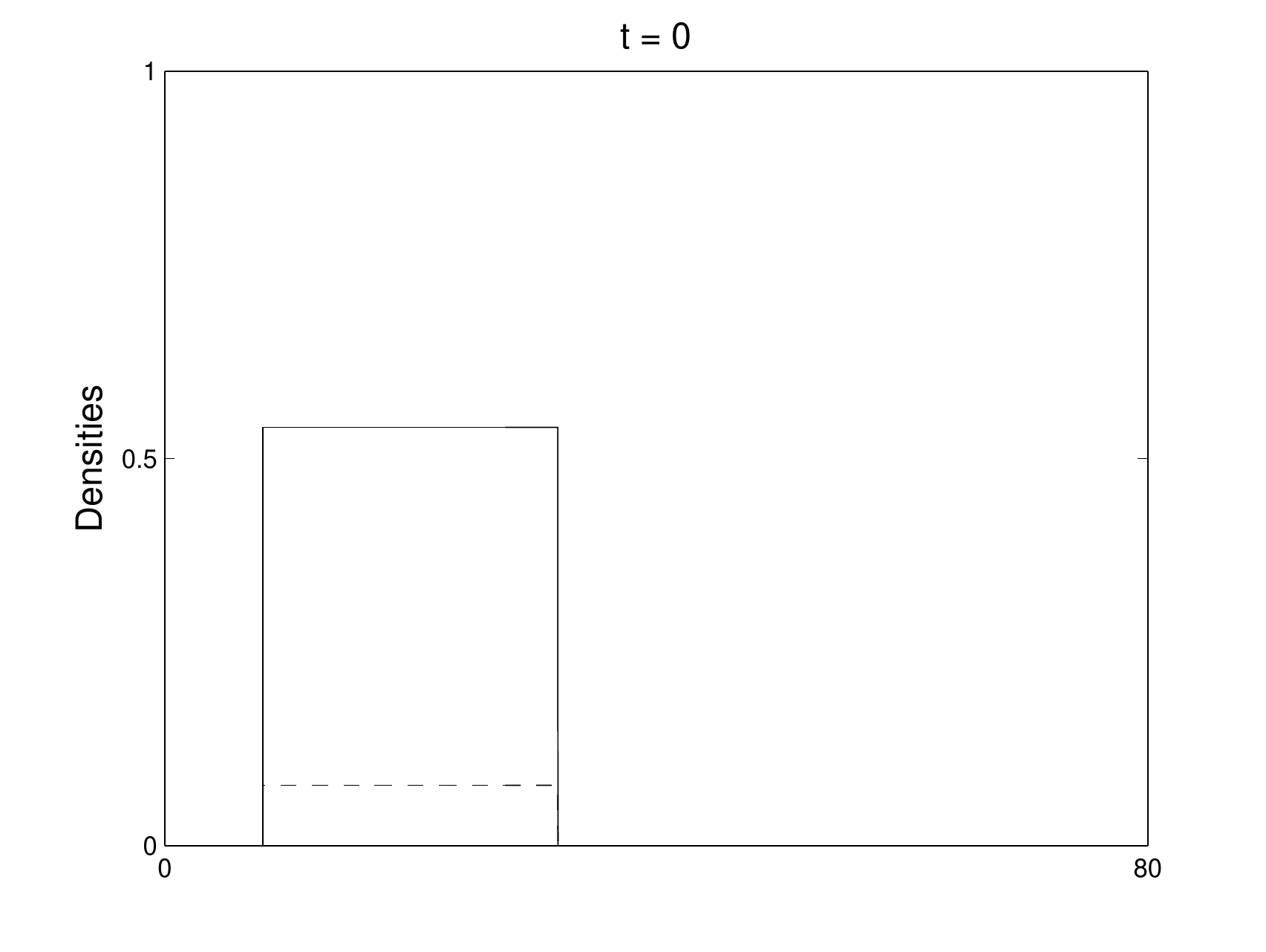} 
\includegraphics[width=.48\textwidth]{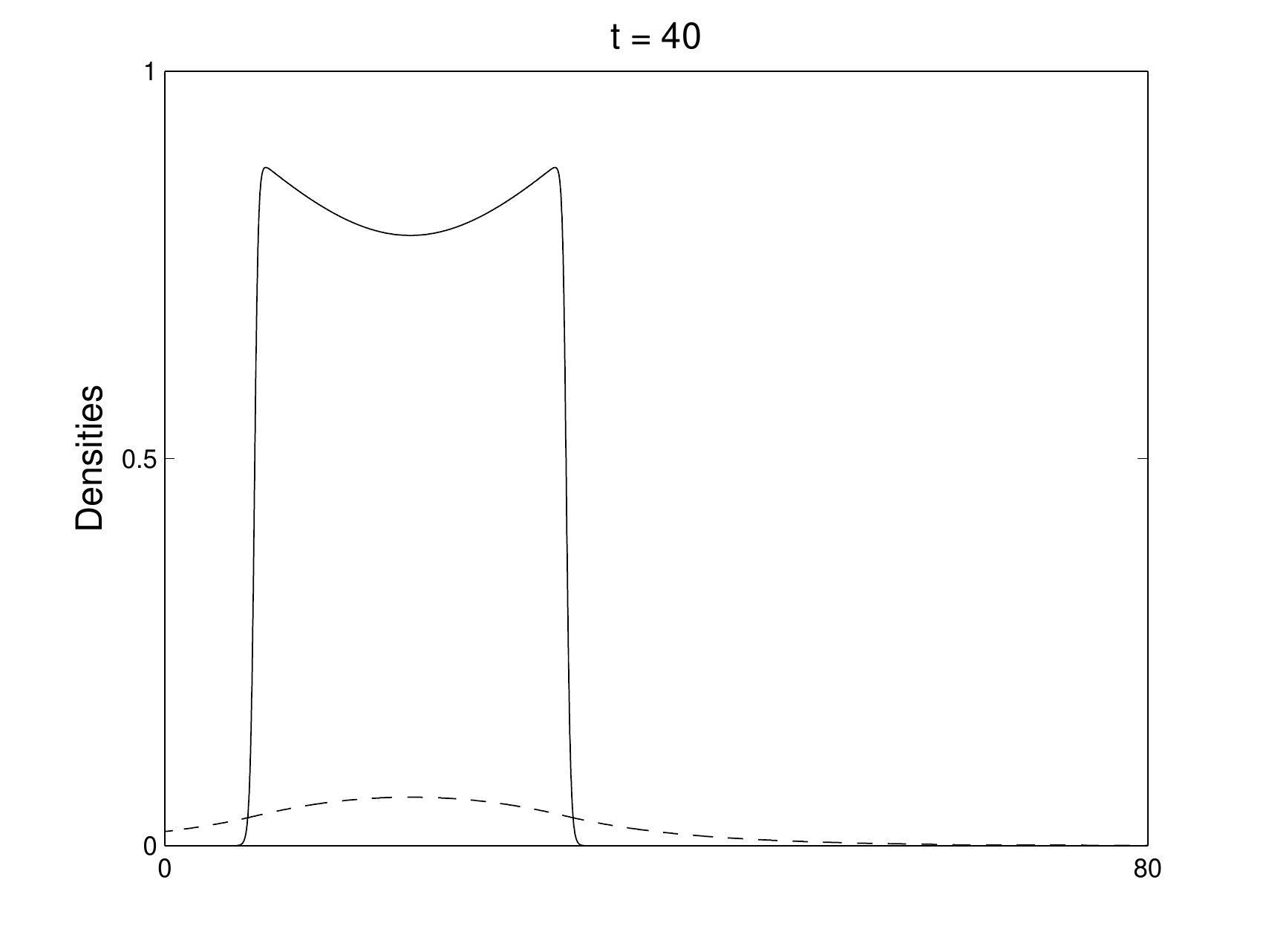}\\
\includegraphics[width=.48\textwidth]{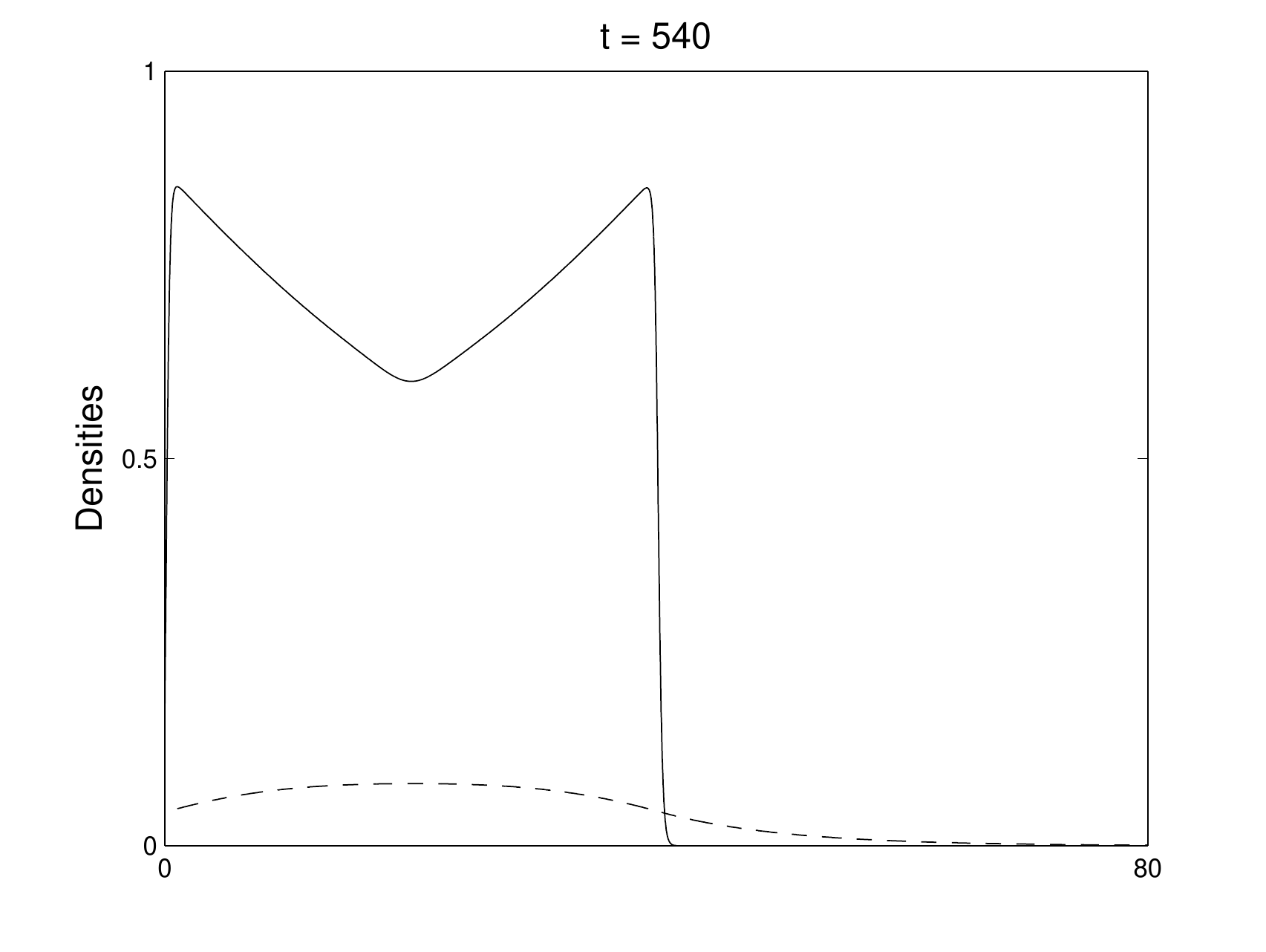}
\includegraphics[width=.48\textwidth]{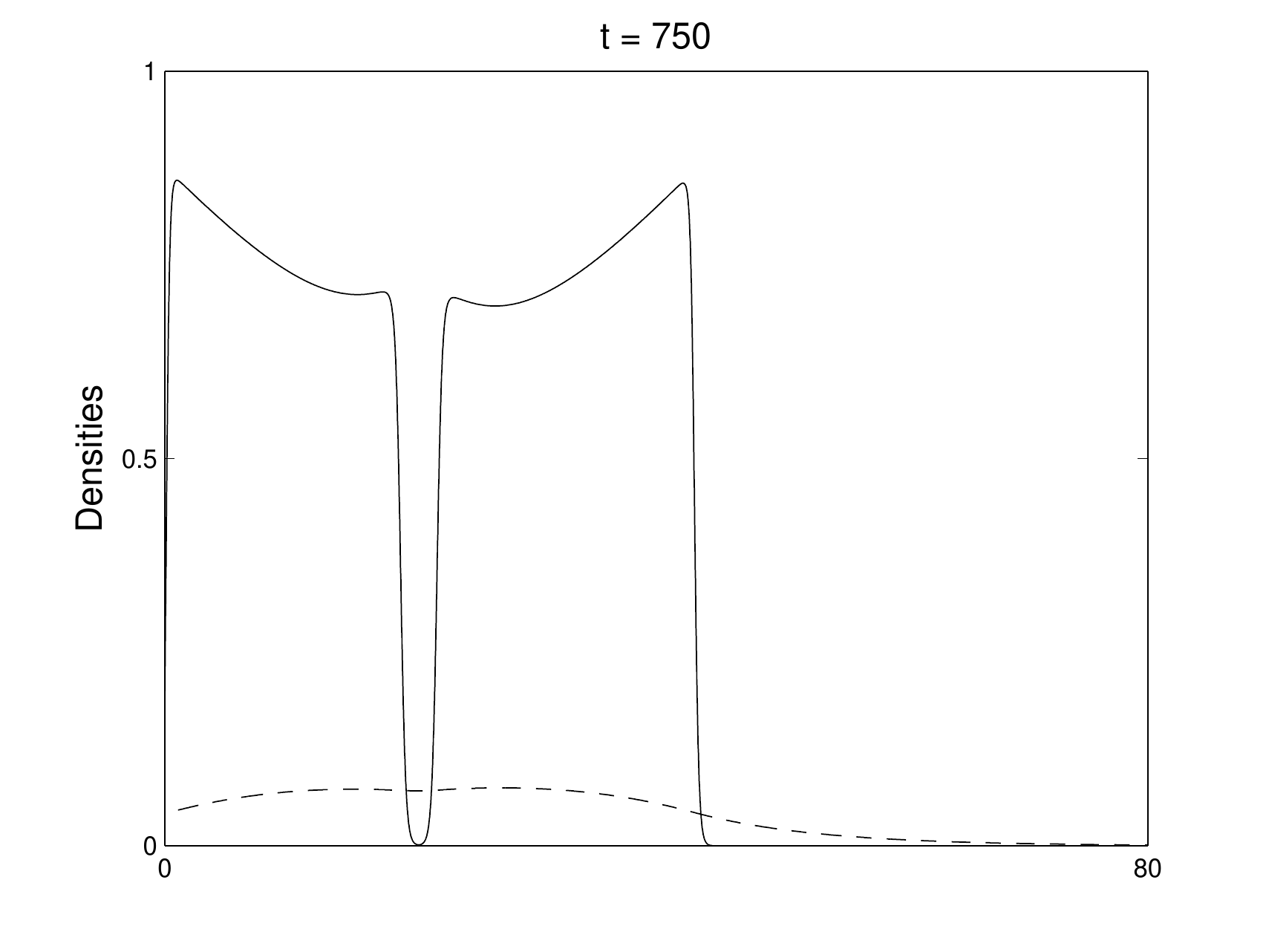}\\
\includegraphics[width=.48\textwidth]{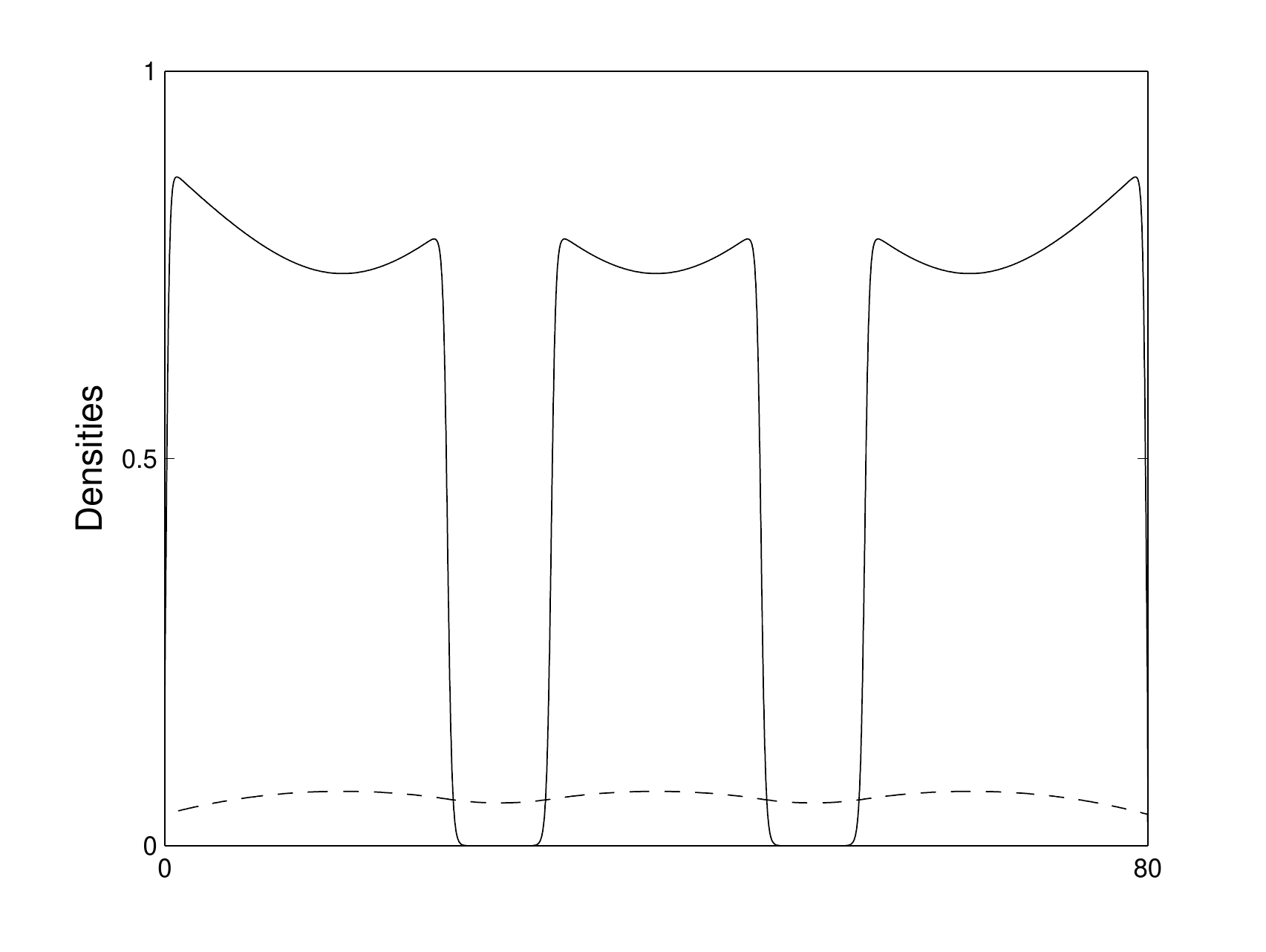}
\caption{Pattern formation for $D=10$, $\varepsilon = 0.1$, $a=2$, $f(N)=(1-N)(N-0.2)$, with initial outbreak of width $24$ in the left part of the domain. Solid and dashed curves represent host and parasitoid densities, respectively.}
\label{num ex}
\end{center}
\end{figure}

\section{Numerical Experiments}
\label{Numerical Experiments}
Our analytical results thus far have focused exclusively on the final state of the system. As will be seen shortly, the dynamics that lead to that state are at least as mathematically interesting and biologically important as the long-term spatial pattern itself. Moreover, the pattern has been described only in general terms up to this point. We now present a visual example of the process of pattern formation in the host-parasitoid model.

\subsection{Laplace Kernel}
We have explored the behavior of~(\ref{model N})-(\ref{model P}) with the Laplace kernel~(\ref{Laplace kernel}) through extensive numerical simulation, using a fast Fourier transform-based convolution algorithm, as described in~\cite{Andersen}, with various grid sizes. Figure~\ref{num ex} shows a typical pattern formation and the steps that lead to it. The initial conditions are shown. After oscillating somewhat, by $t=40$ the densities inside the outbreak closely obey the regular approximation $N(x) = f^{-1}(aP(x))$. This continues as the outbreak spreads until about $t=540$, when $N$ at the center of the outbreak falls below the level predicted by the regular approximation and becomes locally extinct by $t=750$. Eventually another local extinction occurs and the patches arrange into the pattern shown at the bottom of Figure~\ref{num ex}.

Numerical tests with an assortment of parameters and admissible growth functions often yield similar results -- the initial outbreak spreads until it seems to reach a critical width, at which point it divides into two outbreaks, which continue to spread and divide until filling the domain. The movement of the outbreak's front is entrained by the spread of the host, which is slow and steady, as expected from results in~\cite{Kot2}. Changing the magnitude of parasitoid dispersal $d$ is roughly equivalent to changing the length of the domain. The dependence of long-term behavior on other factors is detailed and analyzed below, and the mathematical arguments are fully in agreement with numerical results.

The calculation of~(\ref{matcond}) is analogous to calculations given in~\cite{Conway} and~\cite{Owen} for similar partial differential equation models. However, analogies to the further conditions derived in~\cite{Owen} appear impossible here, given the numerical evidence that interior transition layers form in pairs for this model.

\subsection{Other Kernels}
\label{Other Kernels}
Though our rigorous analysis has focused on the Laplace kernel, one of the strengths of integrodifference models is their adaptability to the varying dispersal behavior of organisms. This variability is reflected in the numerous possible dispersal kernels that can be used in~(\ref{model N})-(\ref{model P}), depending on the organisms under study.

\begin{figure}[htb]
\begin{center}
\includegraphics[width=.48\textwidth]{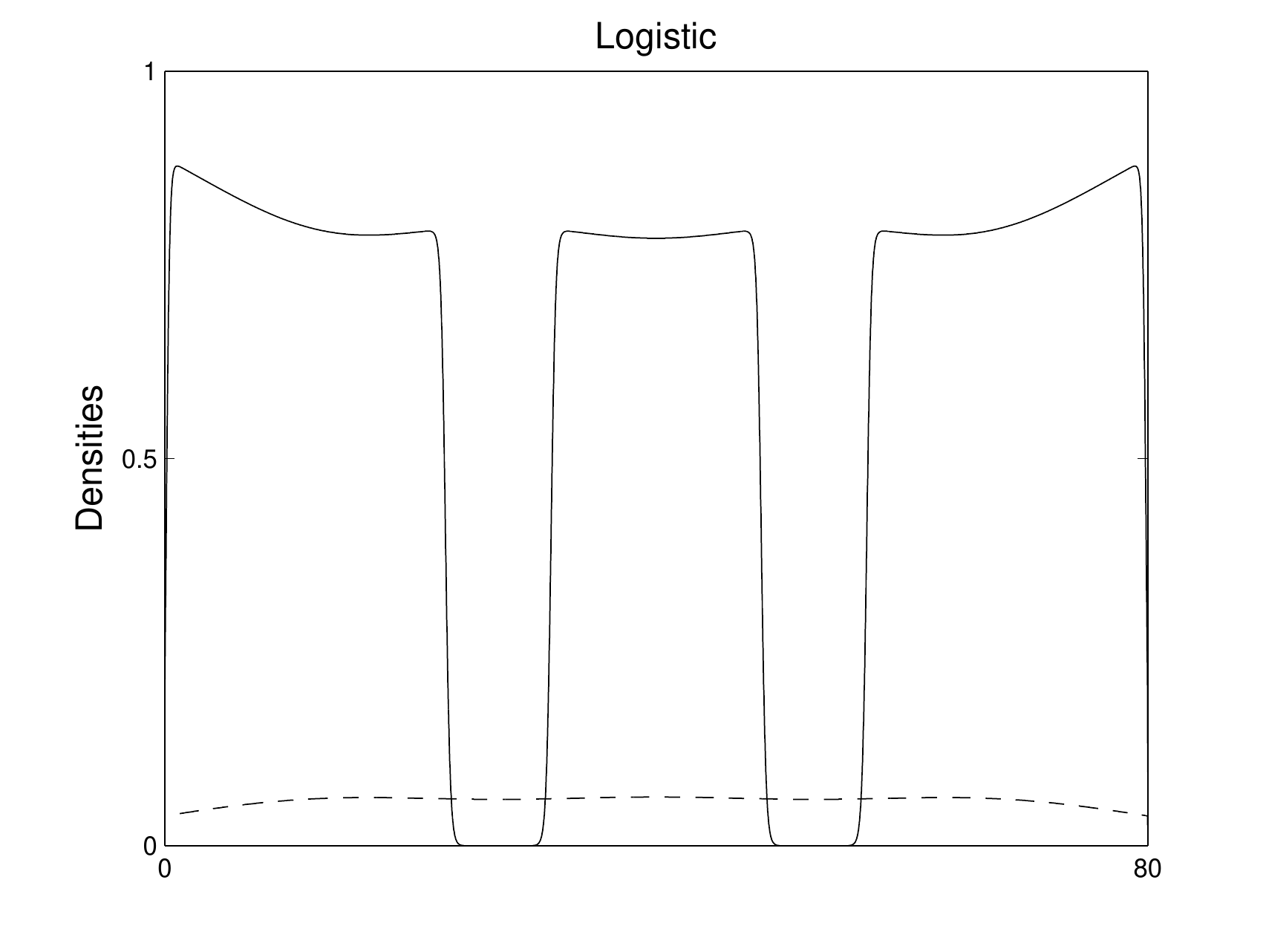}
\includegraphics[width=.48\textwidth]{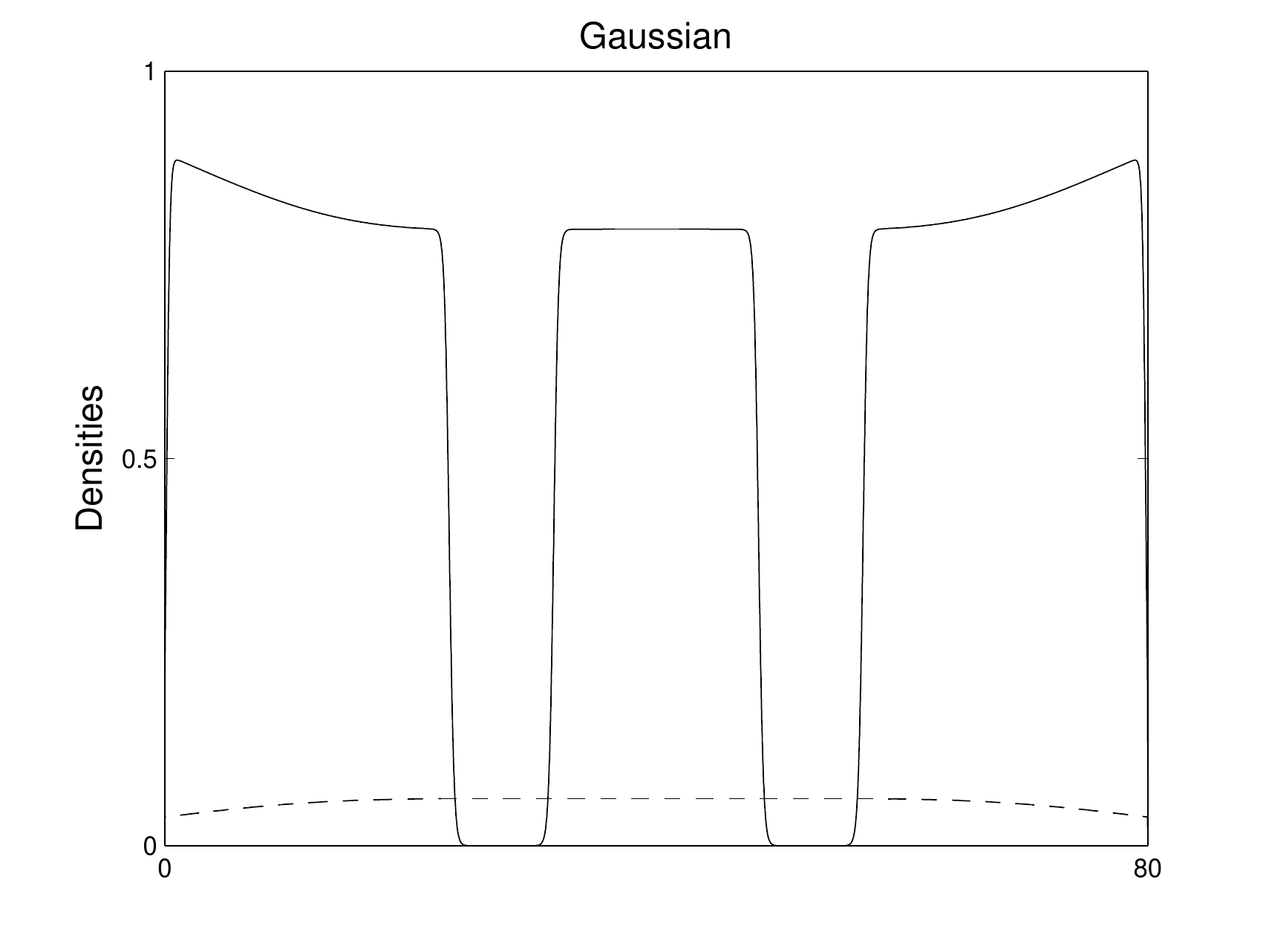}
\includegraphics[width=.48\textwidth]{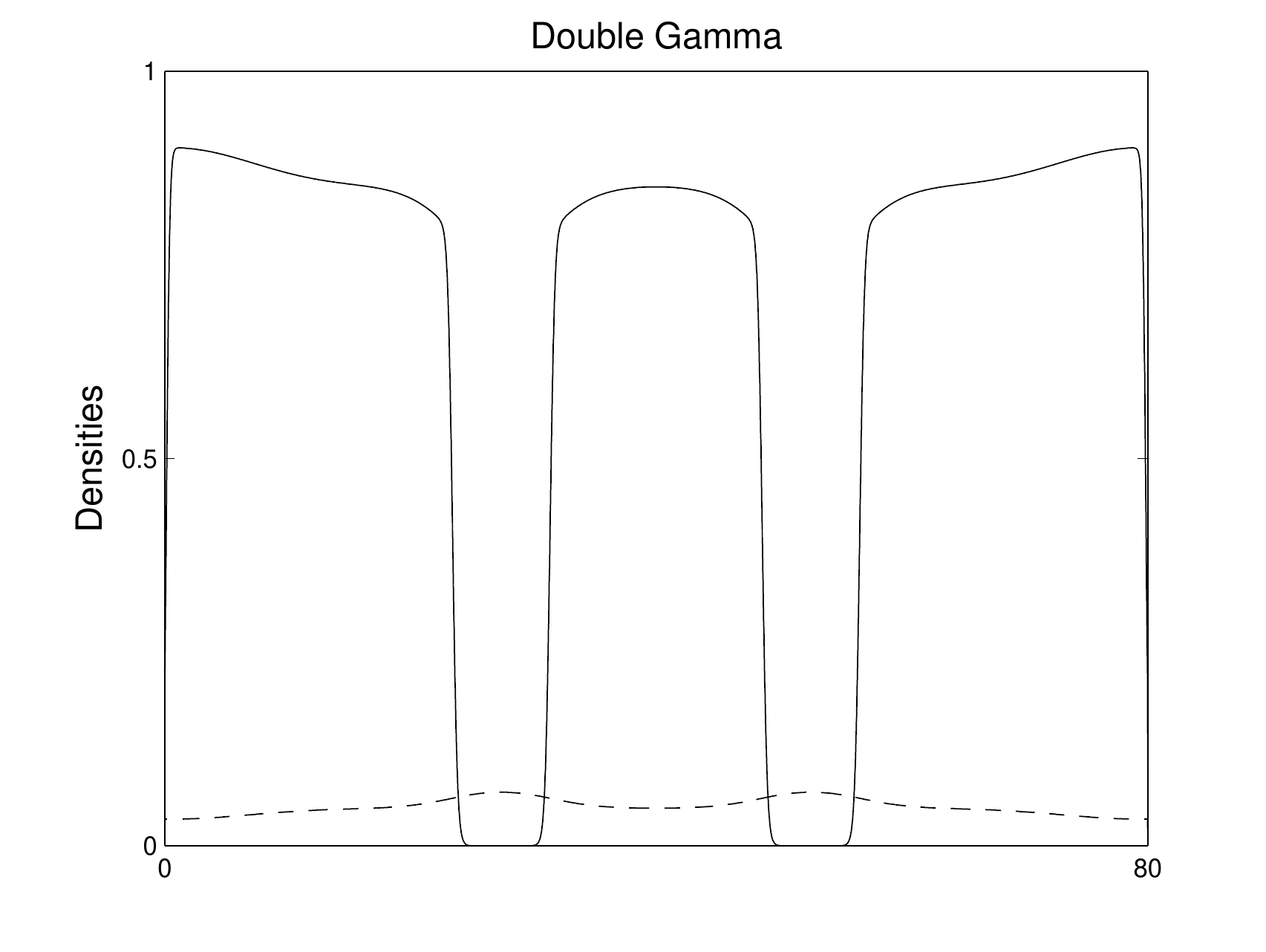}
\includegraphics[width=.48\textwidth]{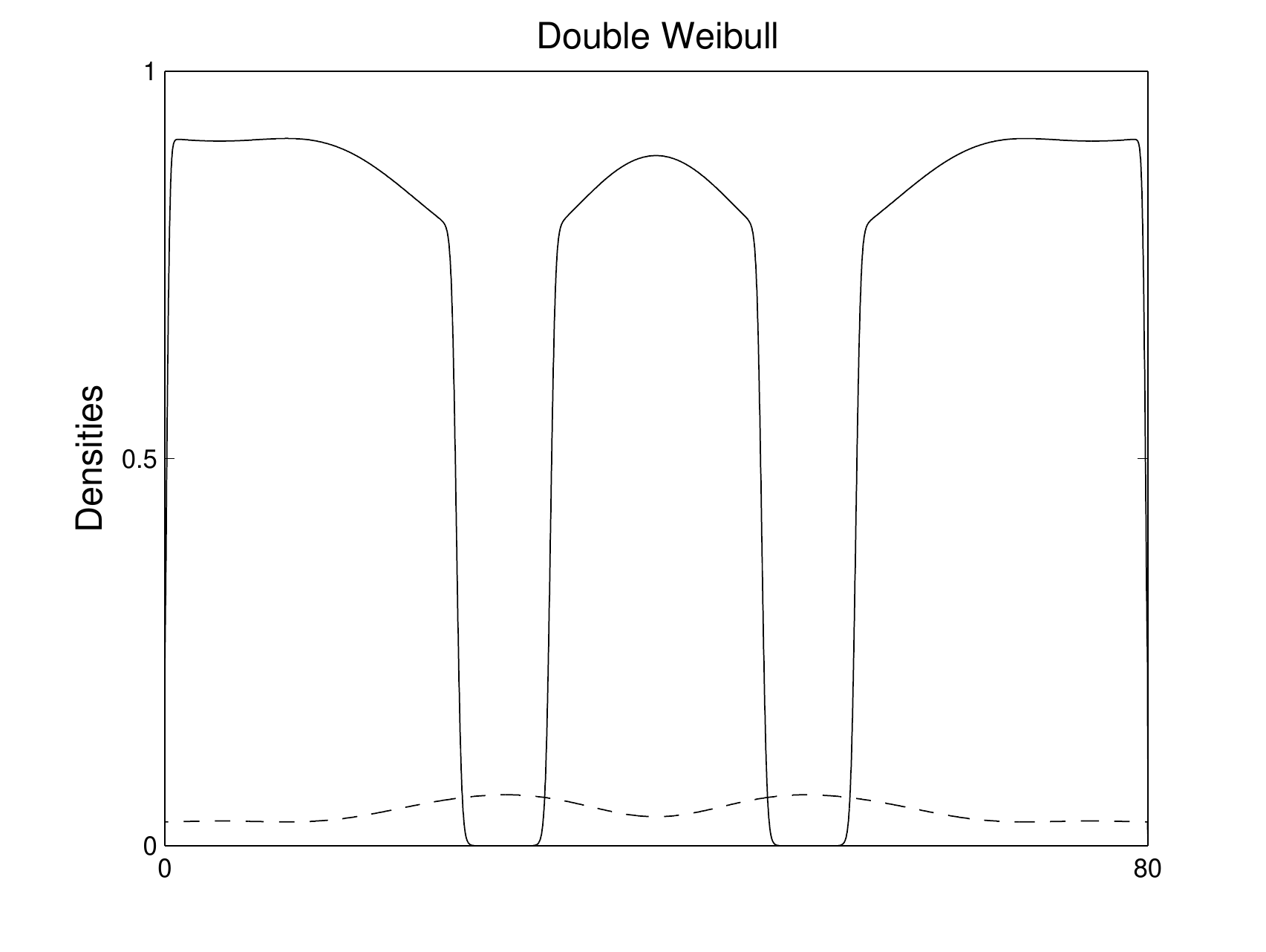}
\caption{Steady states for other dispersal kernels.}
\label{kernels}
\end{center}
\end{figure}

We have simulated the model~(\ref{model N})-(\ref{model P}) with a variety of dispersal kernels using the parameters given for Figure~\ref{num ex}. The variance of each kernel $k_d(x,y)$ was scaled to match the variance of the Laplace kernel~(\ref{Laplace kernel}) for each $d$. Figure~\ref{kernels} shows the numerical steady state result for the logistic kernel \[
k_d (x,y) = \frac{1}
{{4s_d }}\rm{sech} ^2 \left( {\frac{x-y}
{{2s_d }}} \right),
\]
the Gaussian kernel \[
k_d (x,y) = \frac{1}
{{\sigma _d \sqrt {2\pi } }}e^{{{ - (x-y)^2 } \mathord{\left/
 {\vphantom {{ - (x-y)^2 } {2\sigma _d^2 }}} \right.
 \kern-\nulldelimiterspace} {2\sigma _d^2 }}} ,
\]
the double gamma kernel\[
k_d (x,y) = (x-y)^2 \frac{{e^{ - \left| x-y \right|/\theta _d } }}
{{4\theta _d^3 }},
\]
and the double Weibull kernel
\[
k_d (x,y) = \frac{3}{{2\beta _d }}\left( {\frac{{x - y}}{{\beta _d }}} \right)^2 e^{ - \left| {x - y} \right|^3 /\beta _d ^3 }.
\]
For these last two kernels the shape parameter -- referred to as $\alpha$ in~\cite{Lockwood} -- is taken to be 3, yielding bimodal functions with $k_d(x,x)\equiv0$.

The locations of extinction areas seem to depend remarkably little on the specifics of the kernel used. Even the double gamma kernel, derived from biological assumptions leading to dramatically different dispersal behavior~\cite{Neubert}, and also the double Weibull kernel, result in a very similar long-term spatial configuration. The principle difference evident with these and the Gaussian kernel -- in short, the non-leptokurtic kernels -- is that interior patches lack the characteristic increase in host density near the edge of outbreaks, noted in~\cite{Hastings2}.

\section{Routes to Heterogeneity}
We now attempt to understand why pattern formation in our host-parasitoid model~(\ref{model N})-(\ref{model P}) proceeds as described above, through a process of slowly spreading host outbreaks repeatedly punctuated by outbreak divisions. To that end we will employ arguments of varying mathematical rigor.

\subsection{The Nonspatial Model}
\begin{figure}[htbp]
\begin{center}
\includegraphics[totalheight=.4\textheight]{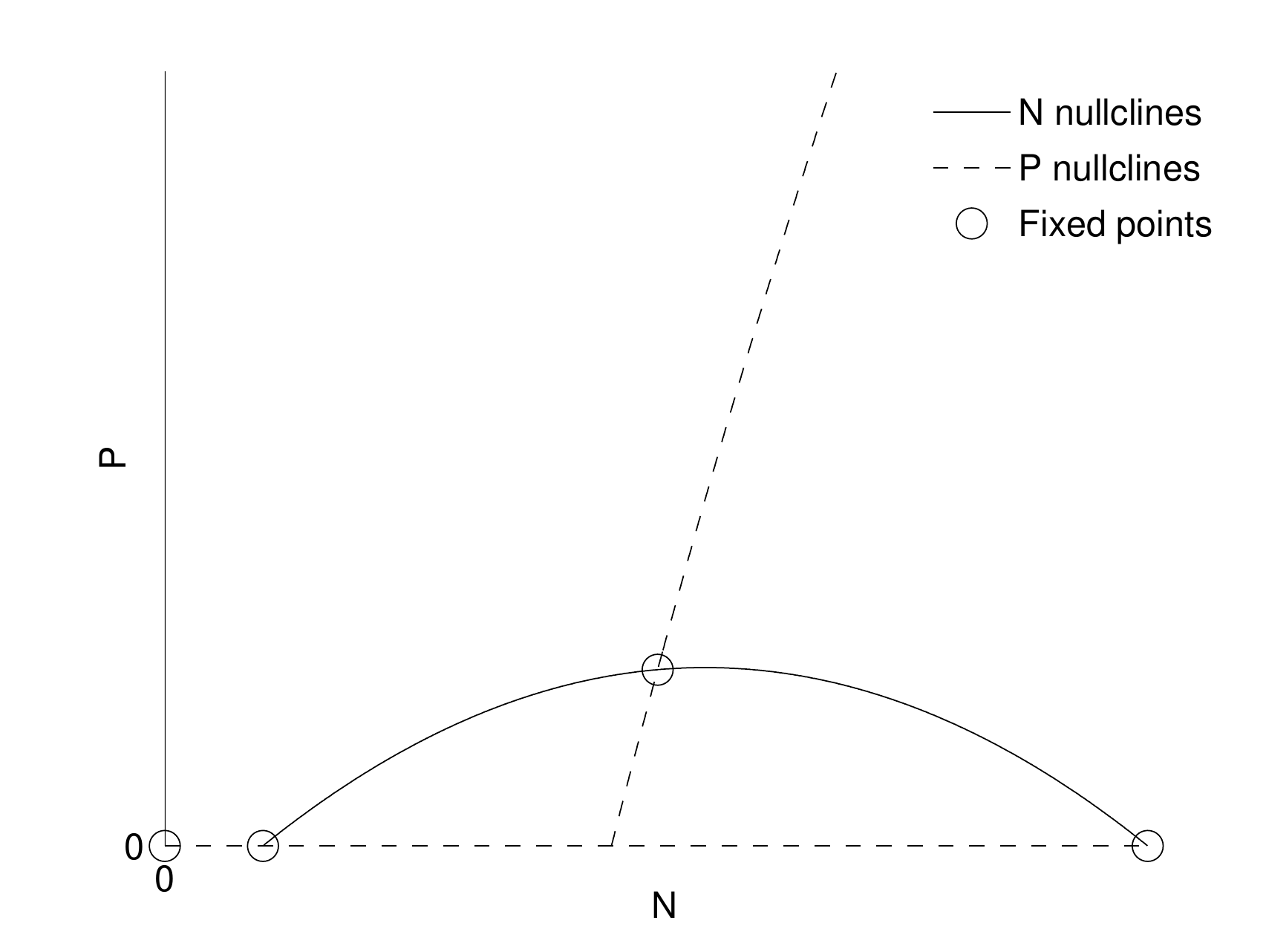}
\caption{Nullclines of the nonspatial model.}
\label{fp}
\end{center}
\end{figure}

Before further consideration of the dynamical behavior of our spatial model, it will be useful to bear in mind some of the properties of the underlying nonspatial equations~(\ref{no dispersal N})-(\ref{no dispersal P}). In the $N$-$P$ plane, there are four nullclines of the difference equations (see Figure~\ref{fp}). Of particular interest is the curve $P=f(N)/a$, a nullcline for $N$, and the curve $N = {P \mathord{\left/
 {\vphantom {P {\left( {1 - e^{ - aP} } \right)}}} \right.
 \kern-\nulldelimiterspace} {\left( {1 - e^{ - aP} } \right)}}$, a nullcline for $P$. 

There are three fixed points along the $P=0$ axis, and we will make the biologically reasonable assumption that there is exactly one more along the nullcline for $N$. This point lies in the positive quadrant only if $a>1$ (we take the carrying capacity of $N$ to be 1). When $a$ is not much larger than 1, the fixed point is oscillatory but stable. As $a$ increases, the nullcline for $P$ is shifted up relative to the nullcline for $N$ and the fixed point moves to the left along the nullcline for $N$ and loses stability in a Hopf bifurcation. The other fixed points on the nullcline for $N$ are never stable. If the Allee effect is strong -- i.e., the lesser zero of $f(N)$ is positive -- then the fixed point at the origin is stable. Otherwise it is not.

\subsection{Evolution to the Steady State}
Returning to the spatial model, consider initial conditions given by a single, narrow patch of host and parasitoid in the interior of an otherwise empty domain. Inside the patch, because of low dispersal the host density approximately obeys~(\ref{no dispersal N}). The parasitoid, on the other hand, is subject to higher dispersal into the empty part of the domain, where it is lost. Referring to Figure~\ref{fp}, the nullcline for $P$ is essentially shifted downward and the interaction at any point inside the outbreak is stabilized, approaching a point on the nullcline for $N$ relatively quickly.

As the outbreak slowly spreads, the effect of parasitoid dispersal on local dynamics in its interior is reduced and the nullcline for $P$ approaches its true position, as determined by the parameter $a$. The interaction at any point inside the outbreak is entrained by the movement of the intersection of the curves, slowly moving left along the nullcline for $N$ toward the nonspatial equilibrium.

If the nonspatial equilibrium lies to the left of the maximum of the nullcline for $N$, as the local interaction moves past the maximum it loses the stability temporarily imparted by the dispersal of the parasitoid. If the Allee effect is strong, as discussed above, the origin is a stable fixed point, and the local dynamics approach it. The first point to approach extinction in this way is the center of the outbreak, since the parasitoid density is least affected by dispersal loss there. As extinction is approached, the parasitoid density is maintained away from zero by an influx from nearby points that are not yet approaching extinction. So the host density at the center of the outbreak is driven to zero; parasitoids that disperse to the center are lost. In effect, two separate outbreaks form, and as they spread the process described above is repeated near the center of each of them. This spreading and dividing continues until spread is halted at the edges of the domain.

\begin{figure}[htbp]
\begin{center}
\includegraphics[totalheight=.4\textheight]{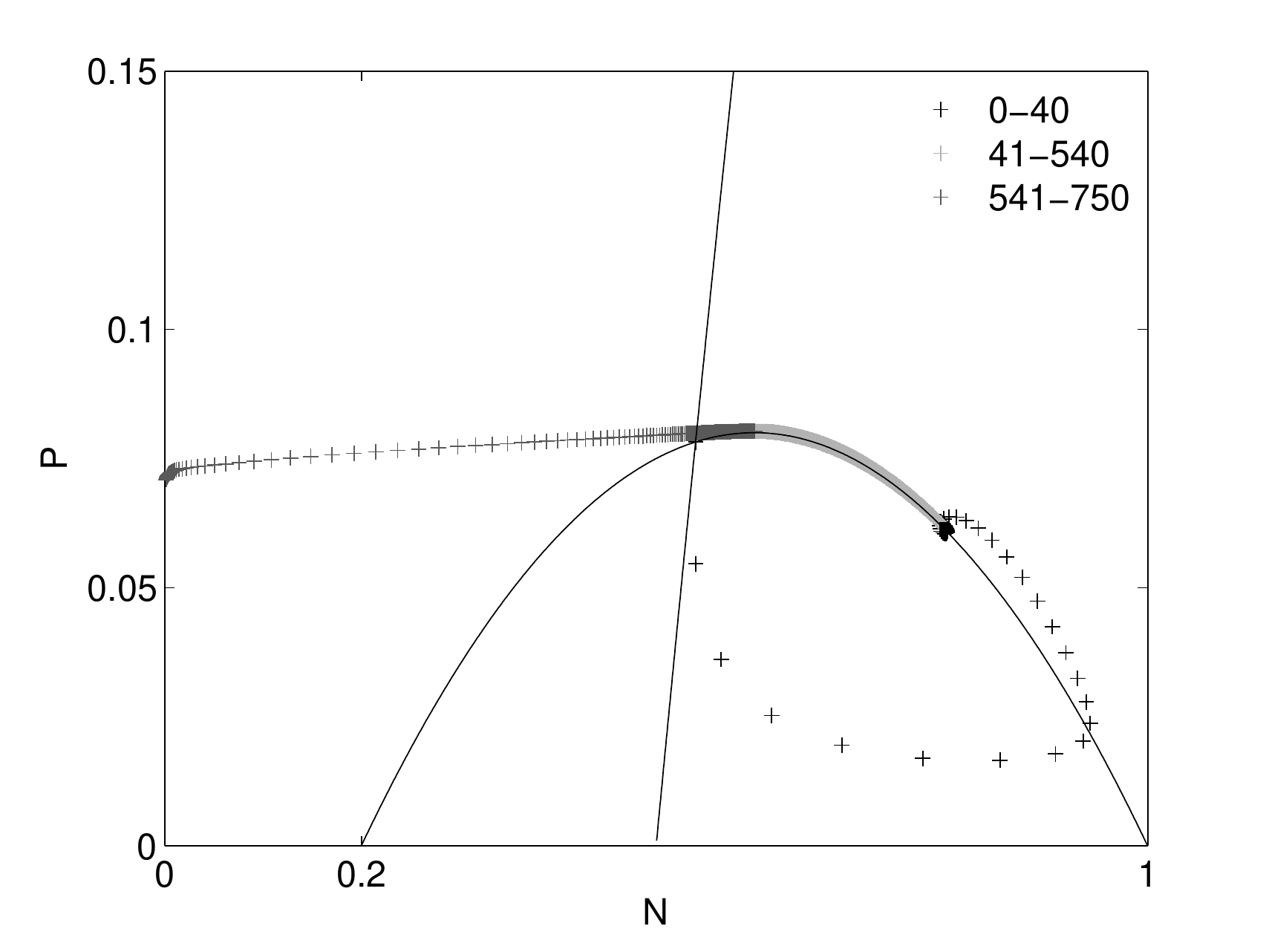}
\caption{Time series at outbreak center for Figure~\ref{num ex}.}
\label{irad_12}
\end{center}
\end{figure}

In Figure~\ref{irad_12}, a time series is plotted for the densities at a single point near the center of the outbreak shown in Figure~\ref{num ex}, from $t=0$ to $750$. 40 time steps are required to reach the temporary fixed point on the nullcline for $N$, after which the densities move along the curve for 500 time steps until reaching its peak and departing for the $P$ axis.

\subsection{Dependence on Parameters and Initial Conditions}
As noted above, the formation of a heterogeneous steady state requires a nonspatial coexistence fixed point to the left of the maximum in the nullcline for $N$ (or equivalently, to the left of the maximum of $f(N)$) and a strong Allee effect. With a weak Allee effect, transient pattern formation is observed during numerical studies, due to the high-amplitude oscillatory nature of the map and its slowing near the origin's saddle.

For a reasonably low Allee threshold, a positive coexistence fixed point results for large values of $a$. It is possible to achieve stable spatial patterns for such $a$, well beyond even the range in which there are limit cycles in the nonspatial model. As is clear from~(\ref{matcond}), $a$ and $P_0$ are inversely proportional for any given $f$. For $P_0$, the density of the parasitoid at the transitions, to decrease, the fraction of the domain in which coexistence occurs must be reduced. This is observed in numerical simulations, as coexistence regions become narrower and farther apart with increasing $a$. In some cases, these regions occur far from the domain edges, with complete extinction near the boundary.

\begin{figure}[htb]
\begin{center}
\includegraphics[totalheight=.4\textheight]{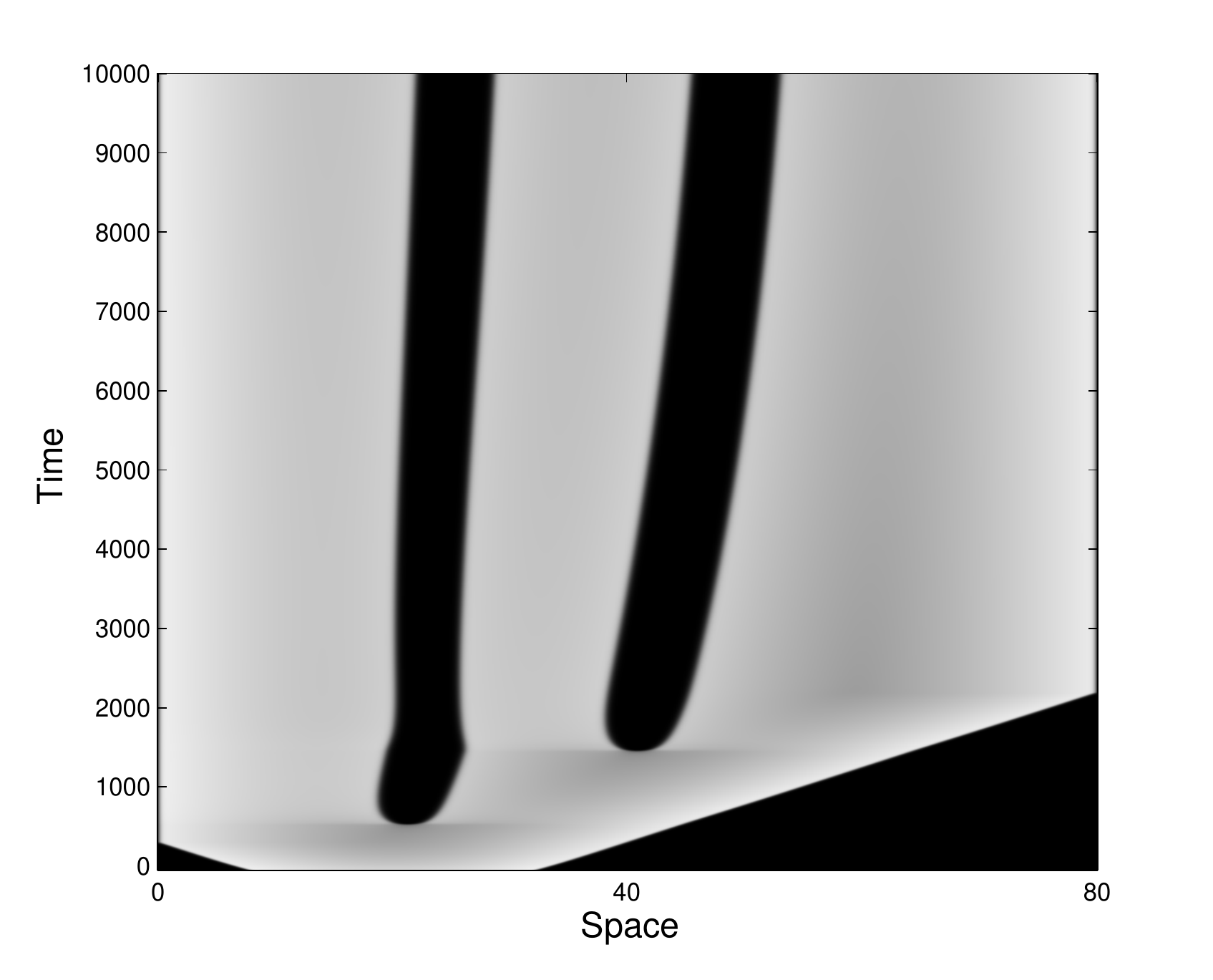}
\caption{Time series for $N$ on the whole domain for parameters and initial conditions from Figure~\ref{num ex}. Host density is shown in shades of gray; i.e., black represents absence of hosts.}
\label{ts}
\end{center}
\end{figure}

Figure~\ref{ts} gives a holistic view of the formation of spatial heterogeneity. Parameters and initial conditions are as before. Note that the essential features of the stable spatial structure are formed within the first $2000$ time steps, although the extinction subdomains take longer -- much longer than we have plotted -- to reach their final positions.

The asymptotic behavior of the model may sometimes depend on the initial conditions. If the initial conditions are extensive enough (for example, a nonzero constant throughout the domain), limit cycles can result. Numerical work indicates that, as one example, with $D=10$, $\varepsilon = 0.1$, $a=2.1$, and $f(N)=(1-N)(N-0.1)$, stable heterogeneity arises for initial conditions of limited extent while stable limit cycles arise for an extensive initial state. 

Certainly for an infinite domain, if the nonspatial dynamics lead to limit cycles, as with the parameters just given, then the spatial result for uniform initial conditions will be the same limit cycle at each point in the domain. It is not altogether surprising, then, that extensive enough contiguous nonzero initial conditions will lead to limit cycles with such parameters. This dependence on initial conditions can be understood somewhat more rigorously by consideration of the linear stability properties of~(\ref{model N})-(\ref{model P}).

\begin{figure}[htbp]
\begin{center}
\includegraphics[height=.48\textwidth,angle=90]{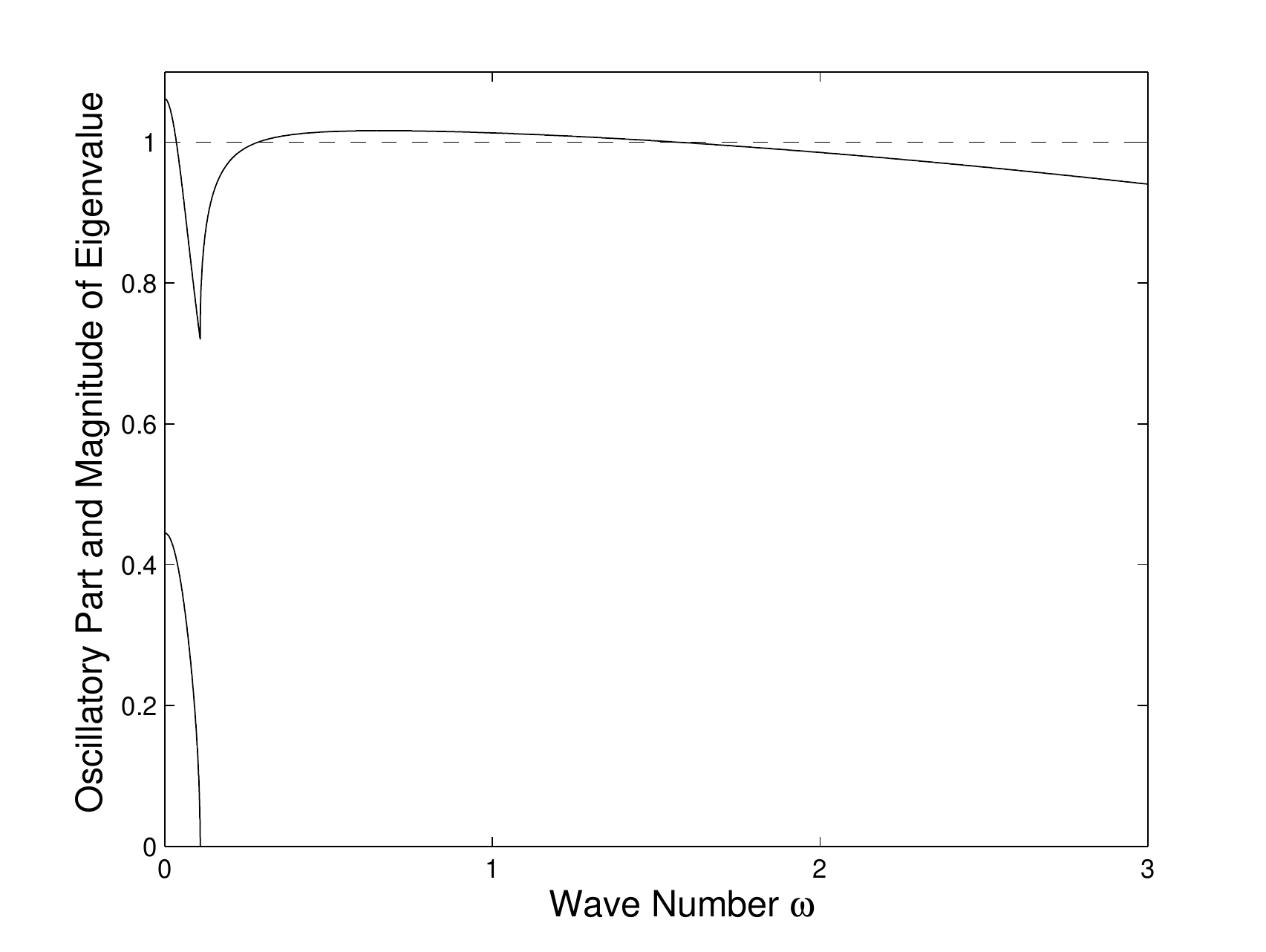}
\includegraphics[width=.48\textwidth]{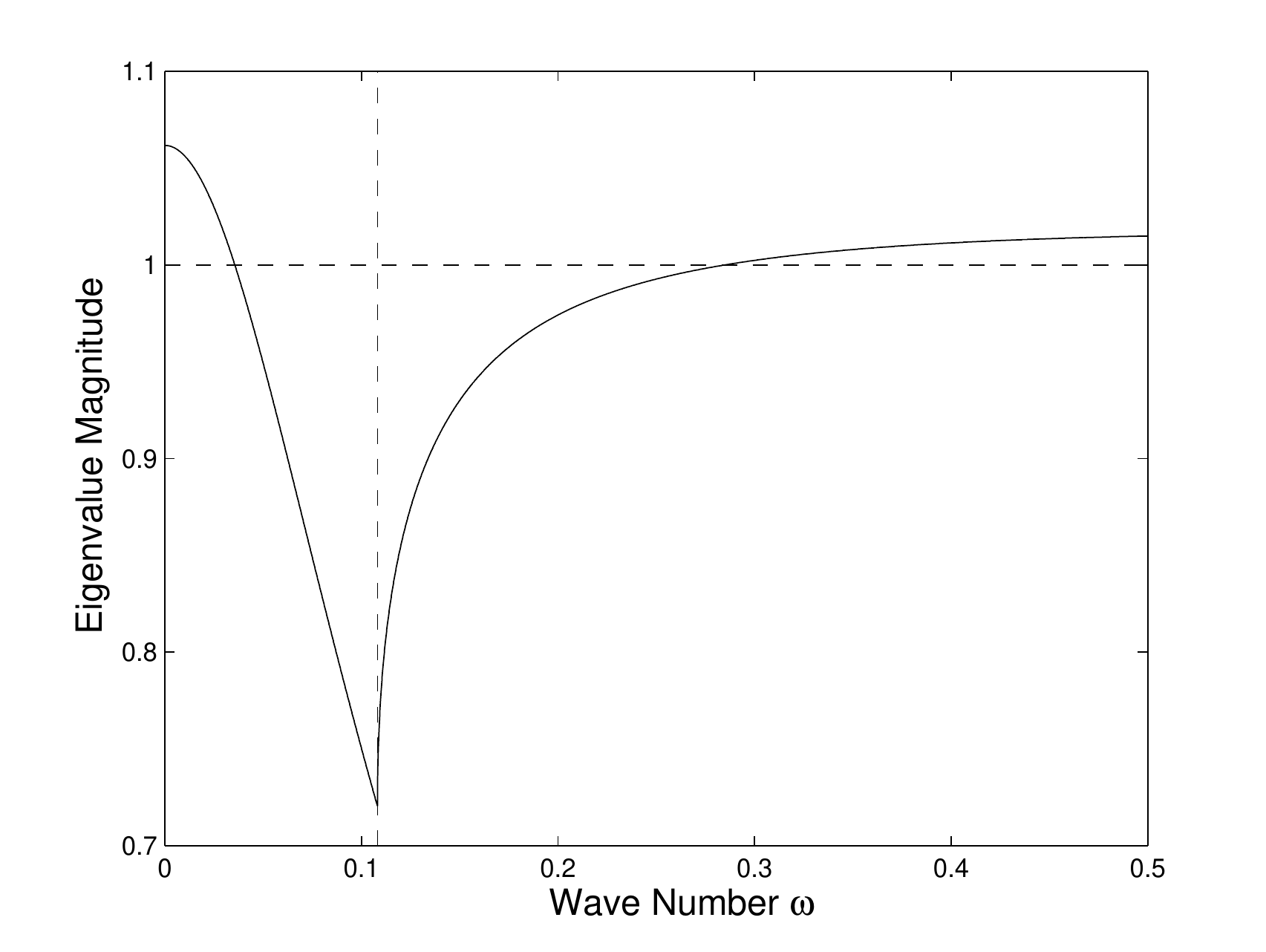}
\caption{Linear stability where outcome is dependent on initial conditions.}
\label{fourier1}
\end{center}
\end{figure}

\subsection{Linear Stability}
\label{Linear Stability}
The linear stability analysis of a system of integrodifference equations is explained and elaborated in~\cite{Neubert}, and carried out on a model similar to ours in~\cite{Wilson}. Briefly, the Jacobian of~(\ref{no dispersal N})-(\ref{no dispersal P}) at the coexistence fixed point is \[
\mathbf J = \left( {\begin{array}{*{20}c}
   {1 + Nf'(N)} & { - aN}  \\
   {1 - e^{ - f(N)} } & {aNe^{ - f(N)} }  \\
\end{array}} \right),
\]
where $N$ is the host density at coexistence. The characteristic function of $k_d(x)=\frac{1}{2d}e^{-|x|/d}$, the Laplace convolution kernel, is $\hat k_d(\omega)={\frac{1}{{1 + d ^2 \omega ^2 }}}$. So, much of the behavior of~(\ref{model N})-(\ref{model P}) is related to the eigenvalues $\lambda$ of\[
\mathbf {KJ} = \left( {\begin{array}{*{20}c}
   {\left( {1 + \varepsilon ^2 \omega ^2 } \right)^{ - 1} } & 0  \\
   0 & {\left( {1 + D^2 \omega ^2 } \right)^{ - 1} }  \\
\end{array}} \right)\left( {\begin{array}{*{20}c}
   {1 + Nf'(N)} & { - aN}  \\
   {1 - e^{ - f(N)} } & {Nf(N)}  \\
\end{array}} \right)
\]
for each wave number $\omega$. Figure~\ref{fourier1} shows the properties of these eigenvalues for the parameters given before, for which long-term behavior depends on the extent of initial conditions. In the left panel of Figure~\ref{fourier1}, the magnitude of the largest eigenvalue (upper solid curve) and the magnitude of its imaginary part (lower left solid curve) are plotted. Eigenvalues of $\mathbf K (\omega) \mathbf J$ are strictly real for $\omega$ greater than about $0.1$. In the right panel, more detail is given for small $\omega$. In each panel the line $|\lambda |=1$ is shown for clarity, and in the right panel the cutoff for complex $\lambda$ is shown.

There are two intervals for $\omega$ in which instability is found. Perturbations away from the coexistence densities with low $\omega$, or equivalently, long wavelengths, grow oscillatorily. Some perturbations with higher $\omega$, or shorter wavelengths, grow monotonically. We now relate this to the behavior of the model. Extensive initial conditions of the kind we have used, viewed as a perturbation from the coexistence densities, have a considerable component of long wavelength. Such components grow much faster for the parameters under consideration than components of short wavelength. But monotonically growing perturbations of short wavelength are precisely the cause of extinction subdomains between outbreaks. For these parameters, with extensive enough initial conditions, the wild oscillation of the long wavelengths disrupts the growth of perturbations with shorter wavelengths predicted by linearizing near the coexistence densities.

\begin{figure}[htbp]
\begin{center}
\includegraphics[width=.48\textwidth]{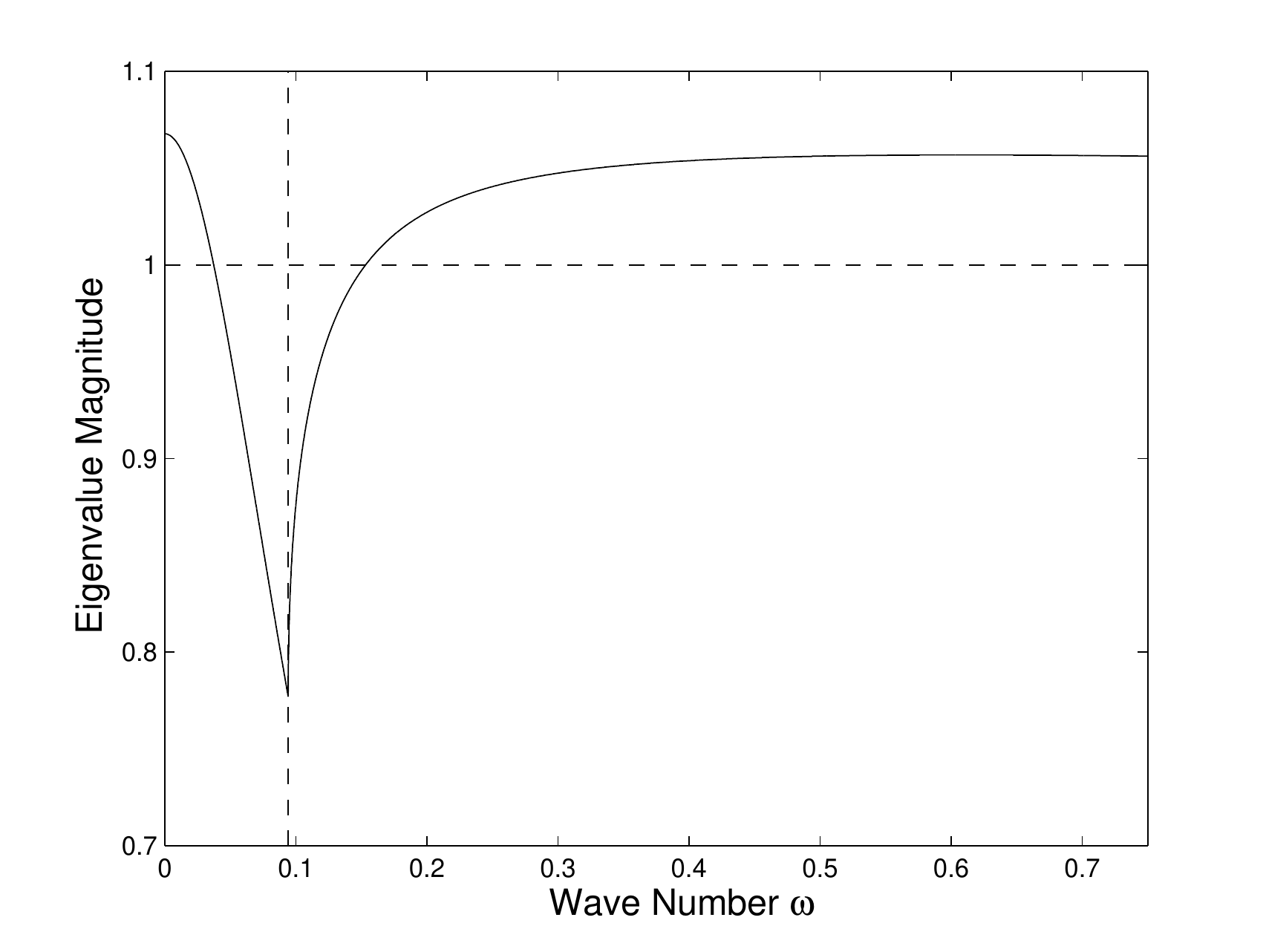}
\caption{Linear stability where spatial patterns are robust.}
\label{fourier3}
\end{center}
\end{figure}

Figure~\ref{fourier3} is analogous to Figure~\ref{fourier1} for the parameters that we have used in the rest of this paper (see Figure~\ref{num ex}). The maximum magnitude of the real eigenvalues is comparable to that of the complex ones, allowing spatial patterns to form more readily.

Note that $\mathbf K (\omega) \mathbf J\to \mathbf J$ as $\omega\to0$ and thus, not surprisingly, the behavior of the model under spatially extended perturbations matches that of the nonspatial model~(\ref{no dispersal N})-(\ref{no dispersal P}). Now, for $\mathbf K (\omega) \mathbf J$ with real eigenvalues, the larger of these, $\lambda_+$, is given by
\[
2\lambda _ +   = \hat k_\varepsilon  (\omega )J_{11}  + \hat k_D (\omega )J_{22}  + \sqrt {\left( {\hat k_\varepsilon  (\omega )J_{11}  - \hat k_D (\omega )J_{22} } \right)^2  + 4\hat k_\varepsilon  (\omega )\hat k_D (\omega )J_{12} J_{21} } .
\]

So if $J_{12} J_{21} < 0$, which is certainly the case in our victim-exploiter system, we have $\lambda_+  \le \hat k_\varepsilon  (\omega )J_{11}$. \label{Limit Properties}Note that $\hat k_\varepsilon (\omega ) \to 1_-$ as $\omega \varepsilon \to 0$, so $\lambda_+  \le J_{11}$, and for fixed $\omega$ this upper bound becomes more accurate as $\varepsilon$ decreases. Note also that $\hat k_D (\omega ) \to 0_ +$ as $\omega D \to \infty$, so the bound becomes more accurate for fixed $\omega$ as $D$ increases. In summary, if $1/D \ll \omega  \ll 1/\varepsilon$, then $J_{11} = 1 + Nf'(N)$ is a tight upper bound on real eigenvalues, so since $N>0$, instabilities leading to pattern formation are only possible if $f'(N) > 0$. This means that, as deduced earlier from simulations, the nonspatial coexistence equilibrium must lie to the left of the maximum of $f(N)$.


The parameters we have used are such that if $\omega\approx1$, then $0.1 = 1/D \ll \omega  \ll 1/\varepsilon = 10$. Hence in Figures~\ref{fourier1} and~\ref{fourier3}, real eigenvalues are largest near $\omega=1$. Also, as the nonspatial equilibrium is moved farther left relative to the maximum of $f(N)$, as in Figure~\ref{fourier3}, these eigenvalues become larger since $f(N)$ is concave.

\subsection{A Bound on Patch Radius}
\label{A Bound on Patch Radius}
Consider a single small outbreak centered around some point $x_c$ in the interior of the domain, far from the boundary, in which local dynamics have reached their temporary equilibrium (i.e. they lie on the nullcline for $N$). As the patch spreads it may reach a width at which it divides at its center. As discussed above, this occurs when the value of $N$ at the center maximizes $f(N)$. Let us call this value $N_{\rm{max}}$, and let $P_{\rm{max}}=f(N_{\rm{max}})/a$. 

We have from~(\ref{model P}) that\[
P_{t + 1} (x_c ) = \int\limits_\Omega  {k_D (x_c ,y)N_t (y)\left( {1 - e^{ - aP_t (y)} } \right)dy}.
\]
Turning again to the Laplace kernel, in order for the patch not to divide it must be that \[
\frac{1}{{2D}}\int\limits_\Omega  {e^{-\left| {x_c  - y} \right|/D} N_t (y)\left( {1 - e^{ - aP_t (y)} } \right)dy}  < P_{\max } .
\]
Since $N_t=0$ outside the patch, we have 
\begin{equation}
\frac{1}{{2D}}\int\limits_{x_c  - R}^{x_c  + R} {e^{-\left| {x_c  - y} \right|/D} N_t (y)\left( {1 - e^{ - aP_t (y)} } \right)dy}  < P_{\max } 
\label{single patch} 
\end{equation}
where $R$ is the radius of the patch. Inside the patch, since the kernel is leptokurtic and we are considering the moment at which the patch divides, we will make the approximation $N_t\approx N_t(x_c) = N_{\rm{max}}$ and $P_t\approx P_t(x_c) = P_{\rm{max}}$. Then \[
\frac{1}{{2D}}\int\limits_{x_c  - R}^{x_c  + R} {e^{ - \left| {x_c  - y} \right|/D} N_t (y)\left( {1 - e^{ - aP_t (y)} } \right)dy} 
\]
\[
 \approx \frac{1}{{2D}}\int\limits_{x_c  - R}^{x_c  + R} {e^{ - \left| {x_c  - y} \right|/D} N_{\max } \left( {1 - e^{ - aP_{\max } } } \right)dy} 
\]
\[
 = N_{\max } \left( {1 - e^{ - aP_{\max } } } \right)\left( {1 - e^{ - R/D} } \right).
\]
So the patch divides approximately when \[
N_{\max } \left( {1 - e^{ - aP_{\max } } } \right)\left( {1 - e^{ - R/D} } \right) = P_{\max },
\]
which is when
\begin{equation}
R = D\ln \frac{{N_{\max } \left( {1 - e^{ - aP_{\max } } } \right)}}{{N_{\max } \left( {1 - e^{ - aP_{\max } } } \right) - P_{\max } }}.
\label{radius eq} 
\end{equation}

Note that this approximation was derived for a single non-oscillatory patch in an otherwise empty domain. A patch with neighbors should have a somewhat smaller radius since the assumption leading to~(\ref{single patch}) does not apply. This is why it is seen in numerical simulations that patches at the boundary are wider than interior patches; at the boundary the assumption that $N_t=0$ outside the patch is partially true. The approximation~(\ref{radius eq}) does not hold for patches with sustained internal oscillations since, as explained previously, such a patch may be arbitrarily large.

\begin{figure}[htbp]
\begin{center}
\includegraphics[width=.48\textwidth]{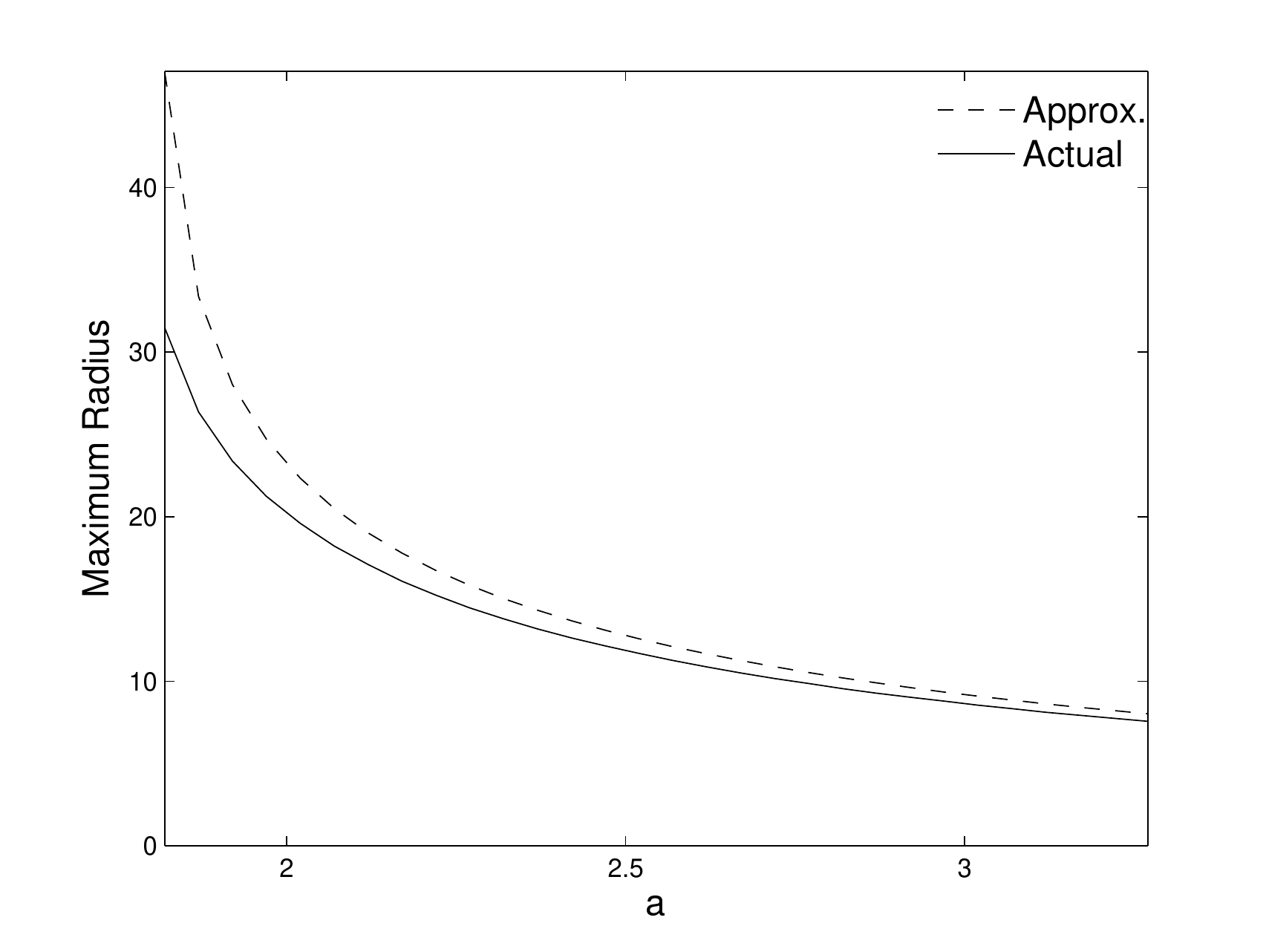}
\includegraphics[width=.48\textwidth]{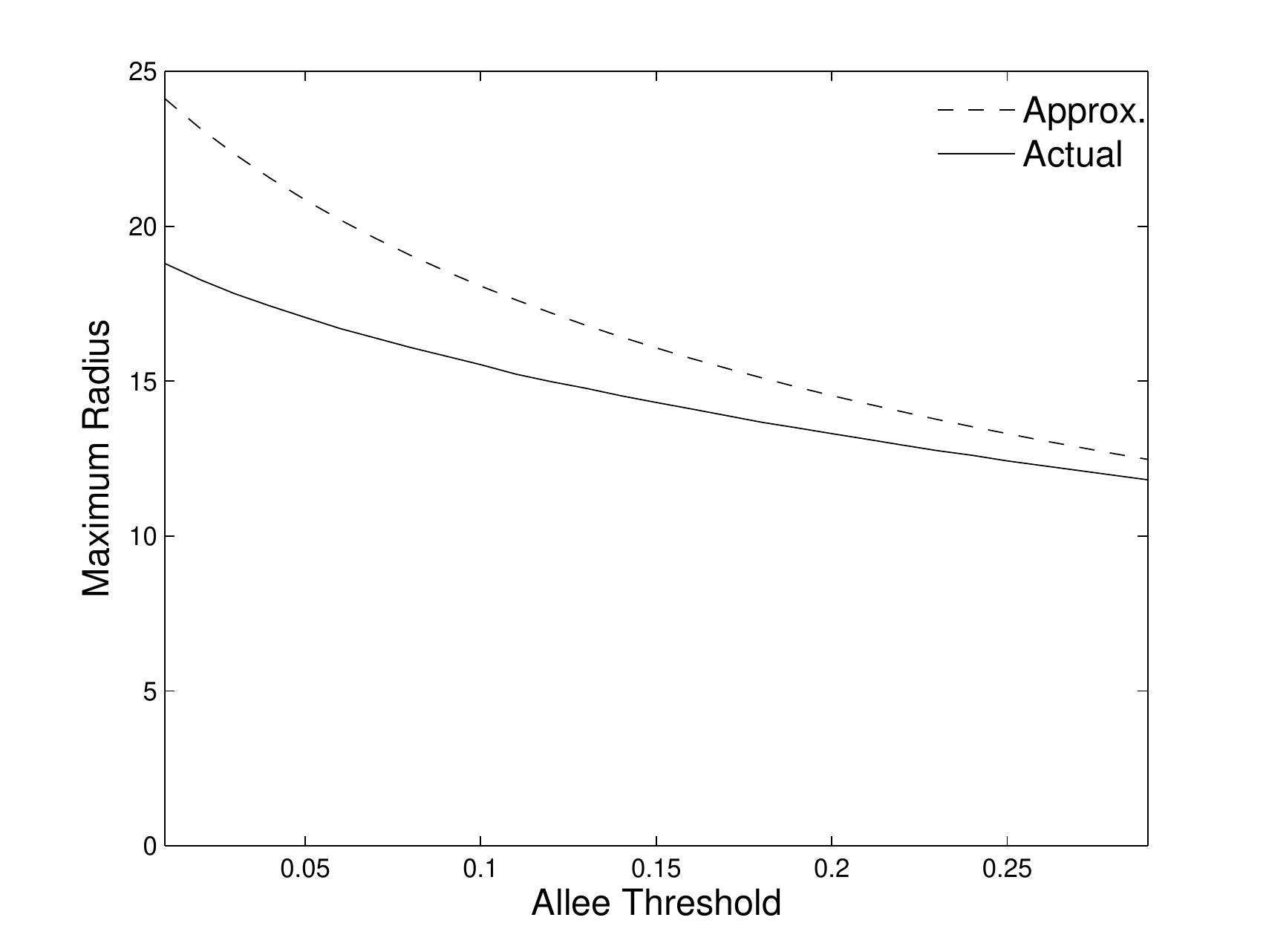}
\caption{Comparison of approximation and actual maximum radius. On the left, $f(N)=(1-N)(N-0.2)$. On the right, $a=2.4$ and $f$ is quadratic and scaled to have maximum $0.2$.}
\label{bounds}
\end{center}
\end{figure}

Figure~\ref{bounds} demonstrates the accuracy of~(\ref{radius eq}) by comparing it to the actual maximum size obtainable in simulations for various values of $a$ and the Allee threshold. It turns out that in most cases~(\ref{radius eq}) should be an upper bound on $R$, because $N_t \left( {1 - e^{ - aP_t } } \right) \ge N_{\max } \left( {1 - e^{ - aP_{\max } } } \right)$ when $N_t$ varies more than $P_t$, as near the outbreak's center during a division (see Figure~\ref{num ex} at $t=540$). 

\section{Discussion}
\label{Discussion 1}
We have observed that our model~(\ref{model N})-(\ref{model P}) for a discrete-time host-parasitoid system can exhibit stable spatial structure. As in previous and somewhat analogous models, the distribution of the host within outbreaks is qualitatively similar to that seen in the field, with increased density near the edge~\cite{Hastings2}. We have shown that this pronounced spatial patterning requires a strong Allee effect in the host and fairly unstable underlying (nonspatial) dynamics.  In such a situation, low host motility actually acts as a survival or stabilizing mechanism. The stabilizing influence of spatial heterogeneity -- as well as the likely local instability of most host-parasitoid systems -- is discussed in~\cite{Keeling}.

Spatial patterns form through a process of spread and division. Their formation does not greatly depend on the specific dispersal behavior of organisms, other than that the dispersal of the host should be comparatively short. In fact, the necessary conditions we derived for internal layers in Section~\ref{Allee result} are independent of the parasitoid dispersal kernel. We saw numerically that patterns may form with other forms of both kernels, although the distribution within an outbreak does depend on the nature of dispersal. This dependence could be grounds for further analytical investigation. It is also worth noting that the dispersal kernel properties used on page~\pageref{Limit Properties} are not unique to the Laplace kernel.

Increased parasitoid efficacy results in more sparse patchiness. That is, coexistence areas become narrower and farther apart as a means of continuing persistence of the overall system. For a given set of parameters, a bound on the possible width of coexistence patches may be found. Beyond that width, a patch will spontaneously divide and the new patches will move apart. In the next chapter we will further explore this behavior and its implications.

While the spatial patterns we have observed and explained are striking and may be unfamiliar in the context of spatial host-parasitoid interactions, very similar patterns are found for a continuous-time predator-prey model in~\cite{Mimura}. A somewhat similar singular perturbation analysis and parameter requirements are given. However, as observed in~\cite{Neubert}, less tenable biological assumptions are invoked to achieve those patterns. Also, the focus of that work is on Turing instability. For the examples given in this paper, Turing instability is impossible because when the nonspatial model has a stable equilibrium, patterns cannot form, and this is true for most reasonable parameters and growth functions. Rather, our examples demonstrate diffusion-mediated stability. Lastly, as noted before, our model is far more relevant to tussock moths and their enemies than is that of~\cite{Mimura}, or other continuous-time predator-prey models such as those in~\cite{Conway} and~\cite{Owen}.

As mentioned at the outset of this paper, a model similar to ours, but with density-independent host growth, is briefly considered in~\cite{Wilson}. The linear stability analysis in that work resembles ours, suggesting the possibility of pattern formation. However, stable patterns of the kind seen in this paper are impossible there. Firstly, depression of the parasitoid nullcline by dispersal cannot achieve stability of local dynamics in that model. Secondly, in that stability analysis there are no real eigenvalues with magnitude greater than $1$. Numerical simulations of that system show the same oscillatory instability that occurs in the underlying nonspatial model.

Another model related to the tussock moth system is given in~\cite{McCann}. It is discrete not only in time, but also in space, with three patches coupled by dispersal. This simple model allows for precise analysis of the conditions required for spatial patterns analogous to those described here. The result of that analysis is that the host nullcline must must have a ``hump," as with an Allee effect, just as we found in Section~\ref{Allee result}.

It is fruitful to compare our model's behavior to the dynamics of tussock moths in the field. High tussock moth density near the edge of outbreaks~\cite{Hastings2} suggests that dispersal of parasitoids is leptokurtic, with a dispersal kernel shaped roughly like that given by~(\ref{Laplace kernel}). Failure of experimental moth invasions outside preexisting outbreaks~\cite{Maron} is consistent with an Allee effect and with numerical simulations of our model. Even more solid evidence is given in~\cite{Harrison2}, where it is reported that the nonlinear functional response of predatory ants induces an inverse density dependence in tussock moths. The details of this are considered in the next chapter.

Further questions are raised that have not been answered to date by field observations. The stability of the nonspatial interaction is of interest. That is, would the tussock moth density fall to zero or oscillate if parasitoids were not lost to dispersal? Also, in an ample and unoccupied habitat, how fast do new moth outbreaks spread? How sparse are old outbreaks? Is there spontaneous division of large outbreaks in the field?

In addition to the implications of our model discussed above, there is another detail that should be carefully noted. As seen in Figure~\ref{ts} and discussed elsewhere, because of low host motility, outbreaks spread fairly slowly. But once the basic spatial pattern is formed, extinction areas travel orders of magnitude more slowly. It is unreasonable to expect the final steady state ever to be reached under such conditions, especially in the actual tussock moth and lupine system that motivates our model, where the habitat is threatened by other organisms and may change dramatically in the course of a few years. However, qualitative knowledge of the state toward which the system is moving helps to understand and possibly predict its transient behavior.

%
%
%
%

    \newchapter{The Effects of Habitat Quality} 
{The Effects of Habitat Quality on Patch Formation} 
{The Effects of Habitat Quality on Patch Formation} 
\label{The Effects of Habitat Quality}
\section{Introduction}
The trophic cascade brought about by soil-dwelling nematodes, whereby the productivity of the lupine is protected to some extent from root-feeding enemies, is an important point of interest to the tussock moth system modeled above. We do not endeavor to formulate a unified model for the entire food web, from the nematode to the tussock moth's enemies (and also including the ants which prey on tussock moth pupae), in this work. However, since some of the connections between the various components likely only operate in one direction -- i.e. the tussock moth has no effect on either nematode or ant dynamics -- it may suffice to explore the possible background conditions, in the form of model parameters and other assumptions. While we ignored outside factors in the preceding chapter, certain qualities of the larger natural system have a large influence on the striking patterns observed there.

We found certain general necessary conditions for pattern formation in Chapter~\ref{Spontaneous Patchiness}. We will now see how the feasibility of these conditions is affected by the nature of the tussock moth's habitat. We first investigate in detail the shape of the host nullcline given the environmental factors present in the natural system. We then turn to the possible results of varying the habitat quality, which determines, in part, the position of the host nullcline.

\section{An Allee Effect Induced by Saturating Predation}

\subsection{Modeling the Growth of the Host}
Much progress has been made recently in connecting mechanistic reasoning about continuous-time in-year dynamics to apparently phenomenological models of discrete-time growth. In \cite{Eskola}, a population consisting of adults and juveniles is considered. Depending on the kind of interactions assumed to occur between and within these classes in continuous time during the year, various year-to-year maps are derived for the density of adults. Likewise, a continuous consumer-resource interaction for in-year dynamics is used in \cite{Geritz} to derive, by varying the specifics of the consumer, another assortment of discrete-time models. Similarly, a ``semi-discrete" model for a host-parasitoid system is considered in \cite{Singh}; parasitism is modeled as a continuous process throughout the larval stage of the host, and all other processes are modeled in discrete time.

There are a number of natural mechanisms that may lead to growth with an Allee effect~\cite{Courchamp,Stephens}. Some of these mechanisms, such as the difficulty in finding a mate at low densities, have been considered explicitly in discrete-time models~\cite{Dennis,Boukal,Schreiber}. The methodology of using a continuous-time model for in-year dynamics to derive a discrete map is extended to Allee growth mechanisms in \cite{Eskola2}, where various mate-finding behaviors are considered, along with cannibalism. 

It has become apparent that predator-prey interactions can induce Allee effects in a variety of situations. In one example, a model predator that attacks only a certain stage of its prey has been shown to exhibit a growth threshold~\cite{vanKooten}. As outlined in \cite{Gascoigne}, saturating predation can induce an Allee effect in a prey population. This general mechanism is considered in \cite{Schreiber} for a discrete-time growth model.

In this section, we focus on a population which is victim to stage-specific, generalist, saturating predation. In particular, we model a species with a pupal stage during which it is susceptible to such a predator. This describes certain members of the family Lymantriidae, such as the tussock moth studied in Chapter~\ref{Spontaneous Patchiness} and the gypsy moth. As mentioned previously, pupae of the tussock moth in coastal California have been found to be subject to attack by ants, a generalist predator~\cite{Harrison2}; pupae of the gypsy moth are preyed upon by mice~\cite{Elkinton}. We will formulate two models for such a victim population, the first an alteration of that given in \cite{Schreiber}, and the second using an approach similar to \cite{Eskola}, \cite{Geritz}, and \cite{Singh}. 

We wish to find a map of the form $N_{t+1}=F\left(  N_t  \right)$, where $N_t$ represents the density of the adult female population at the beginning of the winter before year $t$. These adults lay eggs which overwinter and become larvae. The larvae feed and are subject to density dependent survival as a consequence of limited resources. The survivors pupate; during the pupal stage, they are victim to a generalist, saturating predator. The females that survive this stage to become adults comprise the population $N_{t+1}$, lay eggs, and so forth.

The functional response of the saturating predator, or the rate at which it consumes prey per capita, will be taken as 
\begin{equation}
\frac{m}{1+sx},
\label{sat pred}
\end{equation}
where $x$ is the prey density, $m$ represents the strength of predation (incorporating the constant density of predators), and $s$ is related to the handling time or satiation of the predator.

\subsection{Escape Probability}
A discrete-time model for an Allee effect induced by saturating predation is formulated in~\cite{Schreiber}. Based on the functional response (\ref{sat pred}), the probability of an individual in a population with density $N$ escaping predation for the entire season is \[
 I\left(N_t\right)=e^{-\frac{m}{1+sN}},
\]
so with Ricker density dependence, the year-to-year map is 
 \begin{equation}
N_{t+1} = I\left(N_t\right)N_t e^{r\left(1-\frac{N_t}{K}\right)}=N_t e^{r\left(1-\frac{N_t}{K}\right)  -  \frac{m}{1+sN_t}},
\label{schreiber map}
\end{equation}
where $K$ is the carrying capacity absent predation. This map, however, assumes predation throughout the season, simultaneous with density dependence. In the case considered in the present paper, resource limitation occurs first: \[
Y_0 = N_t e^{r\left(1-\frac{N_t}{K}\right)},
\]
where $Y_0$ is the density of larvae that begin the pupal stage. Then the probability of escaping predation during pupation is of the form \[
e^{-\frac{m}{1+sY_0}}, 
\]
so the year-to-year map is 
\begin{equation}
N_{t+1} = N_t e^{r\left(1-\frac{N_t}{K}\right)  -  \frac{m}{1+sN_t e^{r\left(1-{N_t/K}\right)}}}.
\label{escape map}
\end{equation}

\subsection{Continuous Predation}
The formulation above only evaluates the consumption rate (\ref{sat pred}) at the beginning of the pupal stage. If, instead, predation is considered as a continuous process during the pupal stage, we model it as 
\[
\dot Y =  - \frac{{mY}}{{1 + sY}},
\]
where $\dot Y$ is the rate of change of the pupal population. Integrating and applying the initial condition, at the end of the pupal stage we have $\ln Y + sY = \ln Y_0  + sY_0  - m$. Here we have either taken the length of the pupal stage to be 1, or equivalently rescaled $m$. This relation between $Y$ and $Y_0$ may also be written 
\begin{equation}
Ye^{sY}  = Y_0 e^{sY_0  - m} ,
\label{pupal eq}
\end{equation}
in which form we see that the relation defines an increasing function $Y\left(Y_0\right)$ for positive $Y_0$. In fact, from (\ref{pupal eq}) it immediately follows that $Y\left( Y_0 \right) = \frac{1}{s}W\left(   sY_0 e^{sY_0  - m}    \right)$, where $W$ is the Lambert W function~\cite{Corless} (see also~\cite{Hamback} for a similar ecological application of $W$).

Again we model the larval stage with Ricker dynamics:
\[
Y_0 = N_t e^{r\left( {1 - \frac{N_t }{K}} \right)} .
\]
The Ricker growth equation has been derived in at least two mechanistic contexts -- cannibalism~\cite{Ricker,Eskola} and a limited resource~\cite{Geritz}. We use it here because it is a plausible model for the development of eggs (produced in numbers proportional to $N_t$), through the resource-limited larval stage, to the beginning of pupation. The year-to-year map is 
\begin{equation}
N_{t + 1}  = Y\left( {N_t e^{r\left( {1 - N_t /K} \right)} } \right) = \frac{1}{s}W\left( {sN_t e^{r\left( {1 - N_t /K} \right) + sN_t e^{r\left( {1 - N_t /K} \right)}  - m} } \right).
\label{continuous map}
\end{equation}

\subsection{Properties of the Maps}
\label{Properties of the Map}
Recall the form of the Nicholson-Bailey model (\ref{no dispersal N})-(\ref{no dispersal N}). In Chapter~\ref{Spontaneous Patchiness}, we found certain necessary conditions under which the host-parasitoid integrodifference model (\ref{model N})-(\ref{model P}) can exhibit dramatic and stable spatial patterns, a typical example of which is shown in Figure~\ref{num ex}. To attain these patterns, the host must exhibit a strong Allee effect and the coexistence fixed point (see Figure~\ref{fp}) of the nonspatial model (\ref{no dispersal N})-(\ref{no dispersal P}) must lie to the left of the maximum in the host's nullcline. Since the host nullcline is $P=f(N)/a$, the shape of $f$ directly determines the shape of the nullcline. For small $P$, the parasitoid nullcline can be accurately linearized as $P\approx 2N-2/a$. It crosses the $N$ axis at $N=1/a$ and rises steeply.

\begin{figure}[htp]
     \centering
     \subfigure[Map (\ref{escape map}) with $r = 1$, $m = 1.15$, $s = 2$]{
          \label{a}
          \includegraphics[width=.45\textwidth]{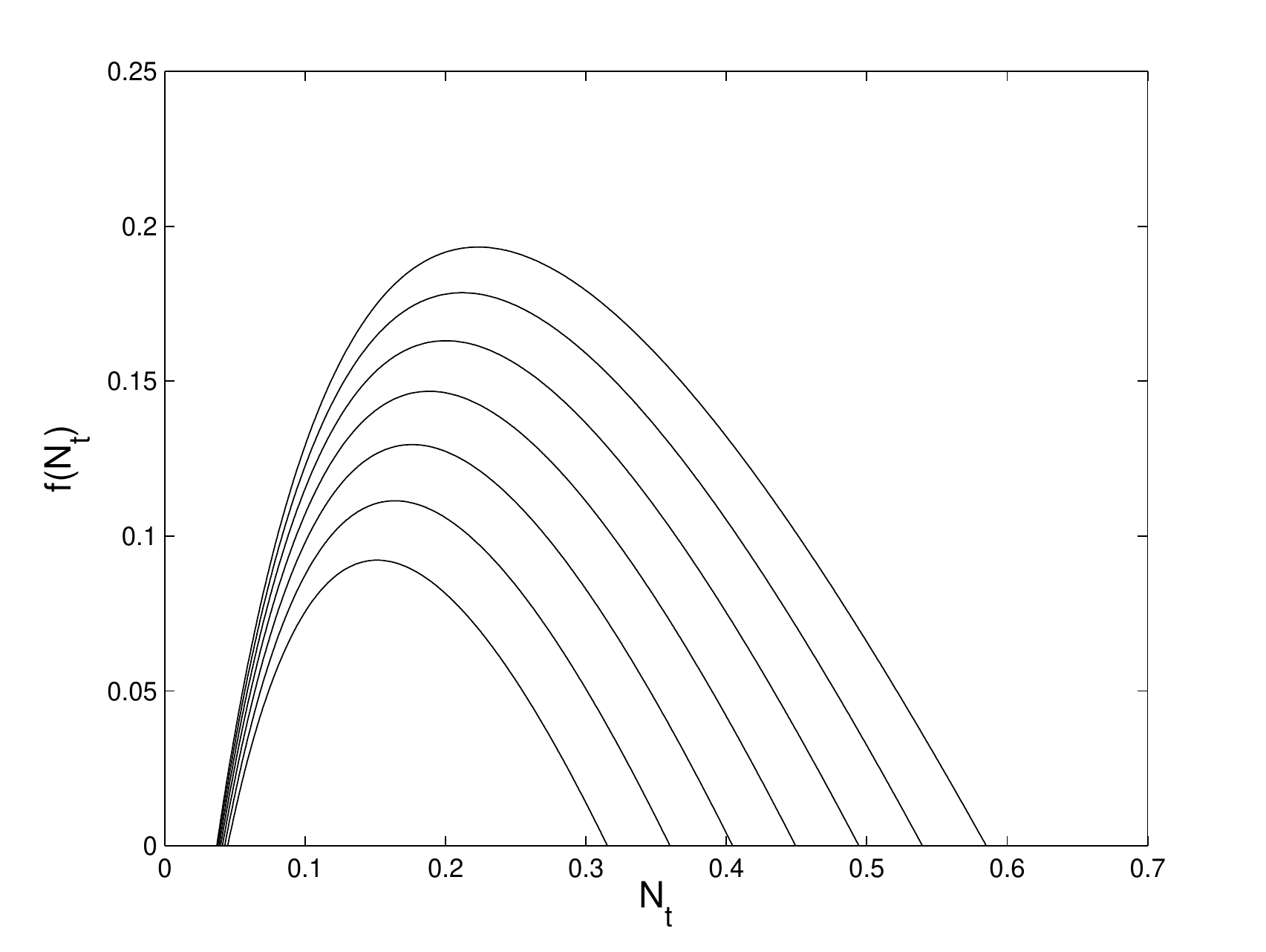}}
     \hspace{.3in}
     \subfigure[Map (\ref{continuous map}) with $r = 1$, $m = 1.15$, $s = 2$]{
          \label{b}
          \includegraphics[width=.45\textwidth]{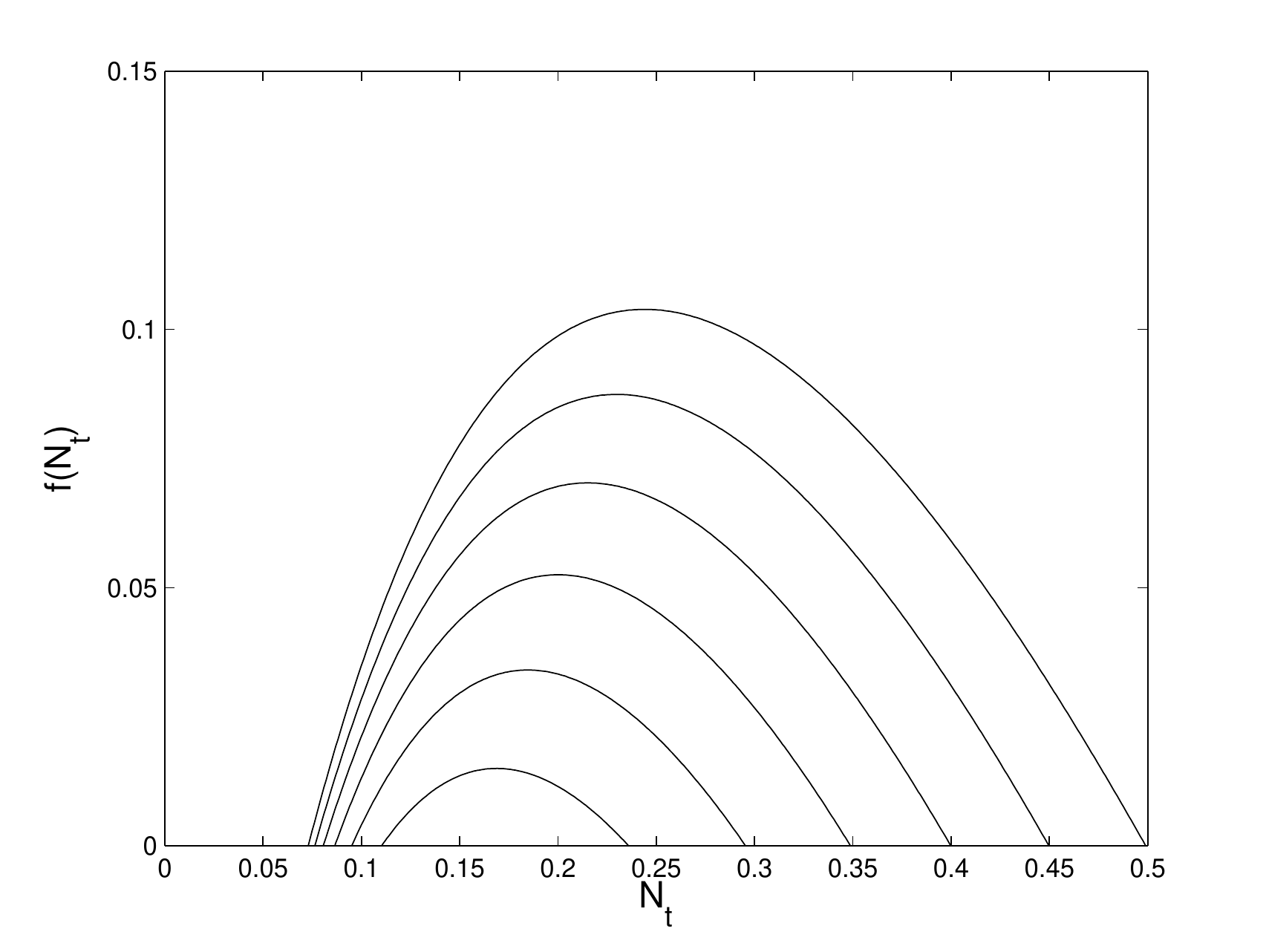}}
     \centering
     \subfigure[Map (\ref{escape map}) with $r = 1$, $m = 1.15$, $s = 4$]{
          \label{c}
          \includegraphics[width=.45\textwidth]{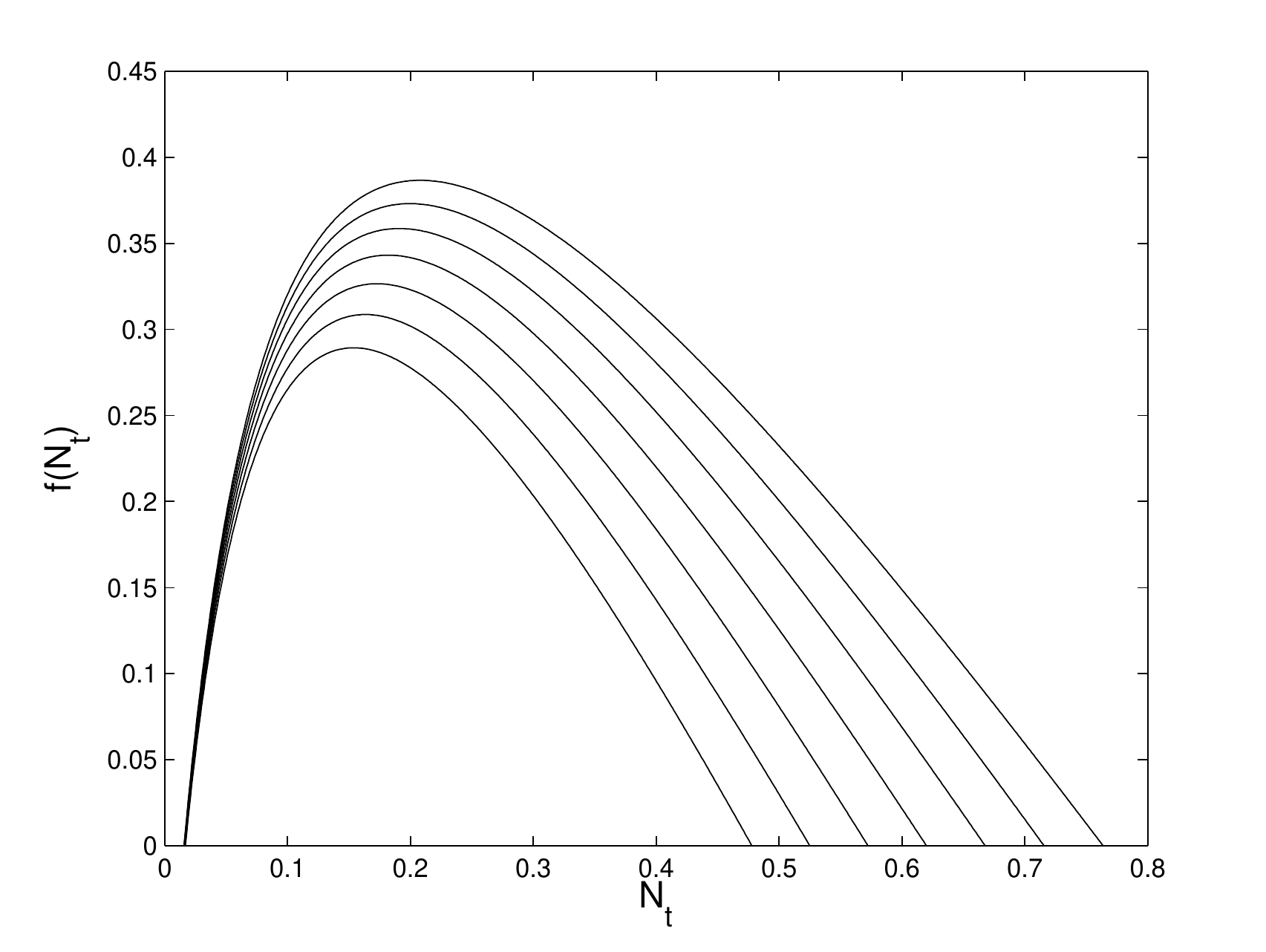}}
     \hspace{.3in}
     \subfigure[Map (\ref{continuous map}) with $r = 1$, $m = 1.15$, $s = 4$]{
          \label{d}
          \includegraphics[width=.45\textwidth]{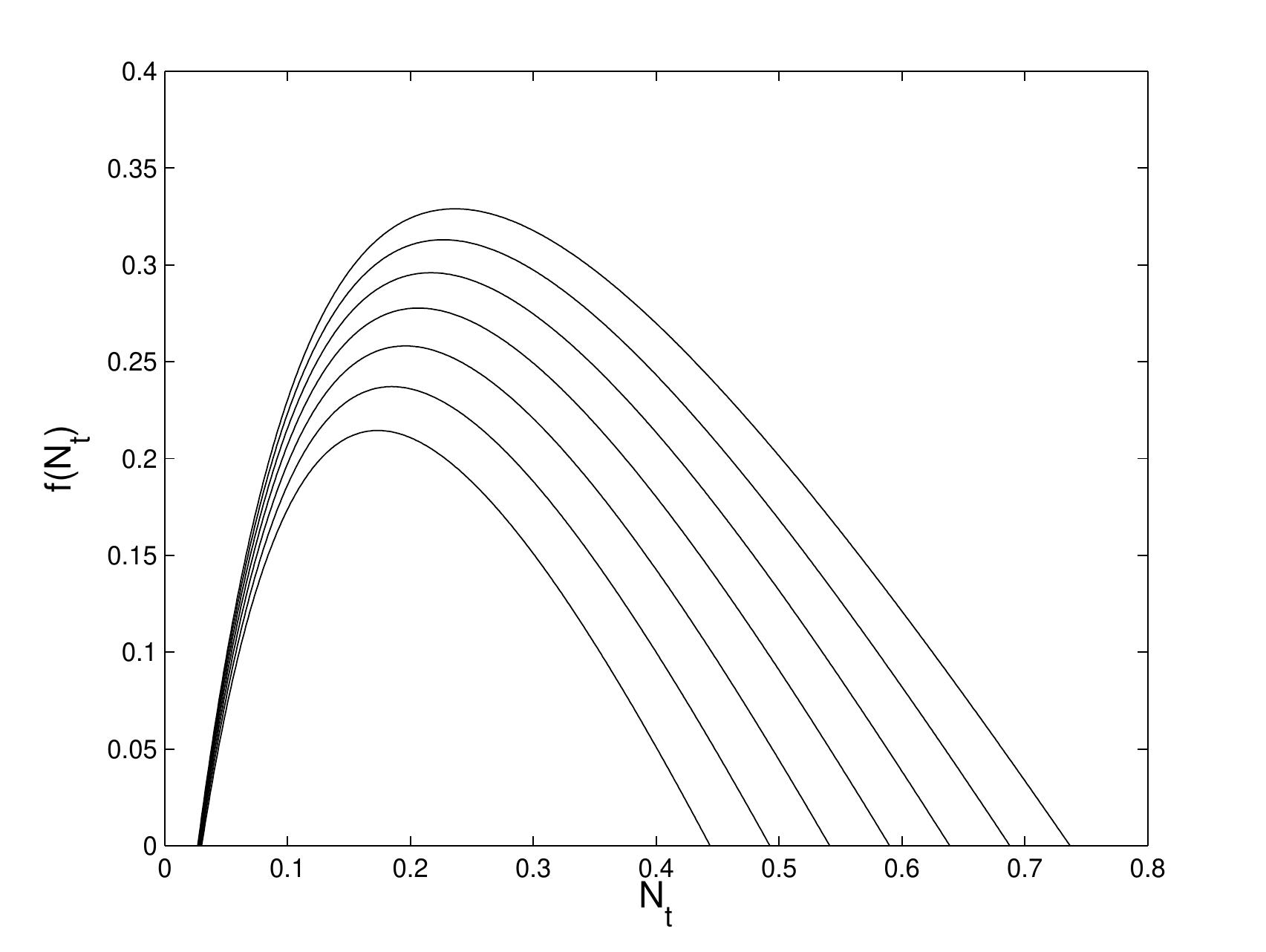}}
     \centering
     \subfigure[Map (\ref{escape map}) with $r = 1$, $m = 0.5$, $s = 2$]{
          \label{e}
          \includegraphics[width=.45\textwidth]{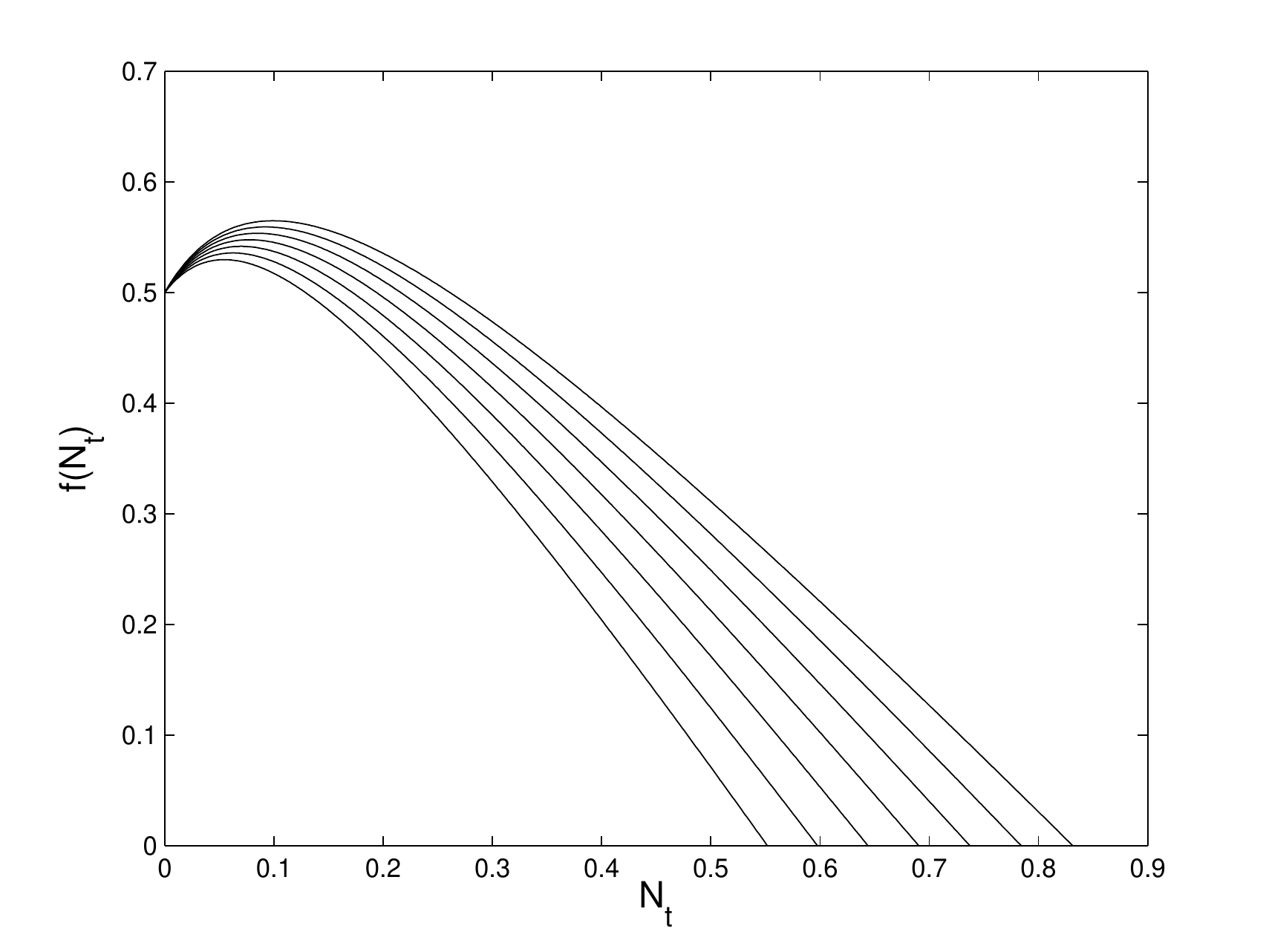}}
     \hspace{.3in}
    \subfigure[Map (\ref{continuous map}) with $r = 1$, $m = 0.5$, $s = 2$]{
          \label{f}
          \includegraphics[width=.45\textwidth]{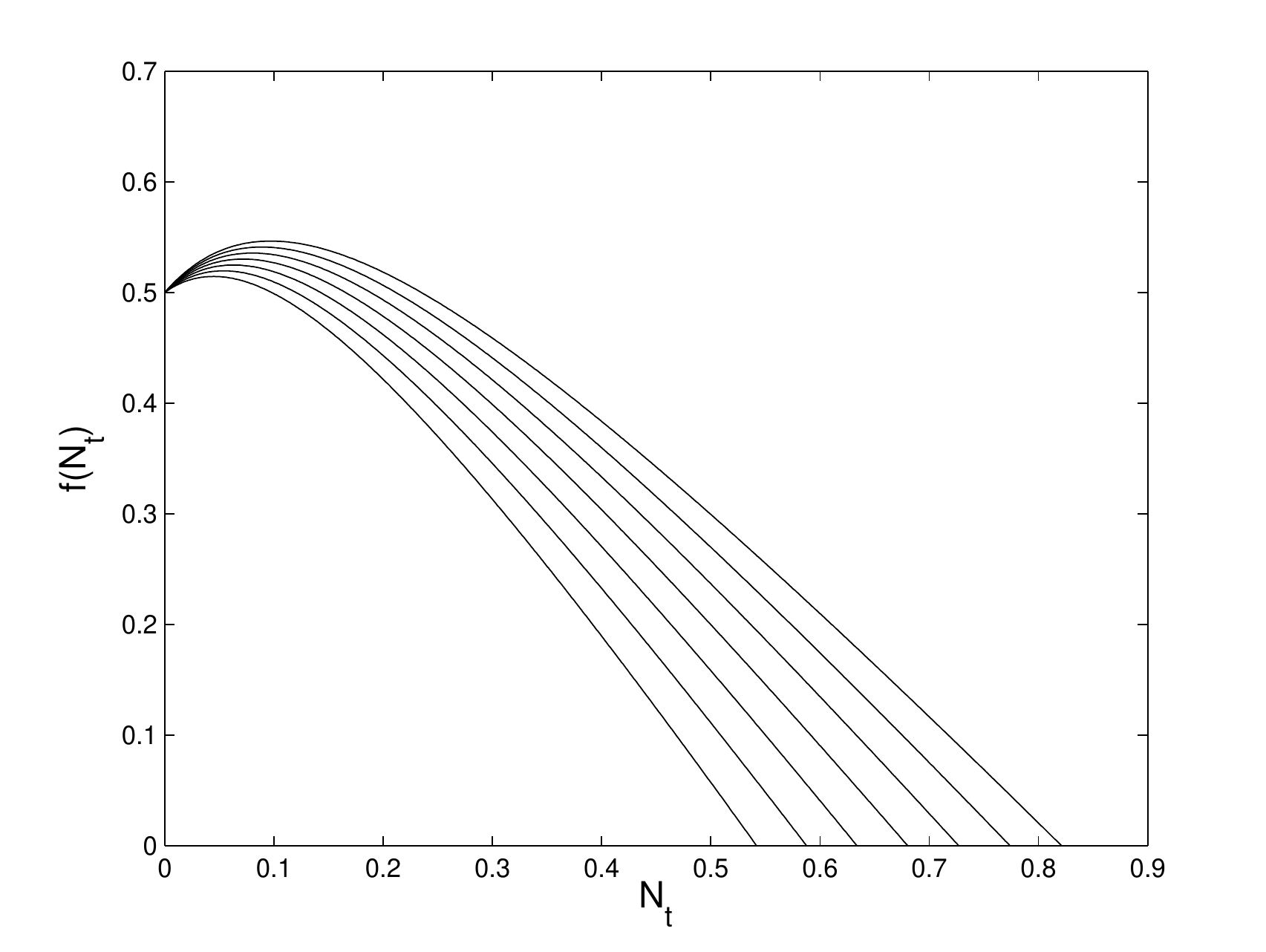}}
     \caption{Growth function $f$ for maps (\ref{escape map}) and (\ref{continuous map}) with varying predation intensity and handling time. In each case, the function is plotted for $K = 0.7, 0.75,...,1.0$.}
     \label{fig}
\end{figure}

In Chapter~\ref{Spontaneous Patchiness}, as in a model for the gypsy moth in~\cite{Liebhold}, a phenomenological growth function of the form 
\begin{equation}
f\left(N_t\right)=r\left(1-N_t/K\right)\left(N_t-c\right)
\label{phenom map}
\end{equation}
 is used to produce a host population with carrying capacity $K$ and Allee threshold $c$. Conveniently, this function produces a host nullcline that is symmetric about its maximum, providing ample opportunity for the steep parasitoid nullcline to intersect to the left, as required for stable outbreaks. However, this cannot be taken for granted in natural systems, given the variety of mechanisms that may shape the left side of the nullcline. 

Many of these mechanisms are reviewed, for example, in~\cite{Courchamp,Stephens}. Some of the most common are those related to the increased individual difficulty of finding a mate at very low population levels. A few models of the resulting Allee growth, derived from underlying individual mating probability, are reviewed in~\cite{Boukal}. These models have in common a kind of singular behavior near zero population -- exemplified by the explicitly boundary-layer model presented and empirically validated in~\cite{Hastings3} -- so that $f\left(N_t\right)=\ln\left(N_{t+1}/N_t\right)$ is extremely steep to the left of its maximum. 

In light of the the results of Chapter~\ref{Spontaneous Patchiness}, a natural means to compare models is by the comparison of their growth functions $f$. Clearly, not all discrete-time single-species models are of the form $N_{t + 1}  = N_t e^{f(N_t )}$, but they can be rewritten as such. For any model $N_{t+1} = G\left( N_t \right)$, we will let $f\left(N_t\right)=\ln\left(G\left( N_t \right)/N_t\right)$. We assume that $G\left( N_t \right)/N_t$ has a limit as $N_t\to 0$, and define $f$ accordingly there.

Note now that \[
\left.\frac{d N_{t+1}}{d N_t} \right|_{N_t =0} = e^{f(0)}.
\]
The derivative at $N_t=0$ for the map (\ref{escape map}) is $e^{r-m}$, just as calculated in \cite{Schreiber} for the map (\ref{schreiber map}). We may differentiate the map (\ref{continuous map}) at $N_t=0$ using the chain rule and the fact that $W'(0)=1$~\cite{Corless}:
\[
\left. {\frac{d}{{dN_t }}\left[ {\frac{1}{s}W\left( {sN_t e^{r\left( {1 - N_t /K} \right) + sN_t e^{r\left( {1 - N_t /K} \right)}  - m} } \right)} \right]} \right|_{N_t  = 0}  = \frac{1}{s}W'(0) \cdot se^{r - m}  = e^{r - m} .
\]
We have $f(0) = r-m$ in each case, so each map exhibits a strong Allee effect if $r<m$. Of course the phenomenological map, with $f$ defined in (\ref{phenom map}), has a strong Allee effect if $c>0$.

It is interesting to note that whether the Allee effect is strong does not depend on the carrying capacity $K$ in any of the models. In the phenomenological model, the threshold density $c$ is set as a parameter independent of $K$. In our other models, the threshold density should not be expected to depend greatly on $K$ (so $\partial c/\partial K \approx 0$), since $f(0)$ does not depend on $K$. The positive fixed point $N^*$ of each of our models, however, moves with $K$ (that is, $\partial N^*/\partial K >0$), as can be seen in Figure~\ref{fig}.

In our mechanistic models, the strength of the Allee effect depends, in a sense, on the predator-related parameters $m$ and $s$. As we decrease $m$, predation becomes less intense; as we increase $s$, the predator is more easily saturated and so predation becomes more density-dependent. It is clear that, as $m\to 0$, the map (\ref{escape map}) approaches the Ricker map. That is, the function $f$ converges to $f\left( {N_t} \right) =  r\left( {1 - N_t /K} \right)$, which is Ricker's growth function. This can be seen by comparing Figures~\ref{a} and ~\ref{e}. Similarly, as $s\to\infty$, for $N_t\ne 0$, $f$ converges to Ricker's growth function. But since $f(0) = r-m$ for the map (\ref{escape map}), $f$ does not converge to the value of Ricker's growth function at zero. That is, $f$ converges pointwise almost everywhere as $s\to\infty$, but it does not converge uniformly. The slope of $f$ near $N_t=0$ grows arbitrarily large as $s$ increases. Also, at every other point, $f$ approaches Ricker's growth function -- which is decreasing in $N_t$ -- so the maximum of $f$ moves toward zero as $s$ increases. This is seen by comparing Figures~\ref{a} and ~\ref{c}.

For the map (\ref{continuous map}), since $xe^x  = W^{ - 1} (x)$, for small $m$ we have\[
\frac{1}{s}W\left( {sN_t e^{r\left( {1 - N_t /K} \right) + sN_t e^{r\left( {1 - N_t /K} \right)}  - m} } \right) \approx \frac{1}{s}W\left( {sN_t e^{r\left( {1 - N_t /K} \right)} e^{sN_t e^{r\left( {1 - N_t /K} \right)} } } \right)
\]\[
 = \frac{1}{s}W\left( {W^{ - 1} \left( {sN_t e^{r\left( {1 - N_t /K} \right)} } \right)} \right) = N_t e^{r\left( {1 - N_t /K} \right)} ,
\]
so again the map converges to the Ricker map for any $N_t$. The above also holds for large $s$, if $N_t\ne 0$.

The behavior of the models as $s$ increases mirrors the singular behavior of models derived from individual mating probability, mentioned above. In the limit of large handling time, then, it becomes unlikely that host and parasitoid nullclines cross in the manner shown in Chapter~\ref{Spontaneous Patchiness} to be necessary for spatial pattern formation. On the other hand, this behavior is not encountered as $m$ decreases. So for reasonably small $s$, the growth function $f$ for each of our models is qualitatively similar to the simple phenomenological function (\ref{phenom map}). The importance of the behavior of the mechanistically-derived function $f$ with changes in $K$ -- roughly the same as for a phenomenological, quadratic function -- will now be shown.

\section{Consequences of Habitat Quality}
\label{Consequences of Habitat Quality}
\subsection{Introduction}
The biological system motivating this work, which centers on the bush lupine, consists of two host-parasitoid interactions, one in the foliage of the lupine (the tussock moth subsystem) and one near the roots. As noted earlier, tussock moths have little to no effect on the year-to-year survival of the lupine; however, root-feeding ghost moth larvae can do great harm to the plant~\cite{Harrison}. We will turn our attention to the particulars of the ghost moths and their parasitoids in Chapter~\ref{Nematode Chapter}. 

The consequences of habitat degradation comprise an important problem in theoretical ecology. A change in the quality of a habitat is usually included in models such as~(\ref{model N})-(\ref{model P}), in which that quality is not dependent on the rest of the system,  by varying a parameter representing carrying capacity. One famous example of this can be found in~\cite{Rosenzweig}. Indeed, the biological system with which we are concerned has been so treated in~\cite{Harrison}, in which the effects of lowering carrying capacity in a reaction-diffusion partial differential equation model of the system are investigated. It is found in that work that the size of an outbreak is positively related to carrying capacity; as the latter increases, outbreaks grow.

We will now see that the opposite result holds for our model (\ref{model N})-(\ref{model P}).

\subsection{Dependence on Carrying Capacity}
As seen in Section~\ref{Properties of the Map}, an Allee growth function derived from the mechanism of saturating predation shares many features with a simple quadratic function, including its behavior upon a change in carrying capacity. Considering this behavior in light of Figure~\ref{fp} and the discussion and results in Chapter~\ref{Spontaneous Patchiness}, some of the consequences of raising or lowering carrying capacity in the model are evident. 

Recall that the $P$ nullcline in Figure~\ref{fp} depends only upon the parameter $a$. As noted in the previous section, the nullcline crosses the $N$ axis at $N=1/a$ and rises steeply. As $K$ decreases, the maximum of the $N$ nullcline moves left and down (Figure~\ref{fig}). If the coexistence point was slightly to the left of that maximum with $K=1$ -- as it is for Figure~\ref{num ex} -- then with lower $K$ it might be on the right. So decreasing $K$ has a stabilizing effect on the underlying, nonspatial dynamics of the model, which means that it reduces or even completely removes the patchiness of the spatial model's steady state. Likewise, increasing $K$ destabilizes the underlying dynamics and can cause patchiness to appear.

This result can be seen more exactly in (\ref{radius eq}), the estimate of maximum outbreak radius. For small $P_\text{max}$, we have
\[
R \approx D\frac{N_\text{max}}{N_\text{max} - 1/a},
\]
where $N_\text{max}$ is the location of the maximum of $f$ (and therefore the $N$ nullcline). If $K$ is lowered such that $N_\text{max}$ decreases toward approximately $1/a$, the estimate of outbreak radius increases without bound. In actuality, as $N_\text{max}$ decreases toward this singular point, the tendency to oscillate (discussed in Section~\ref{Linear Stability}) overcomes the tendency to form steady spatial patterns. As $K$, and consequently $N_\text{max}$, is further decreased, the nonspatial dynamics move through the oscillatory parameter range and eventually become stable, and likewise the spatial model reaches a smooth steady state devoid of patchiness.

\subsection{Numerical Examples}
\begin{figure}[htbp]
\begin{center}
\includegraphics[width=.48\textwidth]{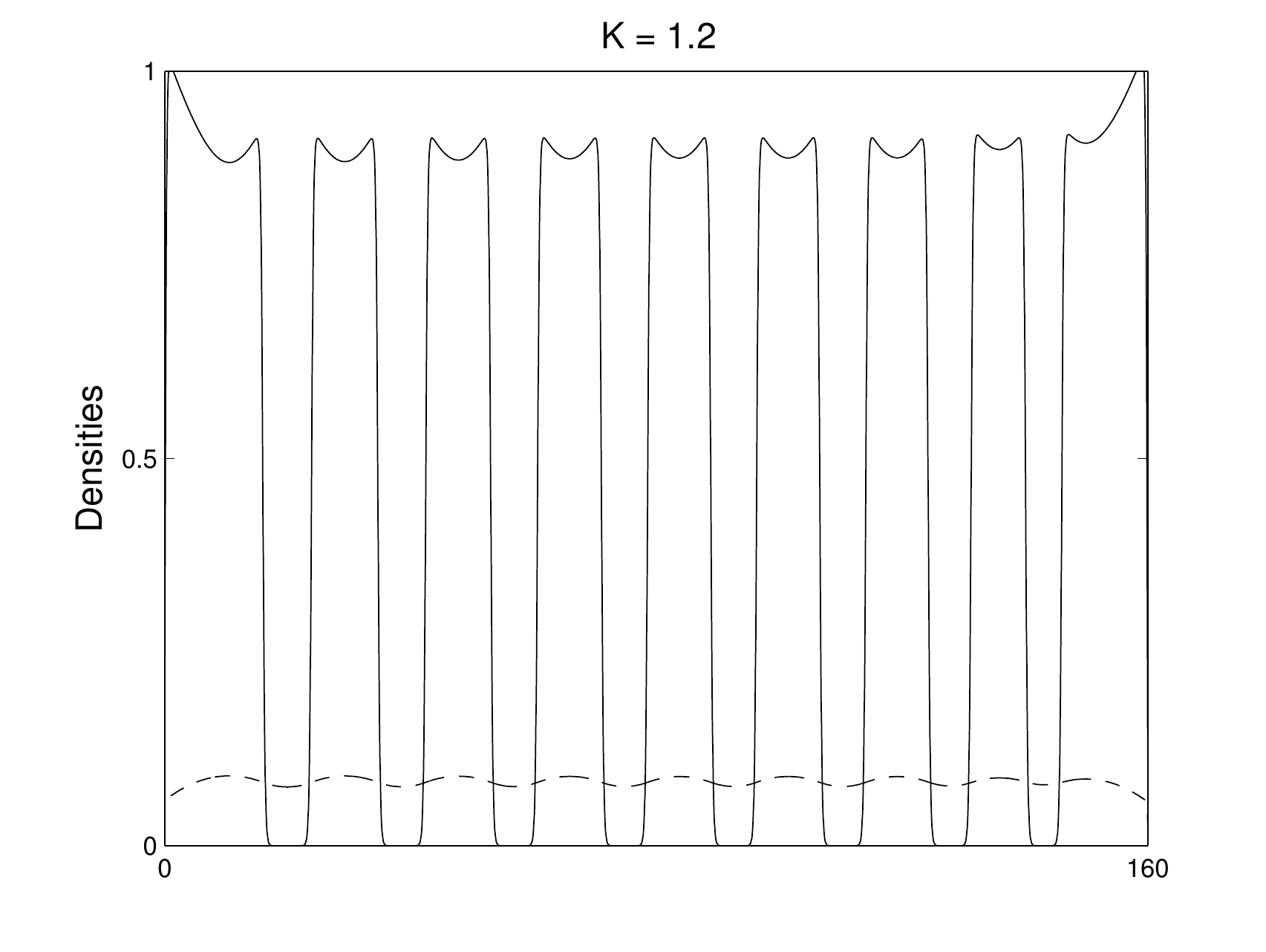} 
\includegraphics[width=.48\textwidth]{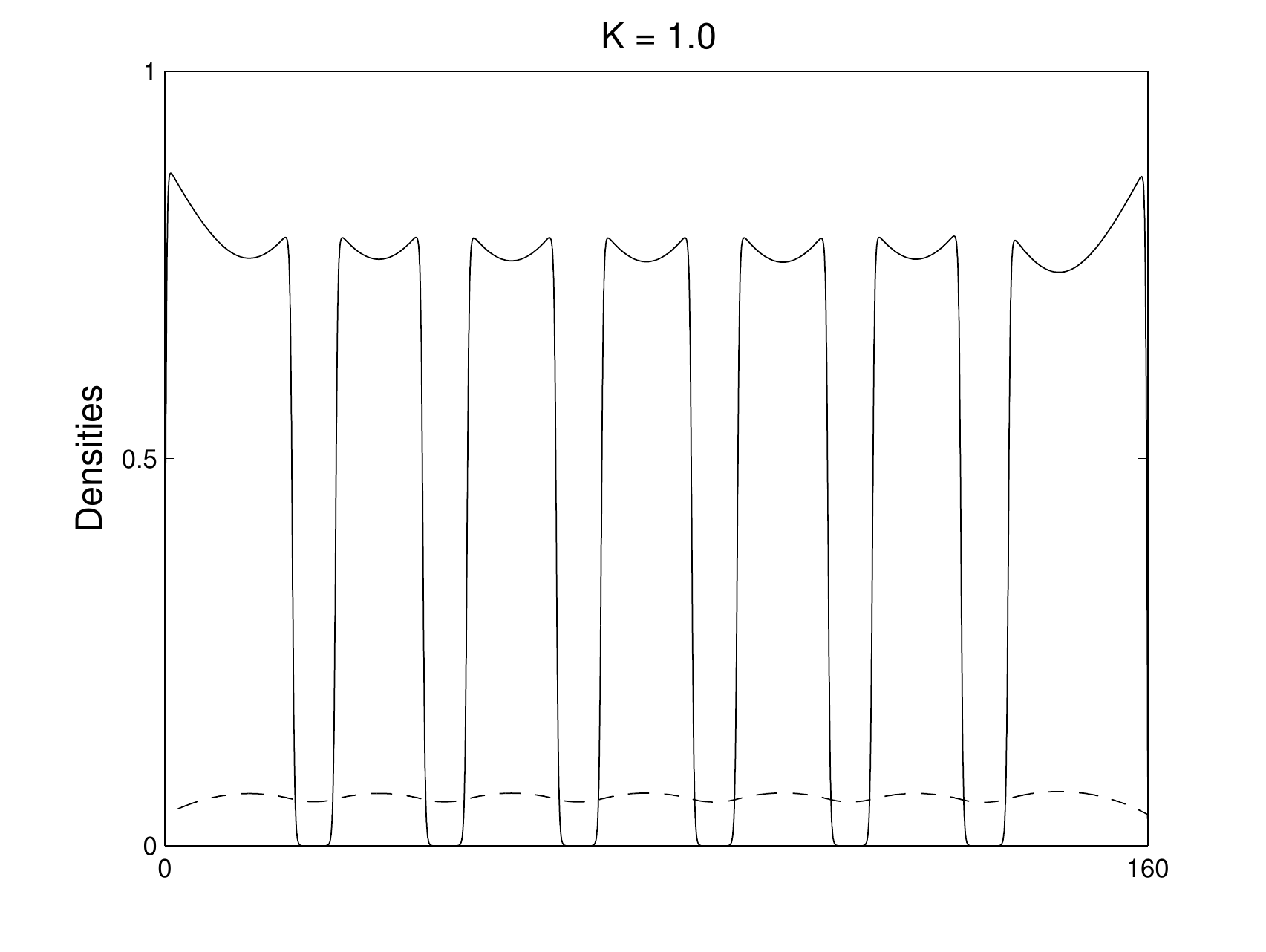} \\
\includegraphics[width=.48\textwidth]{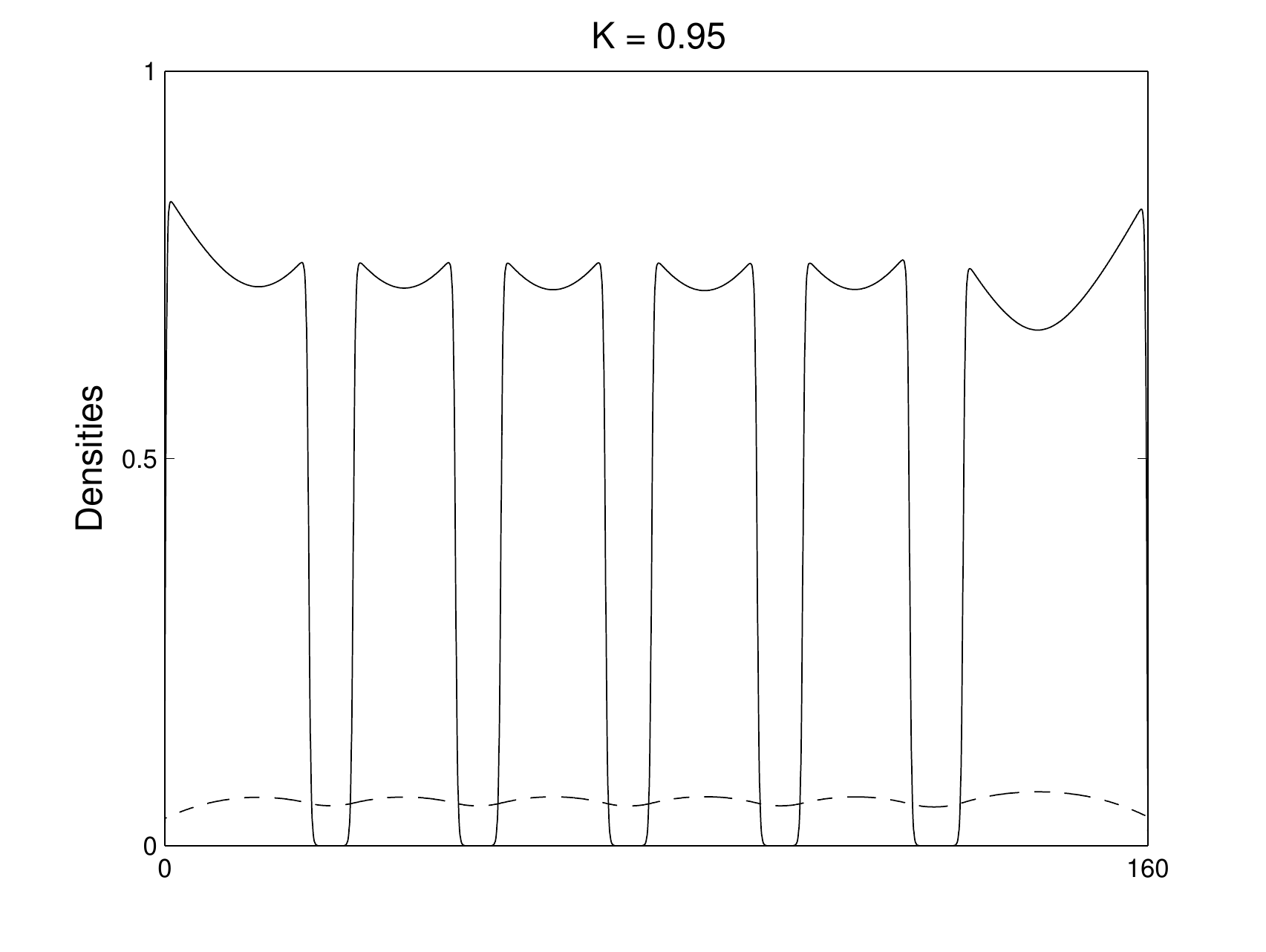} 
\includegraphics[width=.48\textwidth]{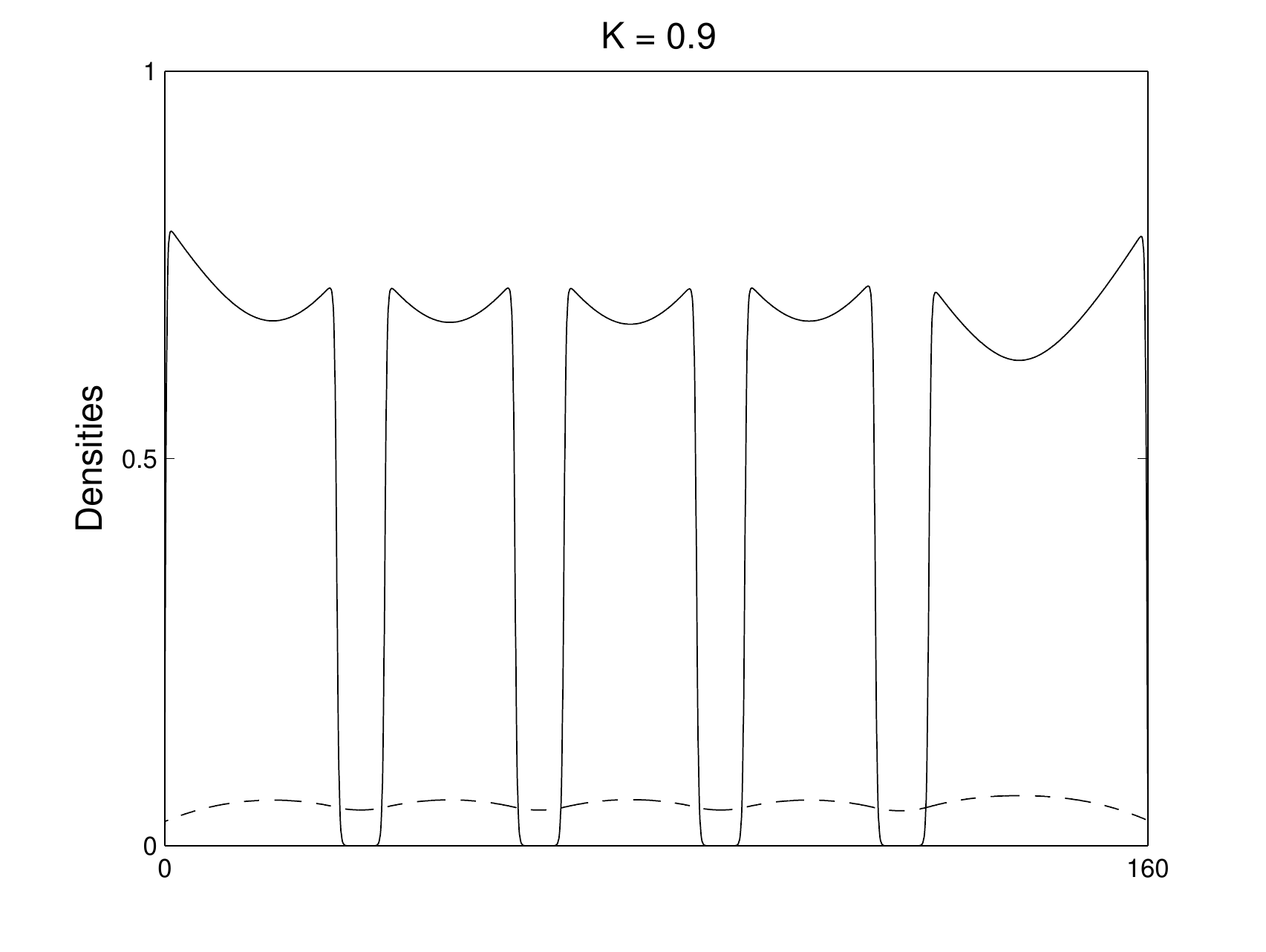} \\
\includegraphics[width=.48\textwidth]{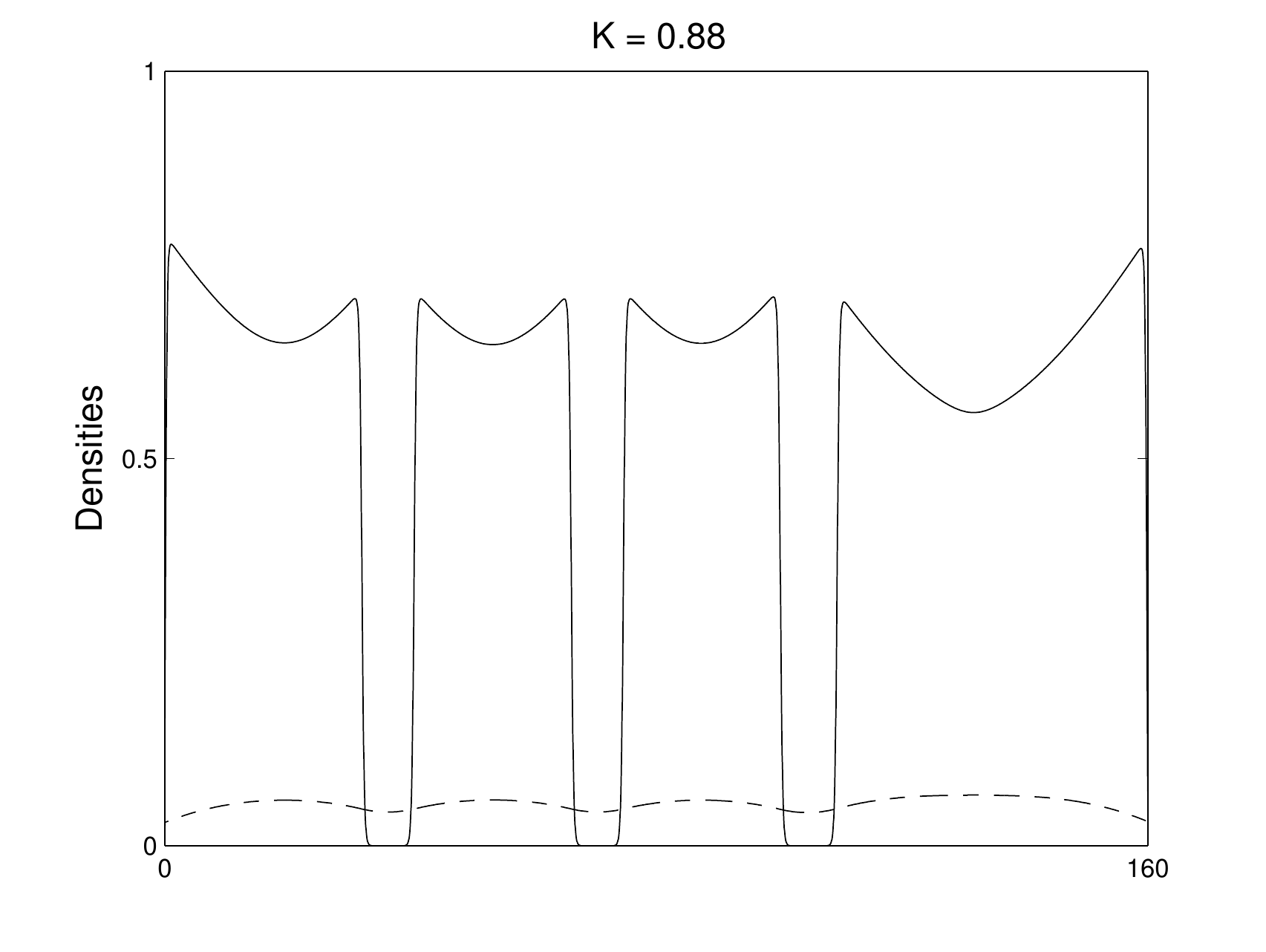} 
\caption{Consequences of varying carrying capacity, with $D=10$, $\varepsilon = 0.1$, $a=2$, $f(N)=(1-N/K)(N-0.2)$, with initial outbreak in the leftmost quarter of the domain. Solid and dashed curves represent host and parasitoid densities, respectively. Each simulation was run to $t=10000$.}
\label{k varied}
\end{center}
\end{figure}
Figure~\ref{k varied} demonstrates the dependence on carrying capacity explained above. Increasing $K$ leads to more patchiness, with smaller outbreaks, and decreasing $K$ has the opposite effect. As $N_{max}$ approaches the singular point, the final state becomes more sensitive to the value of $K$, as expected. Note that, for consistency with Chapter~\ref{Spontaneous Patchiness}, we will continue to use the usual quadratic growth function, since it appears from Section~\ref{Properties of the Map} to be qualitatively valid.

\begin{figure}[htbp]
\begin{center}
\includegraphics[width=.7\textwidth]{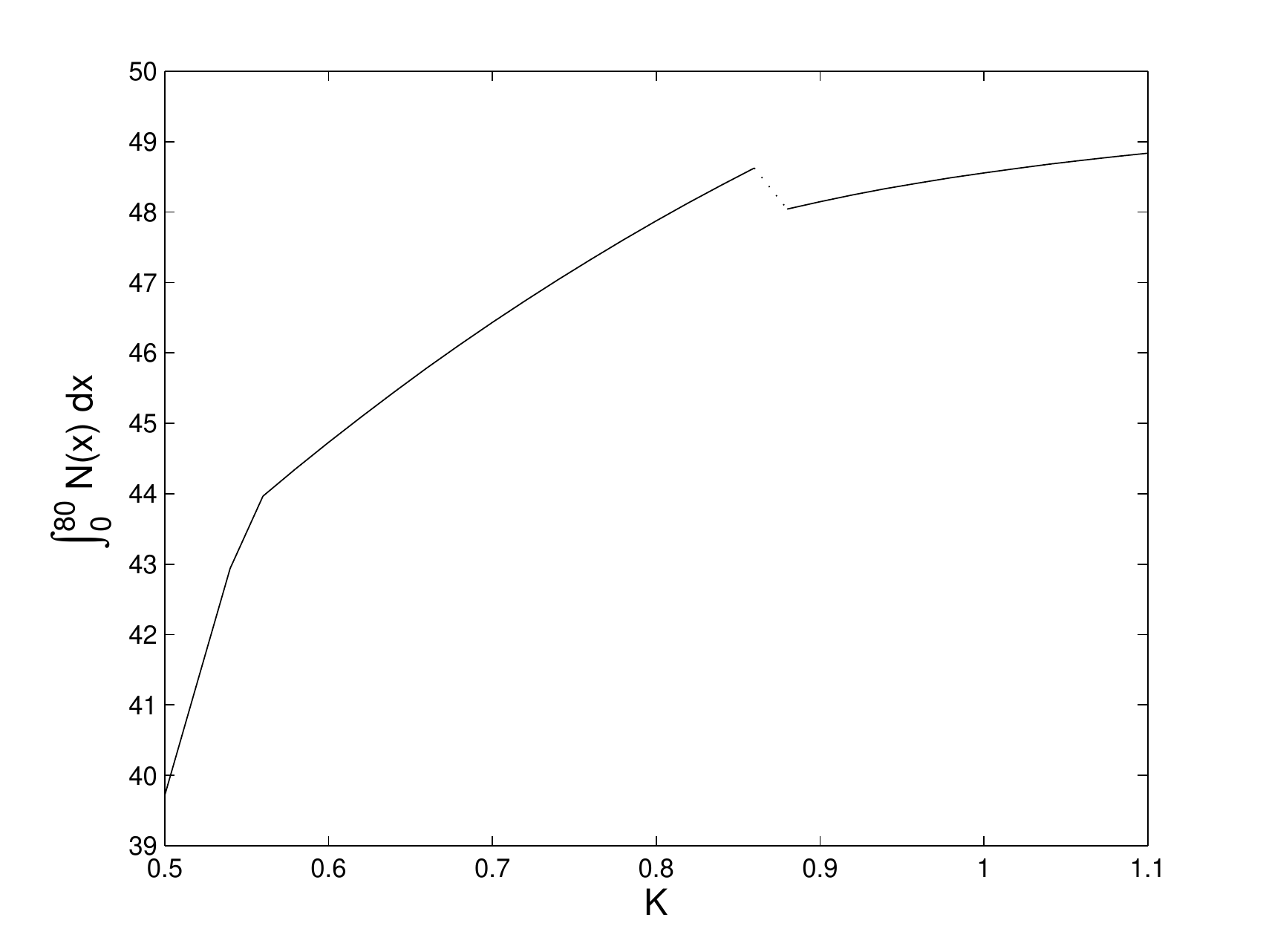} 
\caption{Consequences of varying carrying capacity on overall host abundance in a domain of length 80.}
\label{homo counts}
\end{center}
\end{figure}

Another interesting question about the effects of habitat quality is whether the overall abundance of the host, measured by $\int_\Omega N(x) dx$ (where $N(x)$ is the steady state), increases with $K$. It would be natural to expect this, were it not for the dependence of patchiness on $K$ described above. Figure~\ref{homo counts} shows the numerical results on the domain $\Omega = [0,80]$, with parameters from Figure~\ref{k varied} and $K$ varying from $0.5$ to $1.1$. Results are obtained by running the simulation for a given value of $K$ until $\|N_{t}(x)-N_{t-100}(x)\|_{\ell^2}$ falls below some threshold, then slightly varying $K$ and repeating the process on the previously-obtained steady state. The evident trend is that abundance increases with $K$, except near $K=0.9$ where two extinction regions form. Simulations have not been carried out in enough detail to comment on the exact nature of changes in abundance during the transition to patchiness, and as noted above and in the next section, a steady state may not always be possible in this parameter range. However, it is clear that, while increasing habitat quality will lead to greater host abundance for most values of $K$, this is not always true.

\subsection{The Paradox of Enrichment}
The consequences of varying $K$, the carrying capacity of the host in the absence of the parasitoid, are somewhat counterintuitive but should be familiar to mathematical ecologists. They are a reflection of the so-called paradox of enrichment~\cite{Rosenzweig} whereby a victim-exploiter interaction is destabilized by enriching the habitat of the victim -- or, in a model, by increasing its carrying capacity.

This ``paradox" has been studied for nonspatial models such as (\ref{no dispersal N})-(\ref{no dispersal P}). The addition of spatial dispersal, however, has a profound effect on the stability of the system under changes in carrying capacity:
\begin{center}
\includegraphics[width=\textwidth]{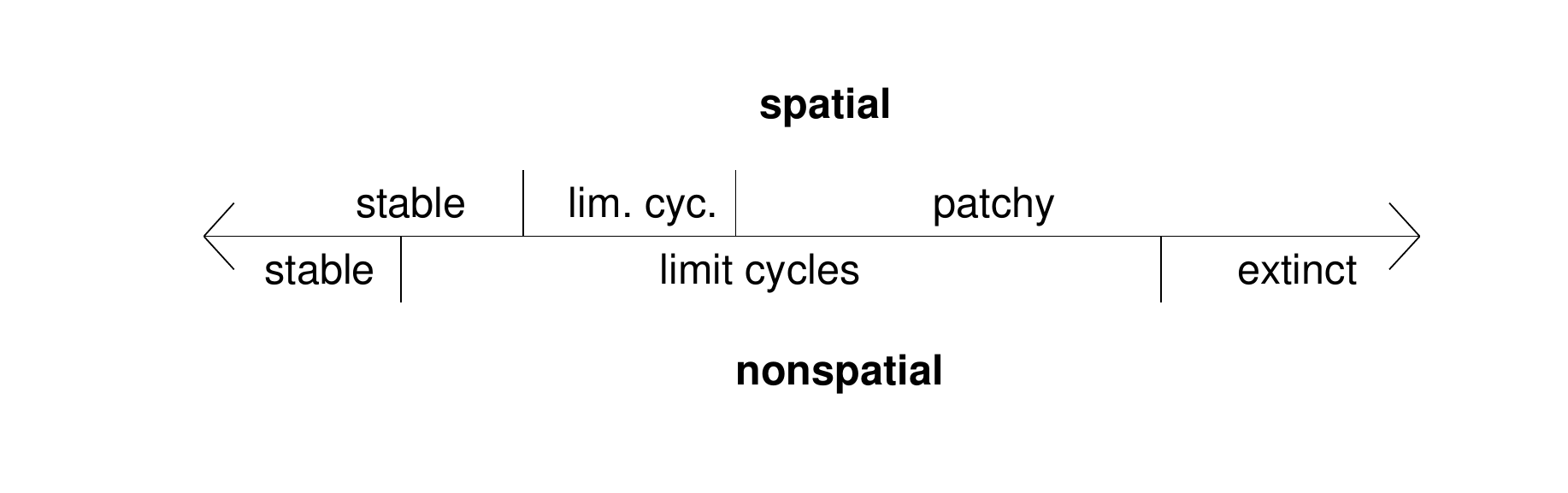} 
\end{center}
The horizontal line represents carrying capacity $K$. As it increases, the nonspatial model loses stability, first to limit cycles and then to extinction. The extent of the oscillatory parameter range is considerable; this diagram was generated for $K$ from 0.7 to 1.1. The spatial portion of the diagram was produced with the parameters used in Figure~\ref{k varied}, on a domain of length 160, with simulations run to 10000 time steps and categorized by eye -- the values of $K$ at boundaries between differing behaviors were not found with any precision and therefore are not labeled. Rather, the importance of this diagrammatic overview is qualitative. Spatial considerations slightly extend the parameter range that produces a long-term stable (not patchy) solution. The limit cycle range is greatly reduced in size, and beyond it the formation of patches prevents global extinction, which numerical simulations show to be the case up to at least $K=2$.

\section{Consequences of Habitat Heterogeneities}
\subsection{Introduction}
A more general situation than the previous discussion of habitat quality is the possibility that carrying capacity $K$ depends on location $x$. As we will see, the most interesting habitat heterogeneities are abrupt changes, or discontinuities, in carrying capacity.

Discontinuities may be a reasonable expectation given the nature of the habitat being modeled. As will be discussed in the next chapter, though the bush lupine habitat may be spatially continuous from the perspective of the tussock moth, it consists of individual plants with taproots set at some distance from each other. The health of each plant depends on conditions in its rhizosphere such as the dynamics of detrimental ghost moth larvae and the parasitic nematodes that exploit them. It is evident that if there is any dynamical coupling between rhizospheres, it is at best weak and sporadic \cite{Dugaw2}; as such, even neighboring bushes may differ greatly in quality. We now consider the consequences of such heterogeneities.

\subsection{Patch Formation}
As in Section~\ref{A Bound on Patch Radius}, consider a point $x_c$ inside the domain of the model (\ref{model N})-(\ref{model P}) near which the local dynamics are settled to the outer approximation $N(x) = f^{-1}(aP(x))$ (see Section~\ref{Regular Perturbation}). Recall that local extinction occurs when the parasitoid density reaches the maximum of the host nullcline, due to -- in a homogeneous environment -- an outbreak growing large enough to sustain that density of parasitoids inside.

In a heterogeneous environment, if the point under consideration is in a region of low quality (low carrying capacity) relative to nearby habitat, there may be interesting repercussions for patch formation. Lowered carrying capacity results in a lower host nullcline, which is more easily overcome by the parasitoid density. In a homogeneous habitat this is countered by the movement of the coexistence fixed point of the local dynamics, which for low enough carrying capacity makes it impossible for parasitoid levels to reach the maximum of the host nullcline. However, a nearby region of higher quality habitat can provide levels of parasitoid influx sufficient, when combined with locally-generated parasitoid densities, to cause local extinction. While homogeneously low-quality habitats are less likely to become patchy through the formation of local extinctions, in a heterogeneous environment low-quality regions are the most likely locations for extinctions.

More interesting effects can be seen if habitat heterogeneities are fairly abrupt. For example, if a domain is divided into one high- and one low-quality region, and the drop in carrying capacity is sufficiently discrete and severe, an outbreak beginning in the high-quality habitat may be halted at the heterogeneity. For a completely discrete (stepwise) drop in carrying capacity, there is clearly a sufficient condition on the magnitude of the drop required to halt the outbreak. At the edge of the outbreak, parasitoid density has some finite value because $D$, the dispersal parameter for the parasitoid, is relatively large. If the low carrying capacity is such that the maximum of the resulting host nullcline ($P_{\text{max}}$ in Section~\ref{A Bound on Patch Radius}) is less than that density, the outbreak should not be able to spread past the location of the drop in habitat quality. In fact, numerical analyses, which we will now present, suggest that the discrete drop need not be even that severe.

\subsection{Numerical Examples}
\begin{figure}[htb]
\begin{center}
\includegraphics[width=.48\textwidth]{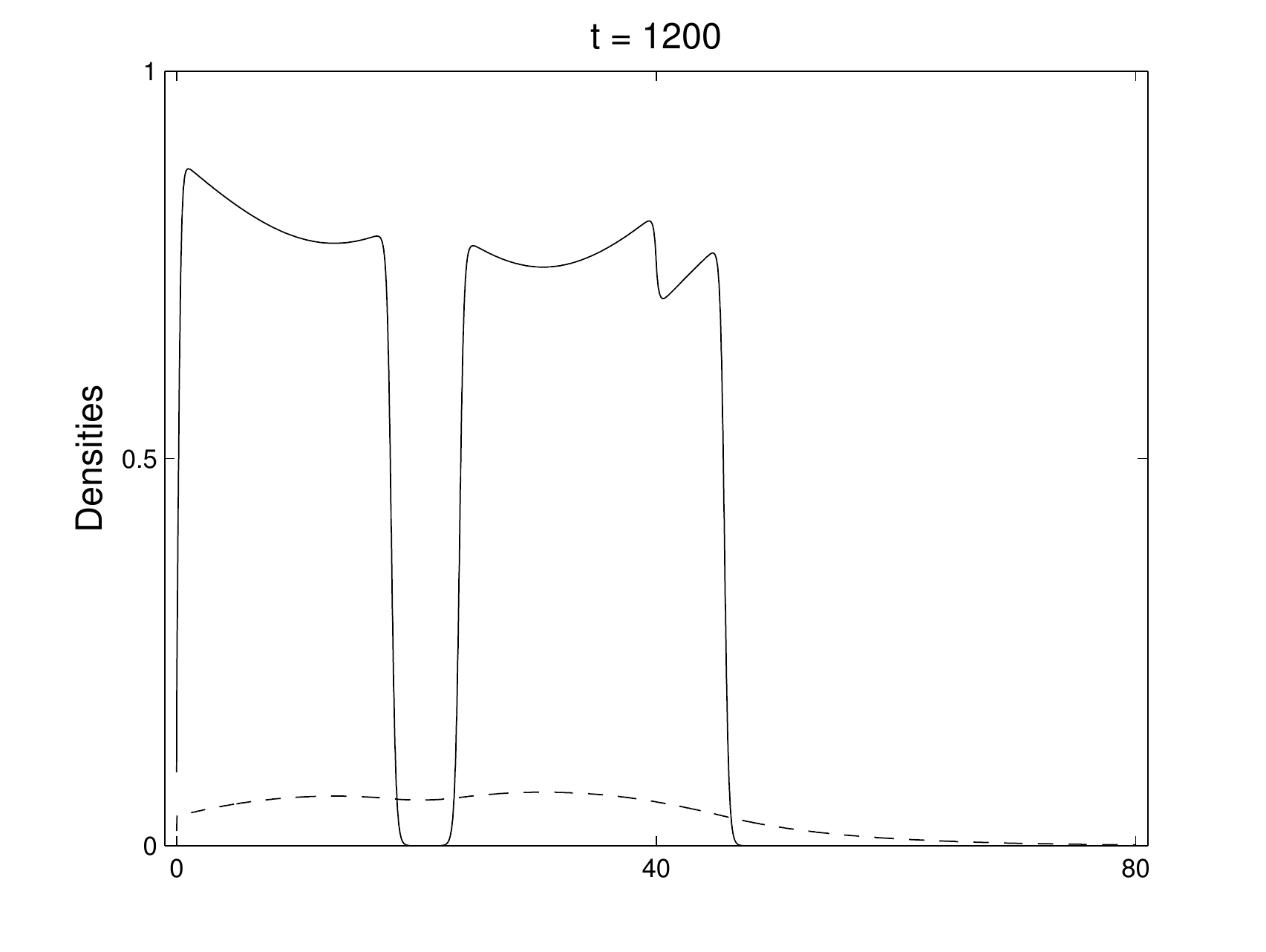} 
\includegraphics[width=.48\textwidth]{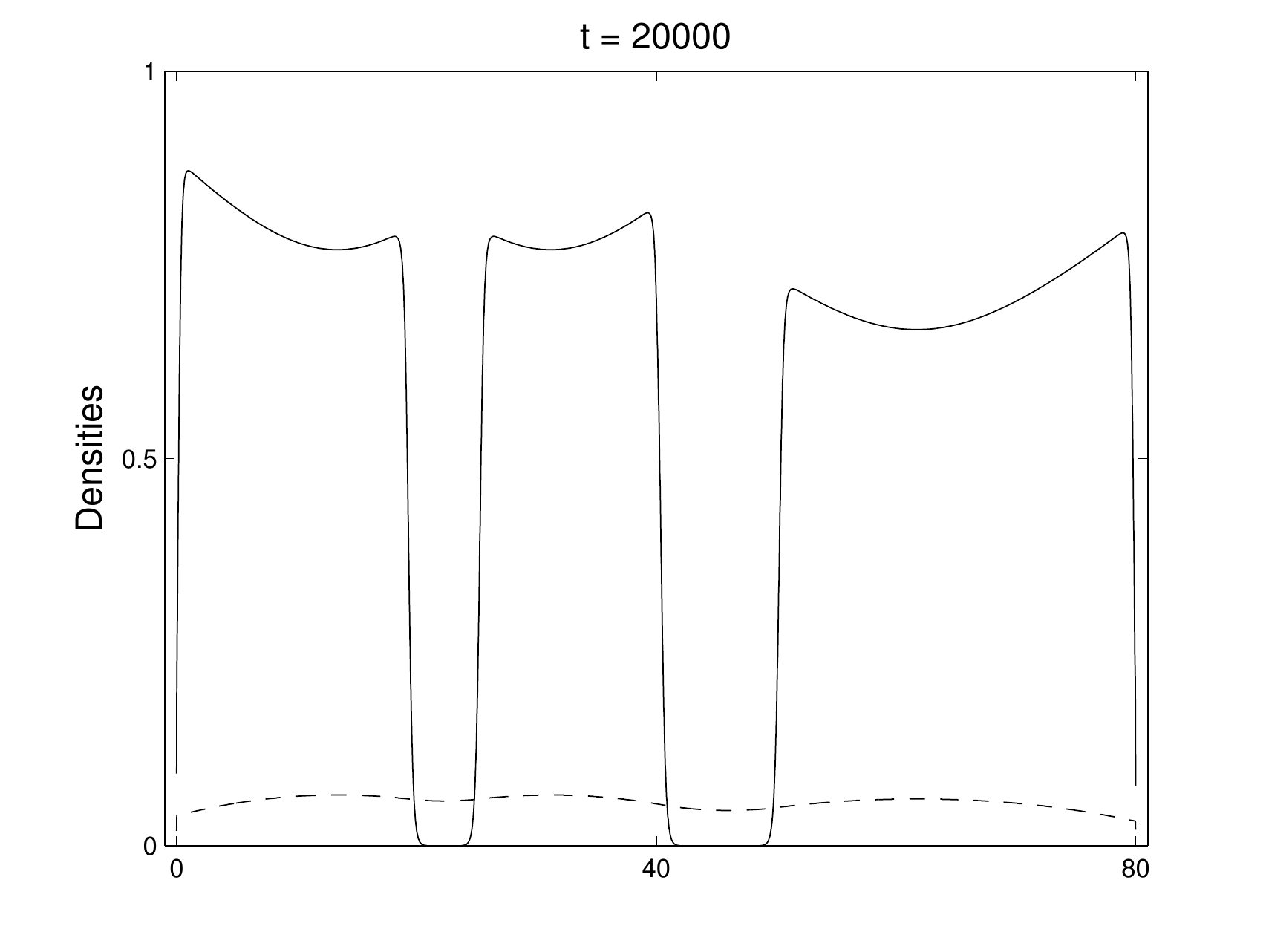} 
\caption{Simulation with $K=1$ in the left half of the domain and $K=0.9$ in the right half. Parameters and initial outbreak size are the same as in Figure~\ref{num ex}.}
\label{K 09 L}
\end{center}
\end{figure}

We investigate the consequences of a habitat heterogeneity by dividing the domain from Section~\ref{Numerical Experiments} into two (equal) regions of higher and lower $K$. The former will have $K=1$. Setting $K=0.9$ in the low-quality region, the numerical results are shown in Figure~\ref{K 09 L}. An outbreak beginning in the high-$K$ subdomain fills it in the usual way and spreads into the area of lower $K$. A local extinction occurs at the sharp boundary between the regions, on the side of lower $K$. The approximate final state is shown. 

Similarly, an outbreak beginning in the lower-quality region spreads to the interface and continues into the subdomain with $K=1$. Shortly thereafter, local extinction occurs at the interface as above, and the final state is identical.

\begin{figure}[htb]
\begin{center}
\includegraphics[width=.48\textwidth]{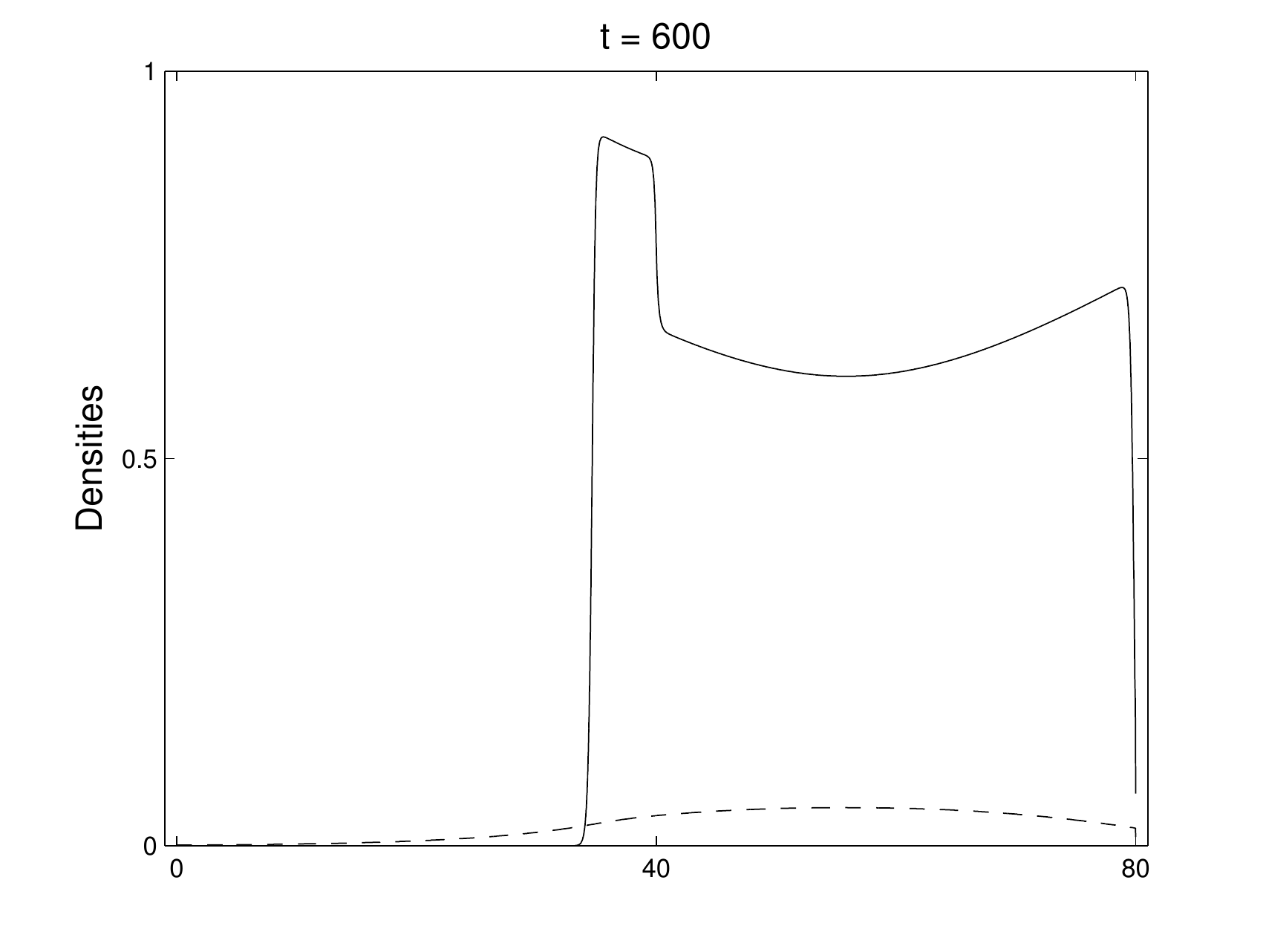} 
\includegraphics[width=.48\textwidth]{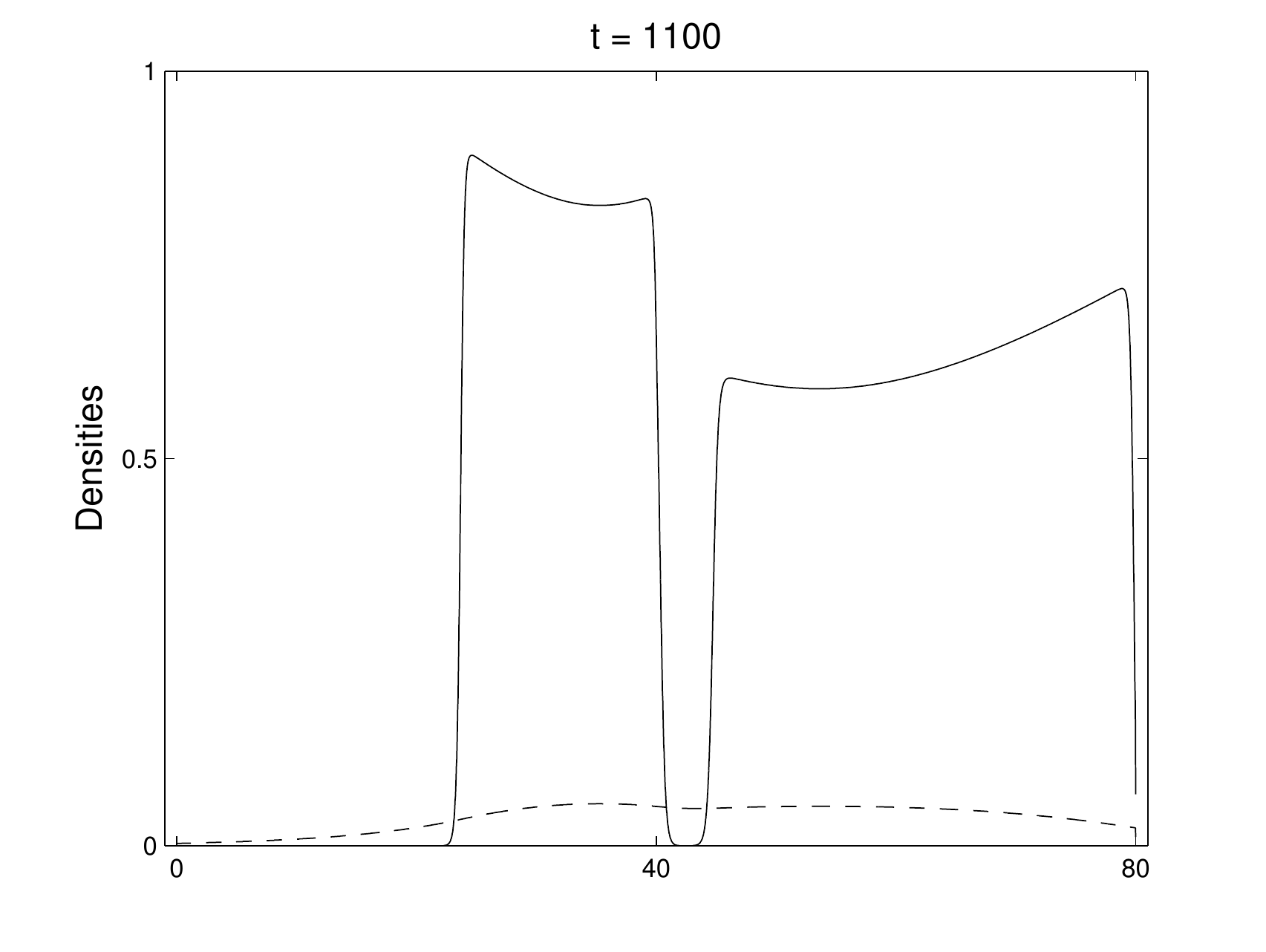} 
\includegraphics[width=.48\textwidth]{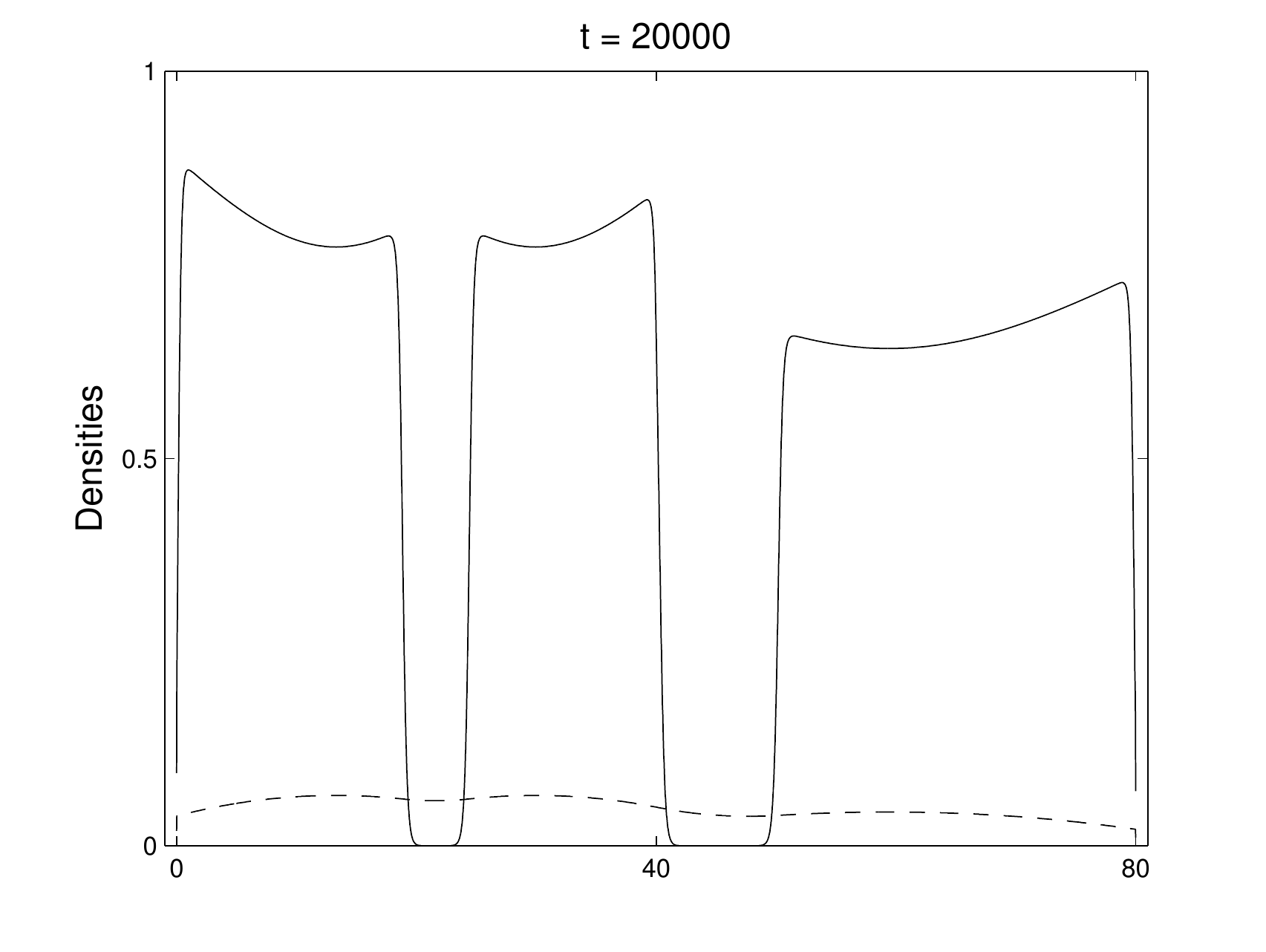} 
\includegraphics[width=.48\textwidth]{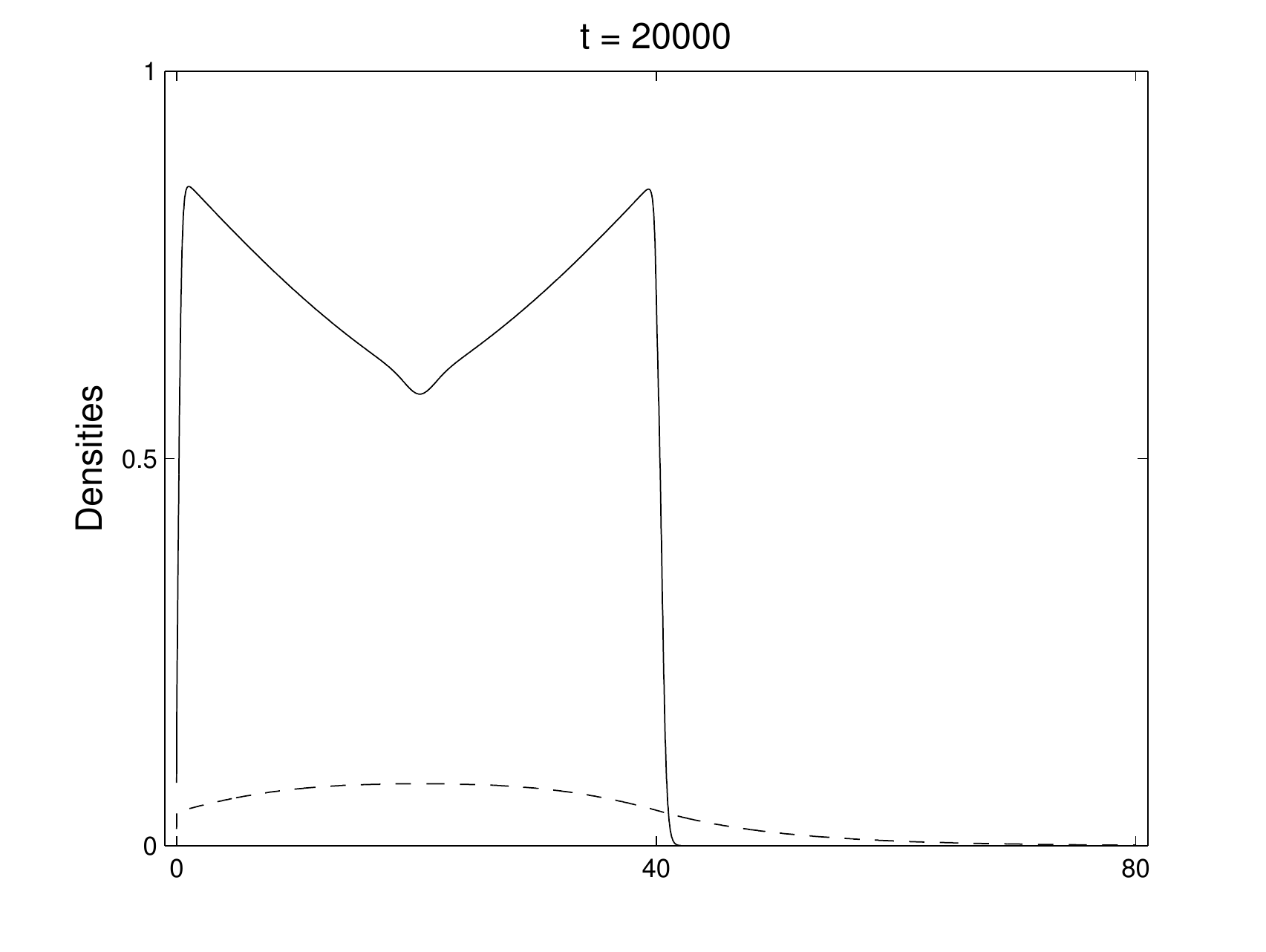} 
\caption{Simulation with $K=1$ in the left half of the domain and $K=0.8$ in the right half. Parameters and initial outbreak size are the same as in Figure~\ref{num ex}. In the bottom right panel, the initial outbreak was placed on the left.}
\label{K 08 RL}
\end{center}
\end{figure}

Further lowering the carrying capacity in the low-quality region leads to the effect described in the previous section -- an outbreak beginning in the high-capacity area may not be able to spread beyond it, even though spread in the opposite direction is still very much possible. Figure~\ref{K 08 RL}  shows spread in the direction of increasing $K$ when $K=0.8$ in the lower-quality region. It is qualitatively similar to the case with $K=0.9$ there. Also shown is the apparently final state -- the result is identical for widely varying grid resolutions -- resulting when the initial outbreak is placed in the higher-quality subdomain. Note that the outbreak does not divide, because it cannot grow quite wide enough. In the previous cases, division was made possible by influx from the low-quality region.

\begin{figure}[htbp]
\begin{center}
\includegraphics[width=.7\textwidth]{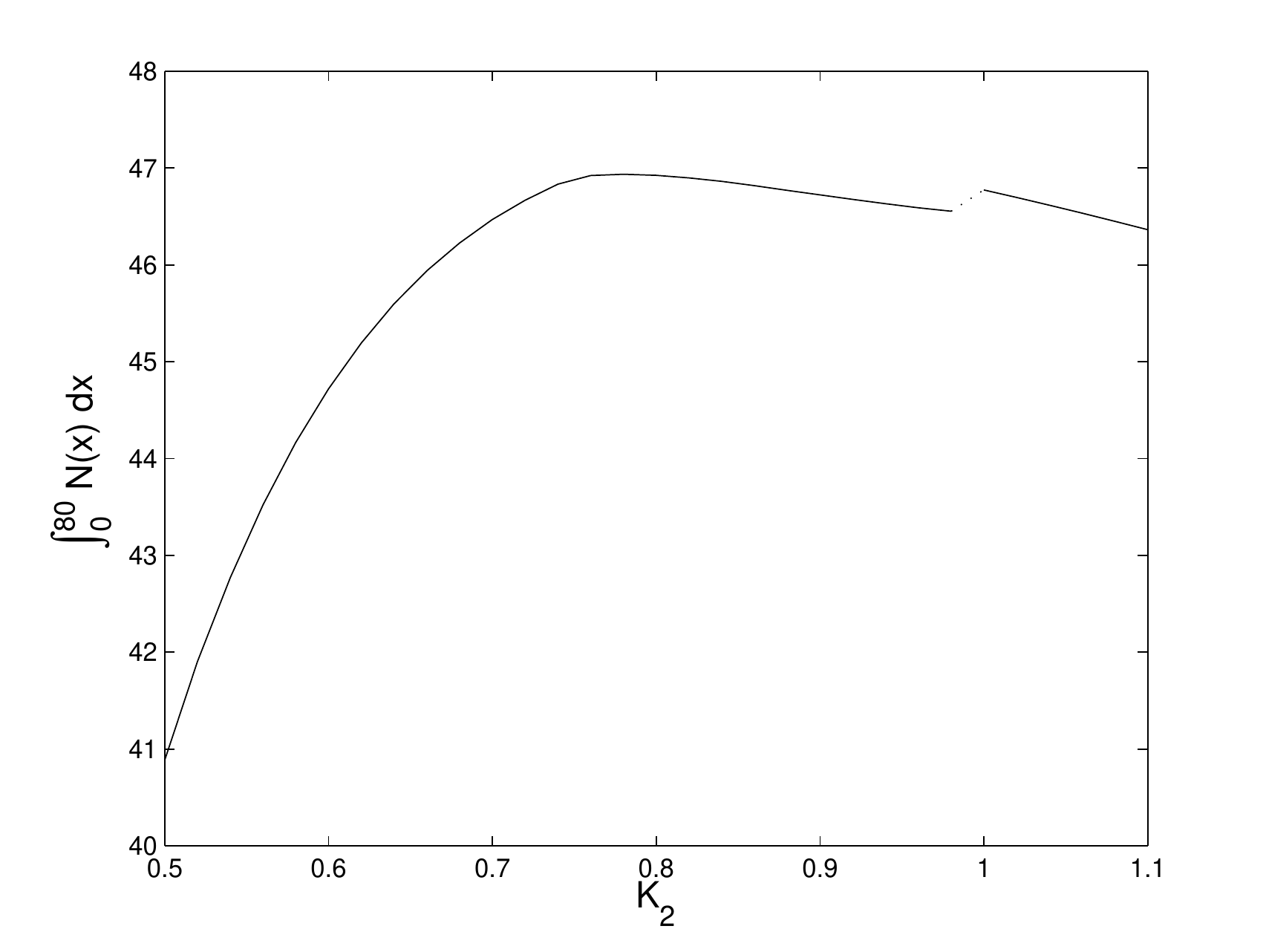} 
\caption{Consequences of varying $K_2$ on overall host abundance in a domain of length 80.}
\label{hetero counts}
\end{center}
\end{figure}

As in Section~\ref{Consequences of Habitat Quality}, we explore the effects of varying the habitat quality -- now heterogeneous -- on the overall abundance of the host. As above, we divide the domain into two (equal) regions, with $K=K_1$ in one and $K=K_2$ in the other. In Figure~\ref{hetero counts}, we set  $K_1=0.75$ constant and vary $K_2$ from $0.5$ to $1.1$ (the same range as in Figure~\ref{homo counts}). As $K_2$ approaches $K_1$, the overall abundance of the host increases; however, as $K_2$ grows larger, the abundance begins to decrease, until $K_2\approx 1$. Throughout this parameter range, the steady state has a single extinction region, on the low side of the heterogeneity as usual. But near $K_2=1$, a second division occurs in the high-$K$ subdomain, so that the steady state is similar to that seen in Figure~\ref{K 08 RL}. Contrary to the result for homogeneous habitat quality, the overall abundance of the host actually increases when this division occurs. 

The two counterintuitive behaviors evident here -- a negative relationship between carrying capacity and abundance, and an increase in overall abundance when a local extinction occurs -- are directly related. An abundance of hosts in the high-quality subdomain drives the formation of an extensive extinction region in the other subdomain by exporting parasitoids. When the productive subdomain experiences a second local extinction, that portion of the subdomain is unproductive with respect to the highly dispersive parasitoids, and the first extinction region narrows accordingly.


\section{Extension to Two Dimensions}
\begin{figure}[tp]
\begin{center}
\includegraphics[width=.4\textwidth,height=.4\textwidth]{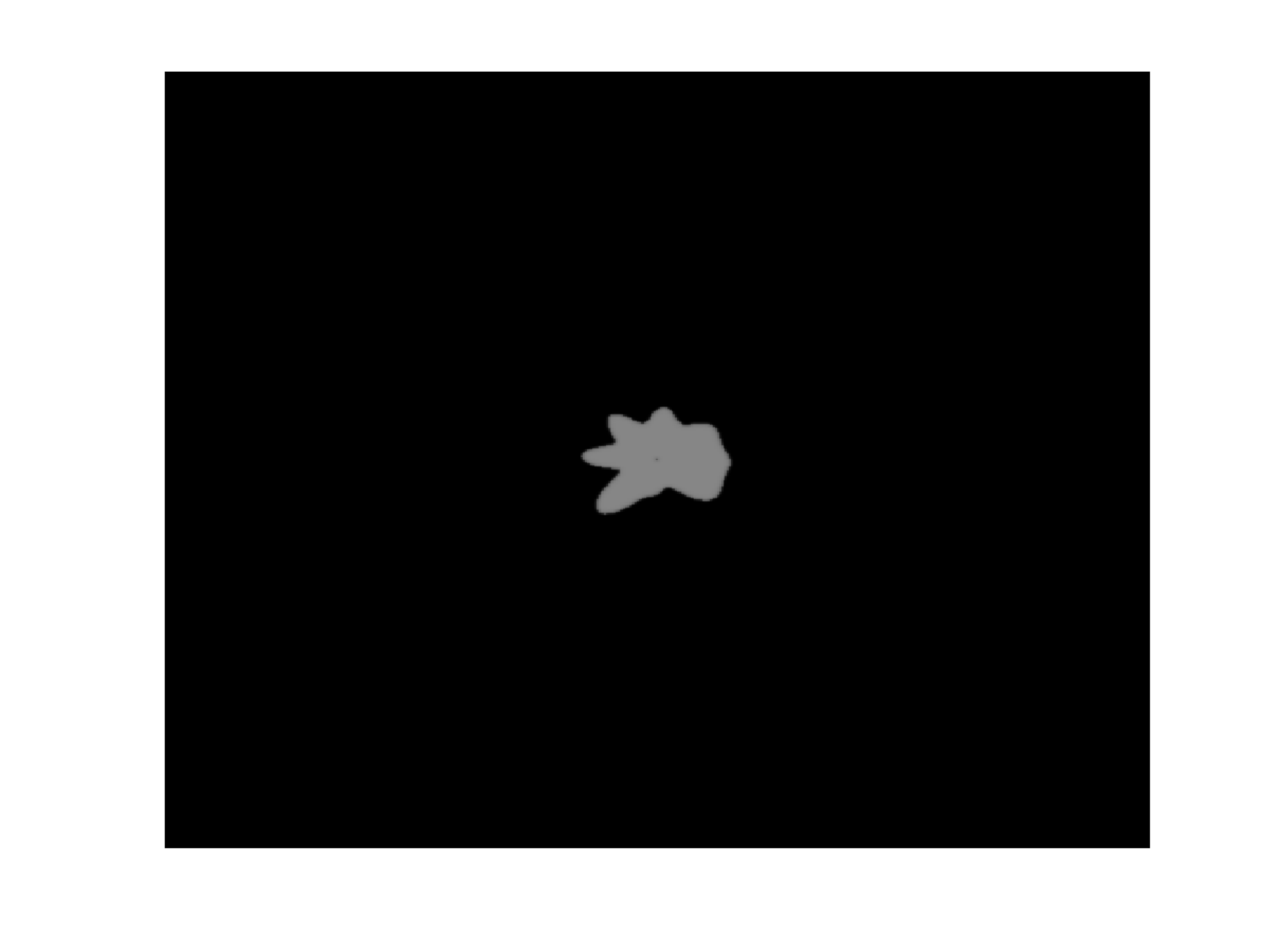} \\
\includegraphics[width=.4\textwidth,height=.4\textwidth]{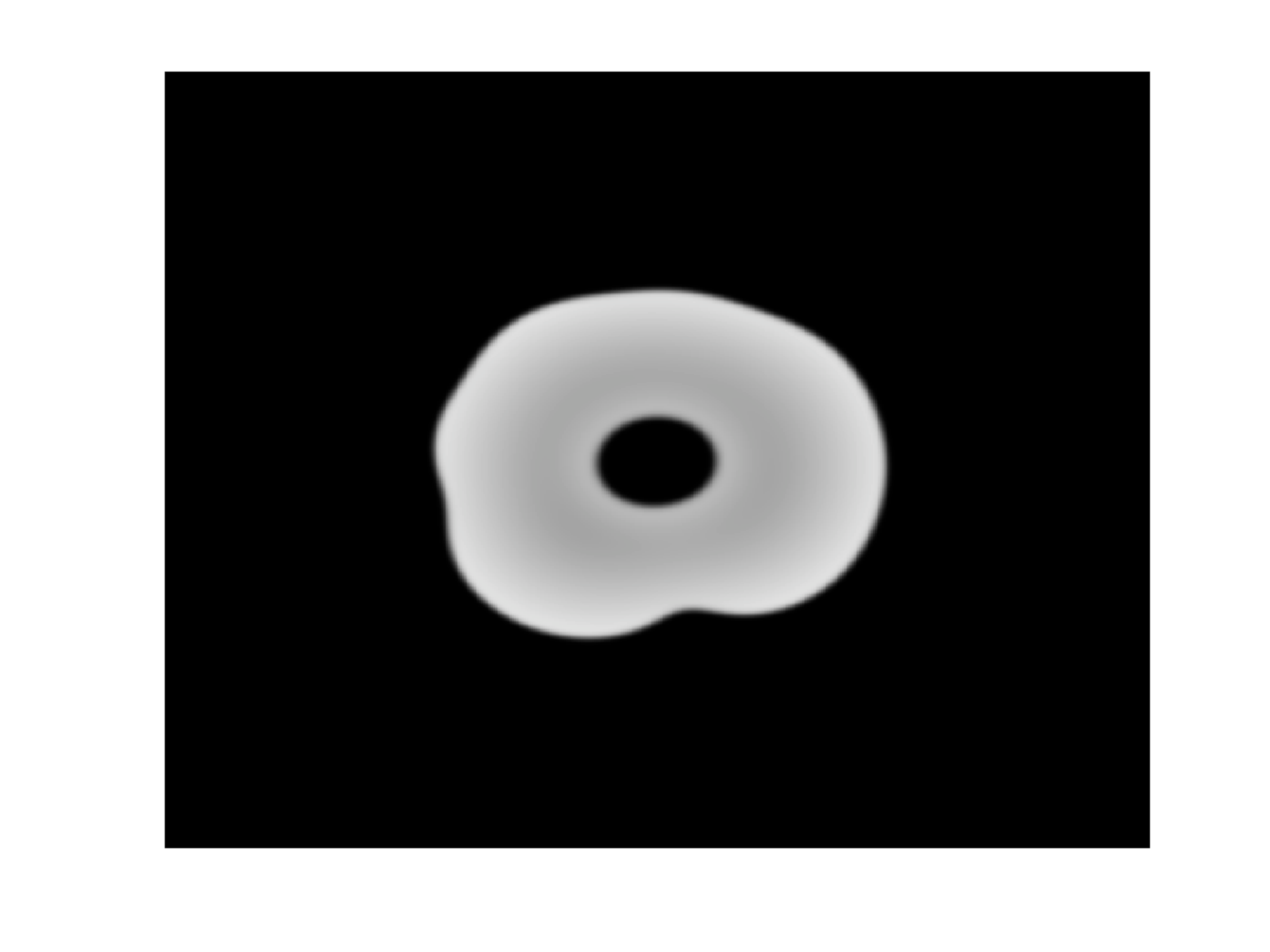} 
\includegraphics[width=.4\textwidth,height=.4\textwidth]{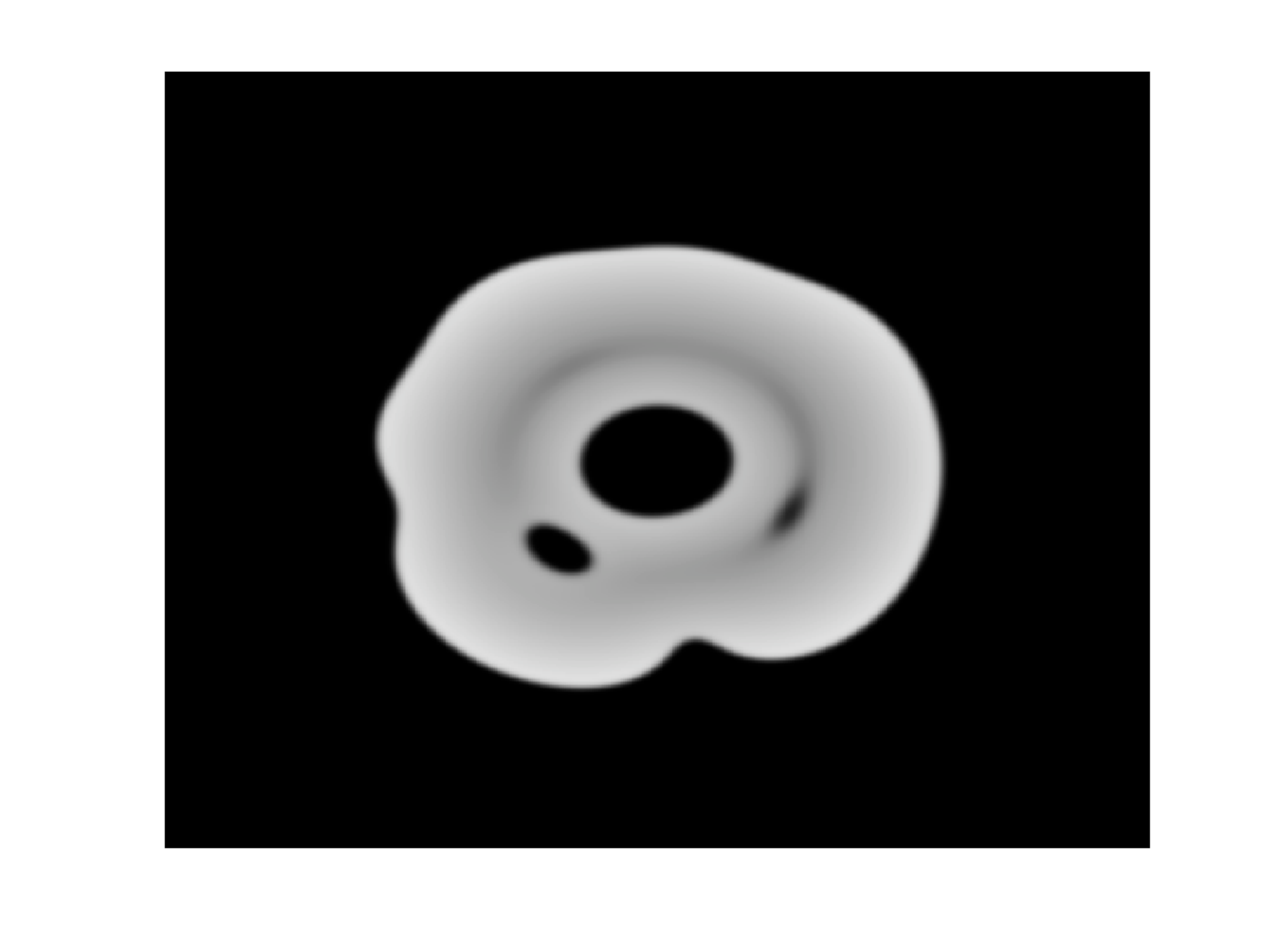} 
\includegraphics[width=.4\textwidth,height=.4\textwidth]{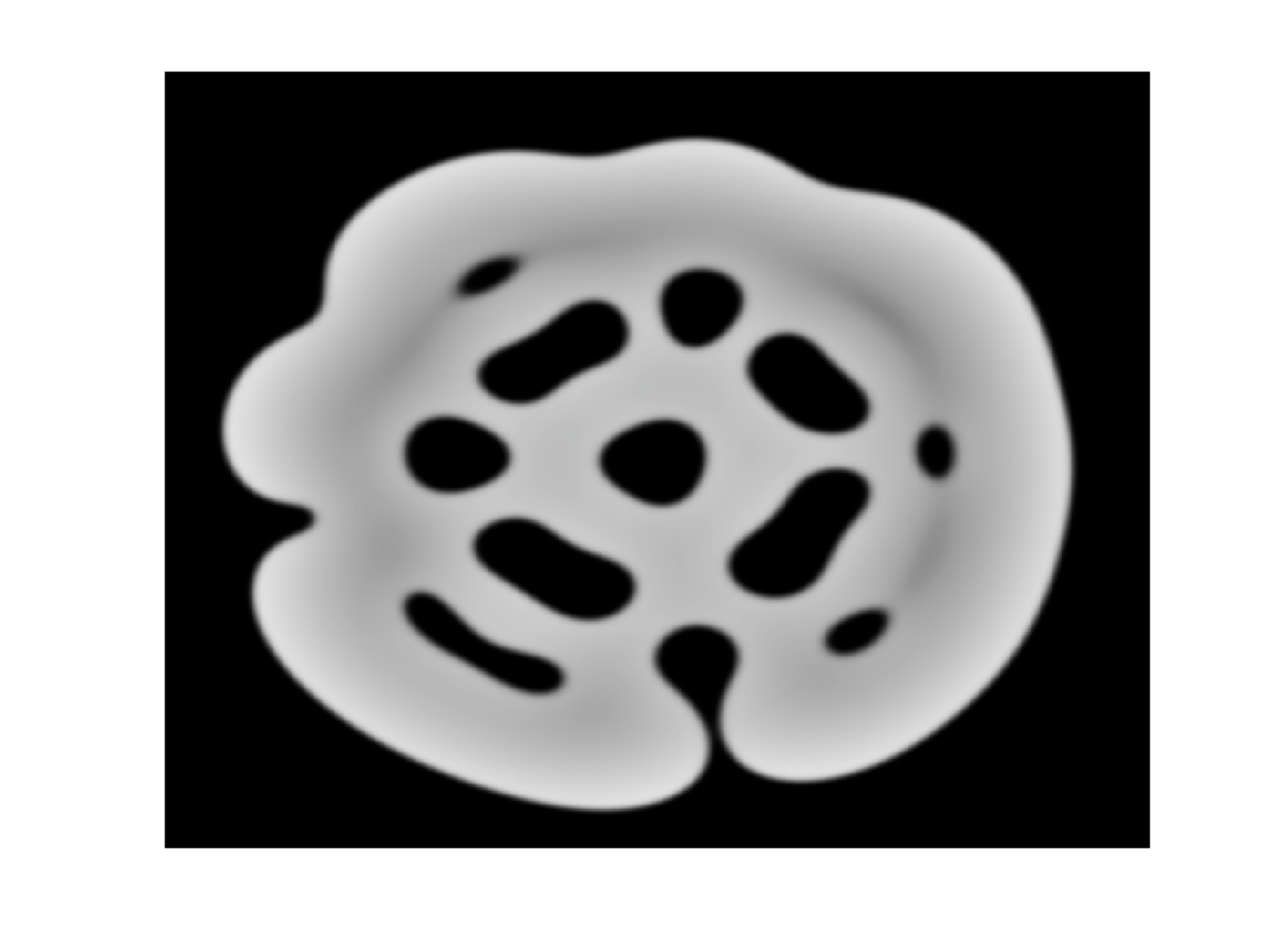} 
\includegraphics[width=.4\textwidth,height=.4\textwidth]{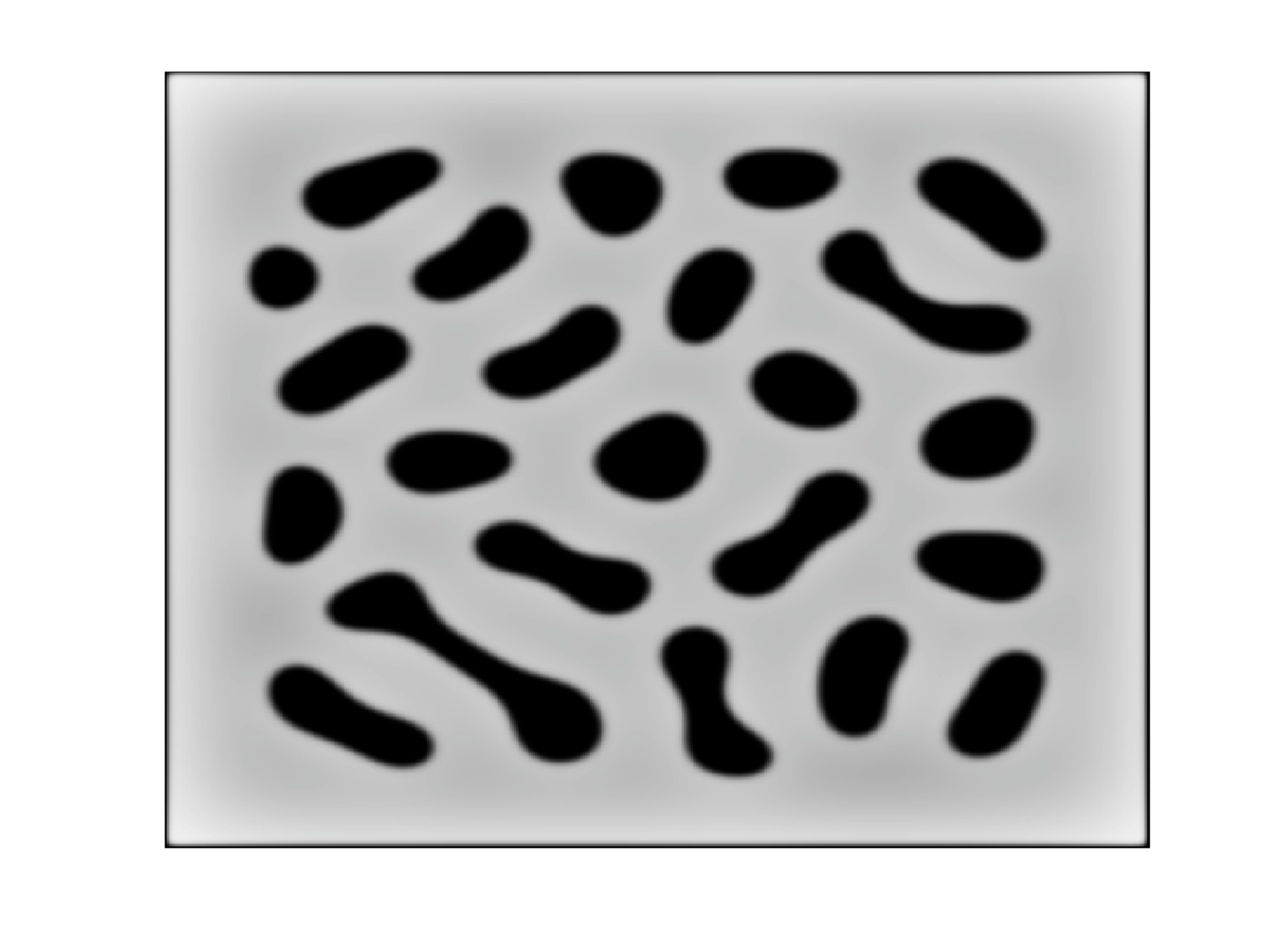} 
\caption{Two-dimensional simulation with $K=1$: Initial condition, $t=450$, $t=680$, $t=1250$, $t=2500$. Host density shown in shades of grey from $N=0$ (black) to $N=1$ (white).}
\label{homo K 1}
\end{center}
\end{figure}
We will now briefly see that all of the qualitative results in one spatial dimension, explored in Chapters~\ref{Spontaneous Patchiness} and~\ref{The Effects of Habitat Quality}, apply analogously in two dimensions. We use the model (\ref{model N})-(\ref{model P}) with Laplace kernel\begin{equation}
k_d(x,y)  =   \frac{1}
{{2\pi d^2}}e^{ - \left\| x-y \right\|_2/d}.
\label{Laplace kernel 2D} 
\end{equation}
We simulate the model with a two-dimensional fast Fourier transform convolution algorithm (very similar to that in Chapter~\ref{Spontaneous Patchiness}) on a square domain. In all of our simulations we use $f(N) = (1-N/K)(1-0.2)$, $a=2$, $D=10$, $\varepsilon = 0.3$, and domain $\Omega = [0,240]\times[0,240]$.

\begin{figure}[htb]
\begin{center}
\includegraphics[width=.4\textwidth,height=.4\textwidth]{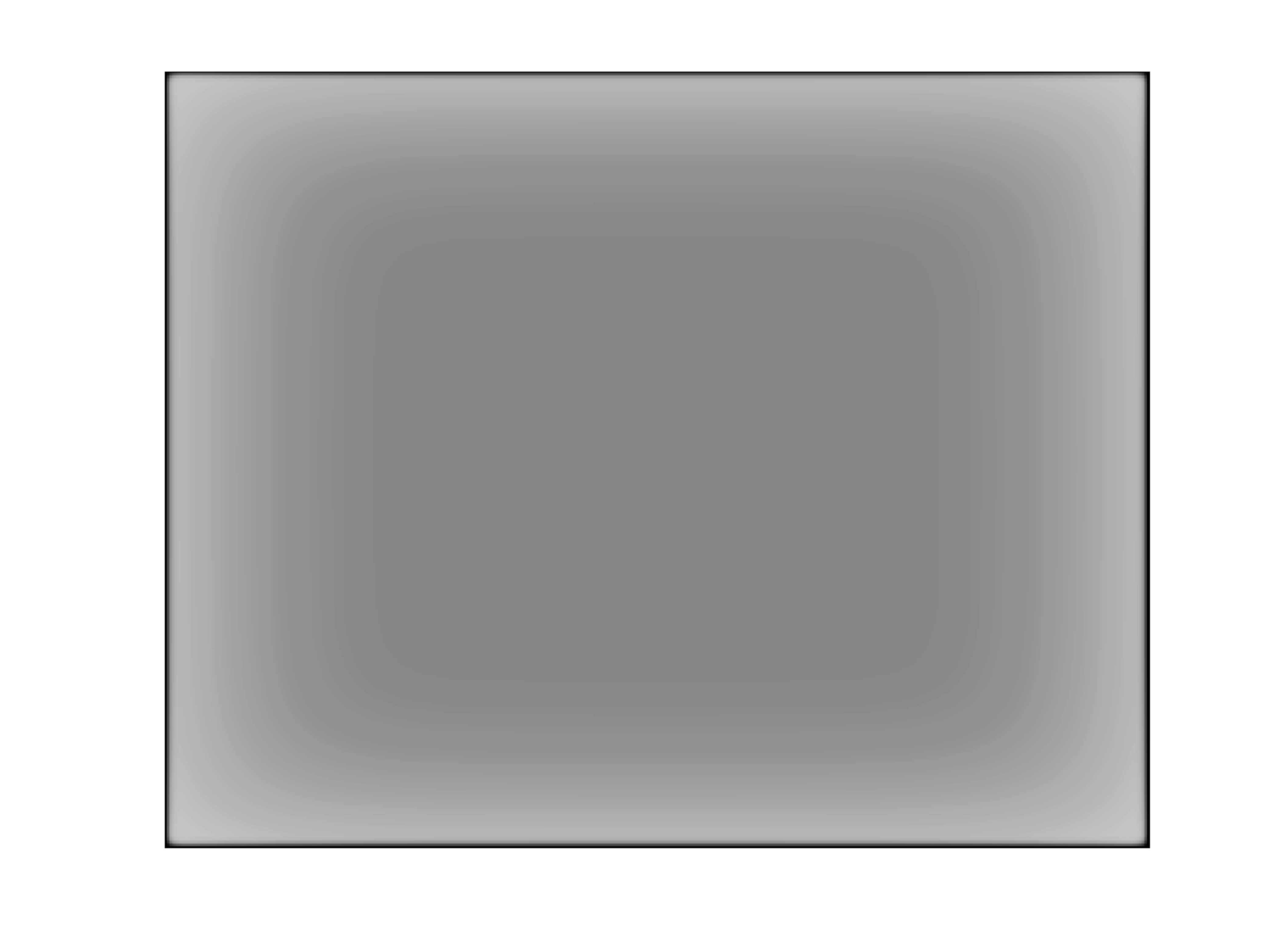} 
\includegraphics[width=.4\textwidth,height=.4\textwidth]{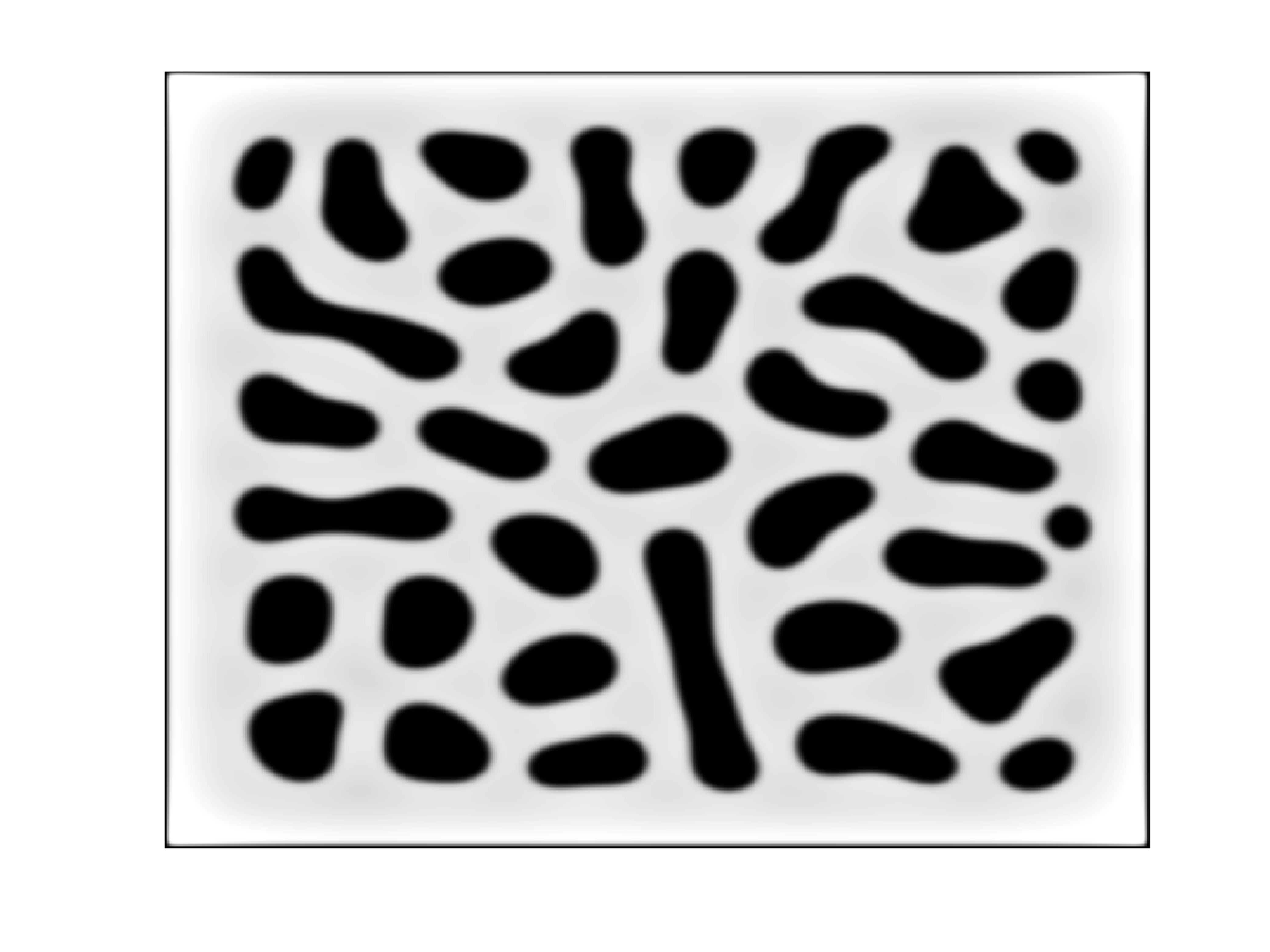} 
\caption{Two-dimensional simulation with $K=0.8$ (left) and $K=1.2$ (right) at $t=2500$. Host density shown in shades of grey from $N=0$ (black) to $N\ge1$ (white).}
\label{homo K 08 12}
\end{center}
\end{figure}

In our first three examples we compare the outcome of a small, random initial outbreak on a domain with homogeneous $K$ of various values. Figure~\ref{homo K 1} shows the initial condition used in all three cases, along with some snapshots of the spread of the host with $K=1$. As the outbreak spreads, a local extinction occurs in its center. Later, a ring-shaped extinction begins to form. However, the initial irregularities in the outbreak's edged lead to inclusions which disrupt its symmetry. Eventually a second near-ring forms, and spread continues to the boundary. With a circular initial condition, extinctions would occur in concentric rings until the outbreak reached the domain edge and disturbances propagated back into the domain, breaking symmetry. 

Figure~\ref{homo K 08 12} shows the effects of raising and lowering carrying capacity on the whole domain. With $K=0.8$, the outbreak fills the domain with no extinction areas. With $K=1.2$, the long-term state is patchier.

\begin{figure}[htb]
\begin{center}
\includegraphics[width=.4\textwidth,height=.4\textwidth]{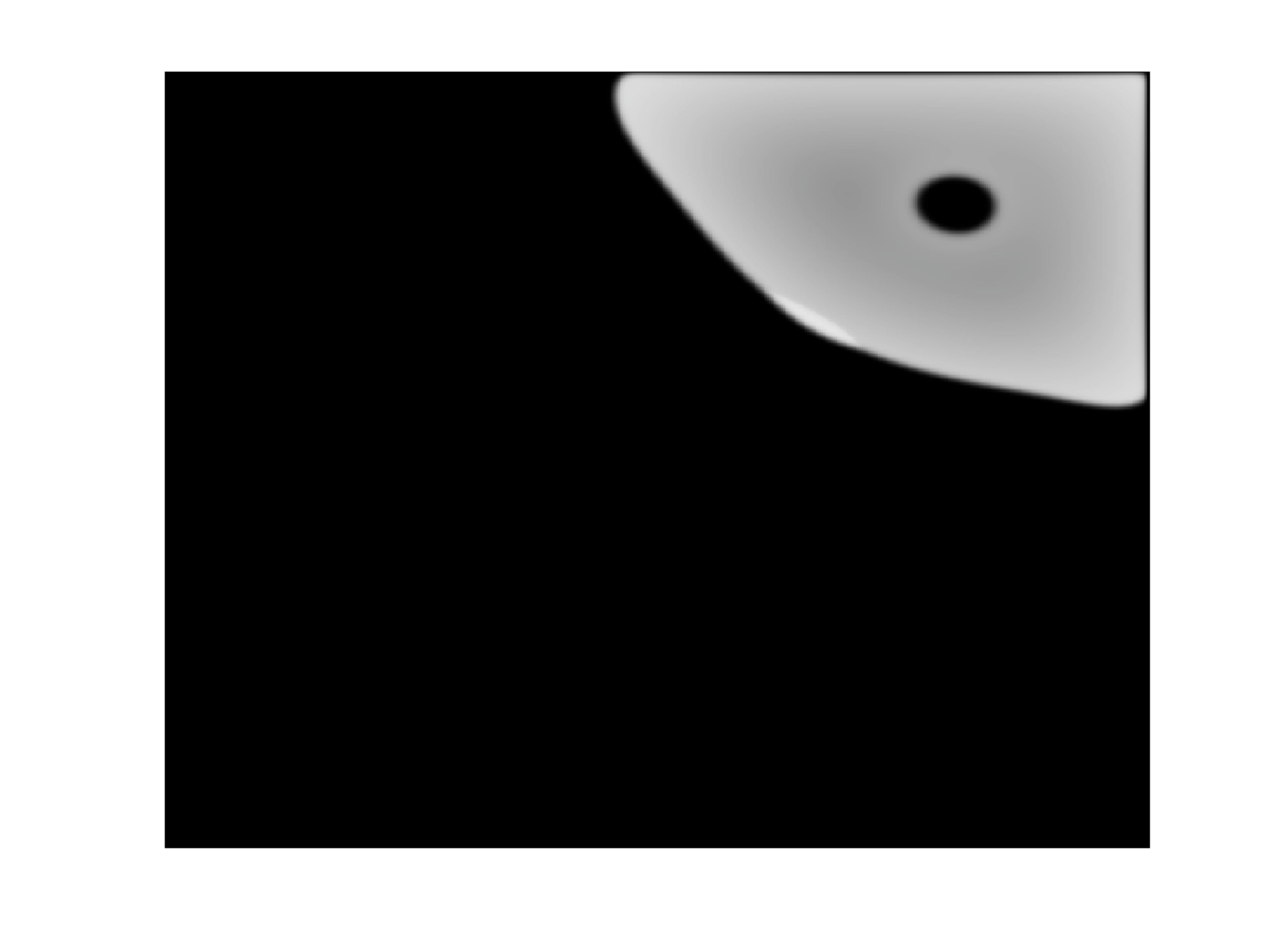} 
\includegraphics[width=.4\textwidth,height=.4\textwidth]{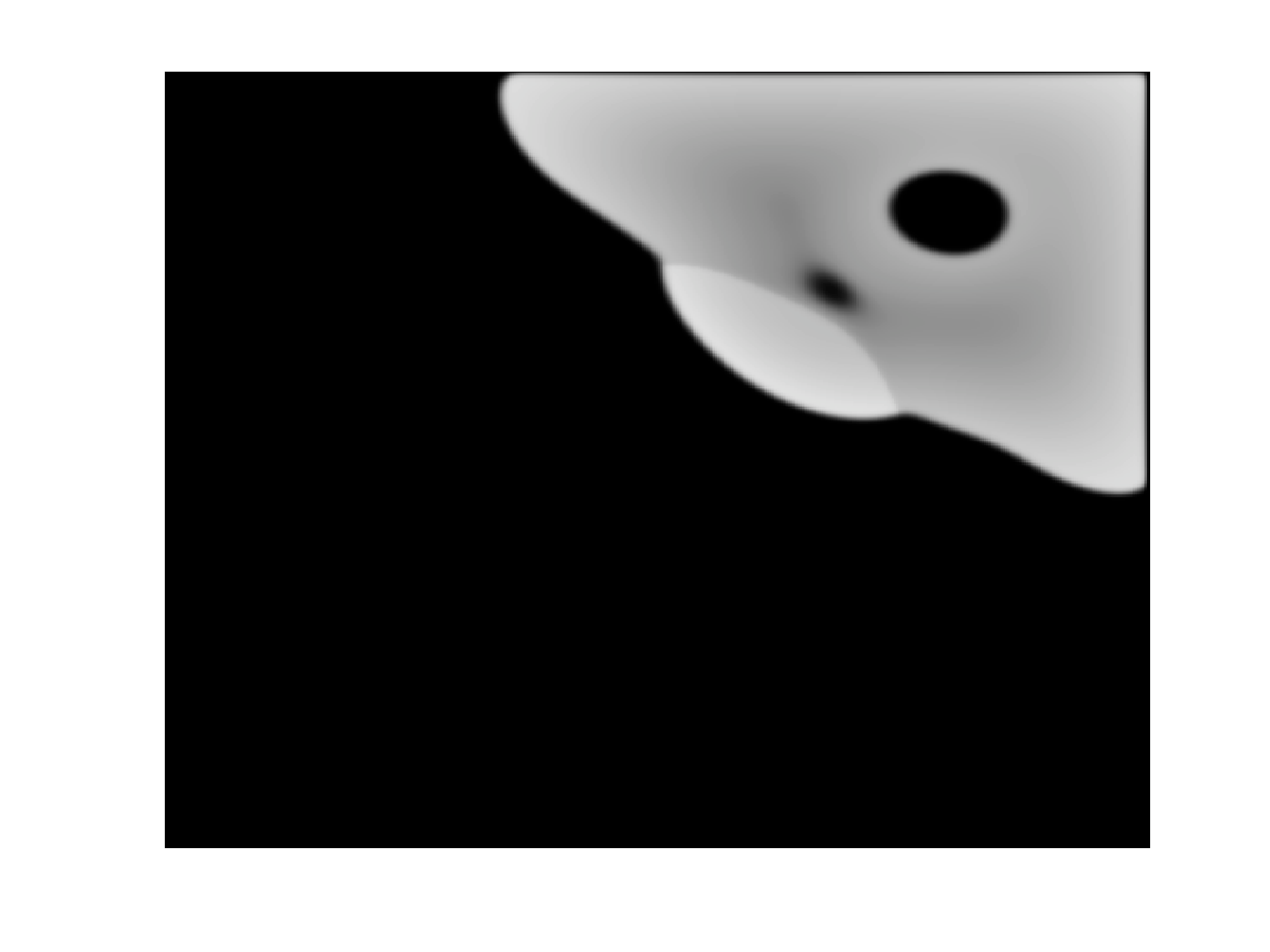} 
\includegraphics[width=.4\textwidth,height=.4\textwidth]{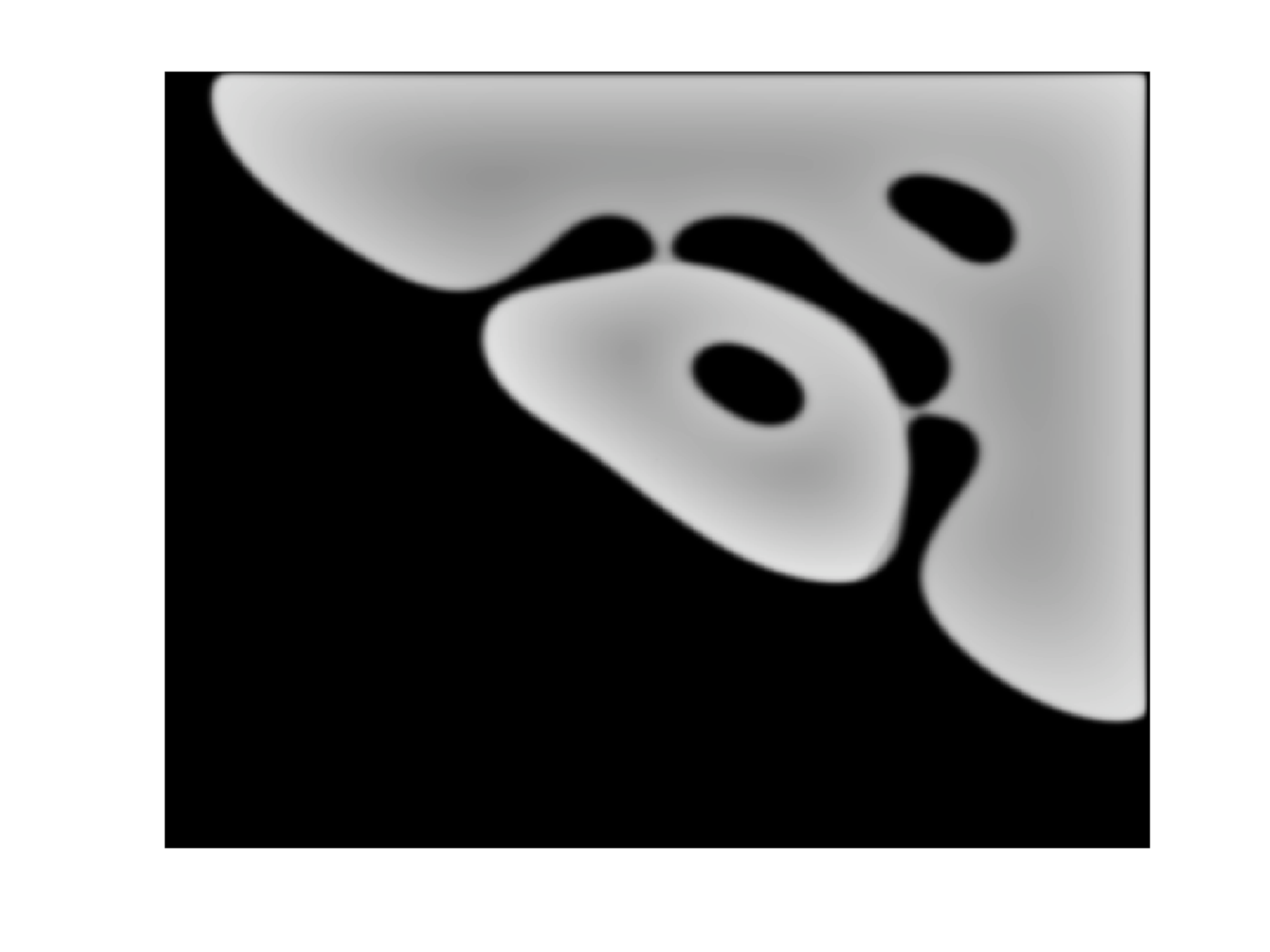} 
\includegraphics[width=.4\textwidth,height=.4\textwidth]{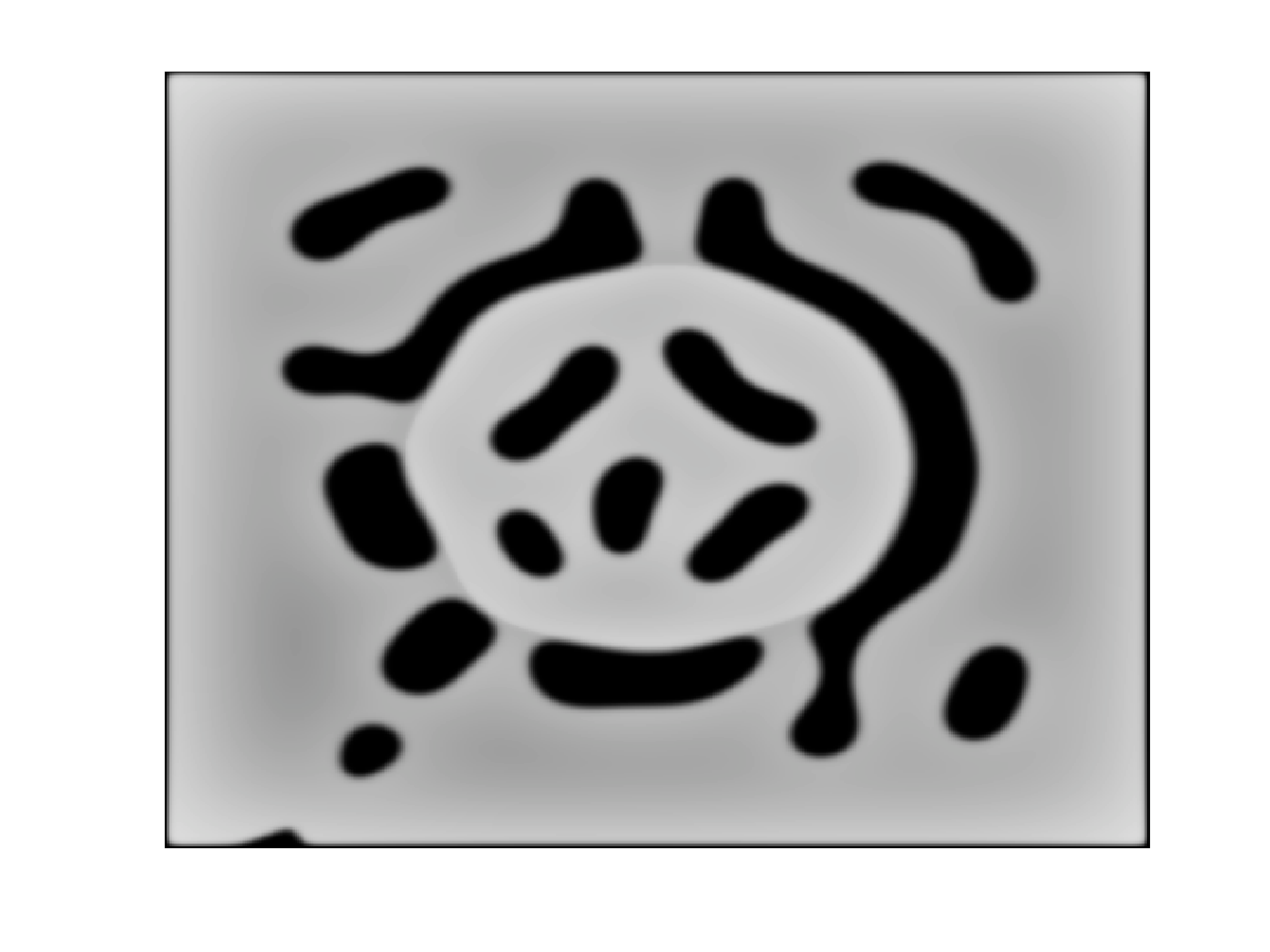} 
\caption{Two-dimensional simulation with heterogeneous domain quality ($K=1$ in the center, $K=0.9$ at the edges): $t=450$, $t=750$, $t=1500$, $t=4000$. Host density shown in shades of grey from $N=0$ (black) to $N=1$ (white).}
\label{hetero K 09 out}
\end{center}
\end{figure}

\begin{figure}[htbp]
\begin{center}
\includegraphics[width=.4\textwidth,height=.4\textwidth]{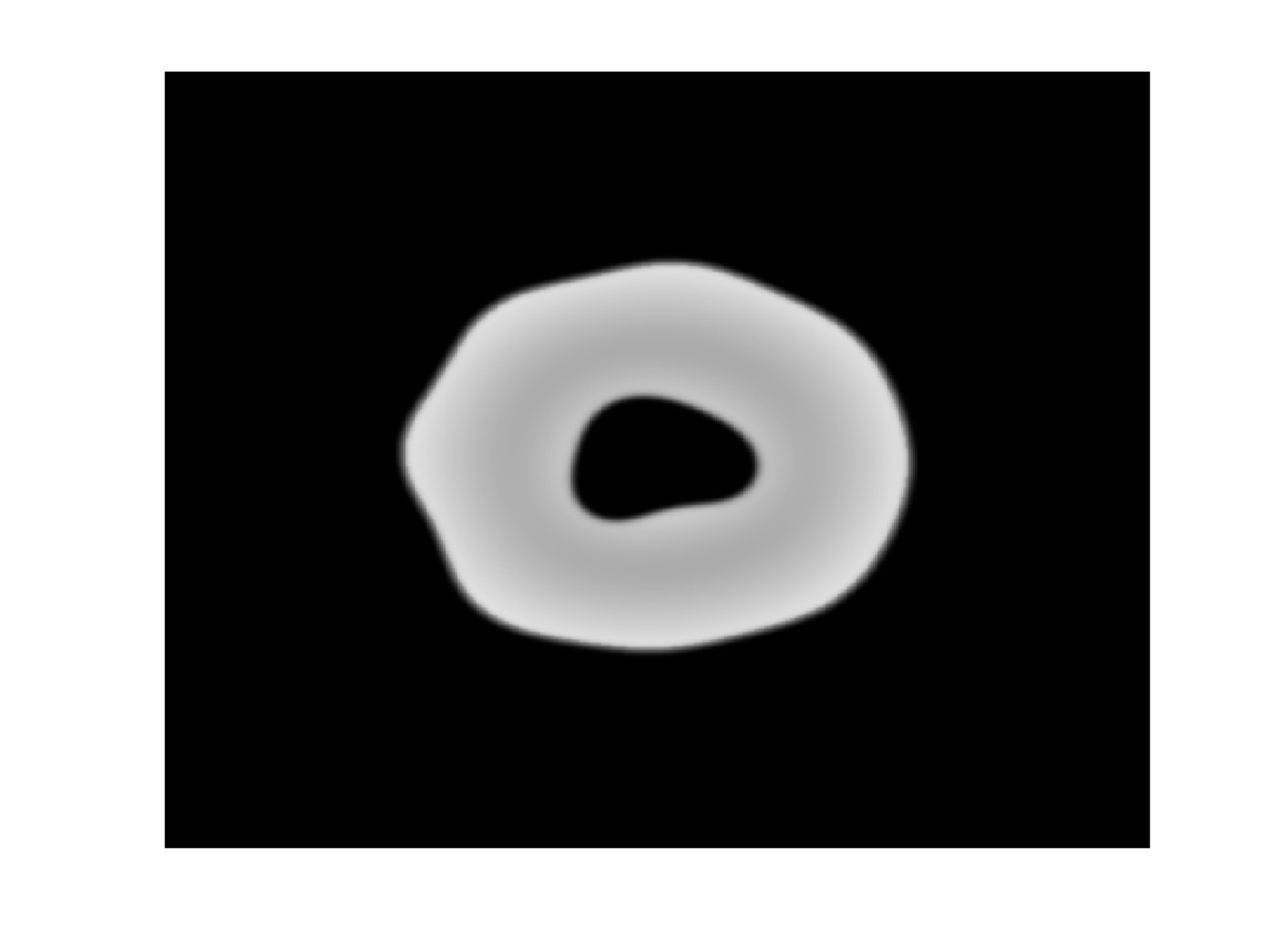} 
\includegraphics[width=.4\textwidth,height=.4\textwidth]{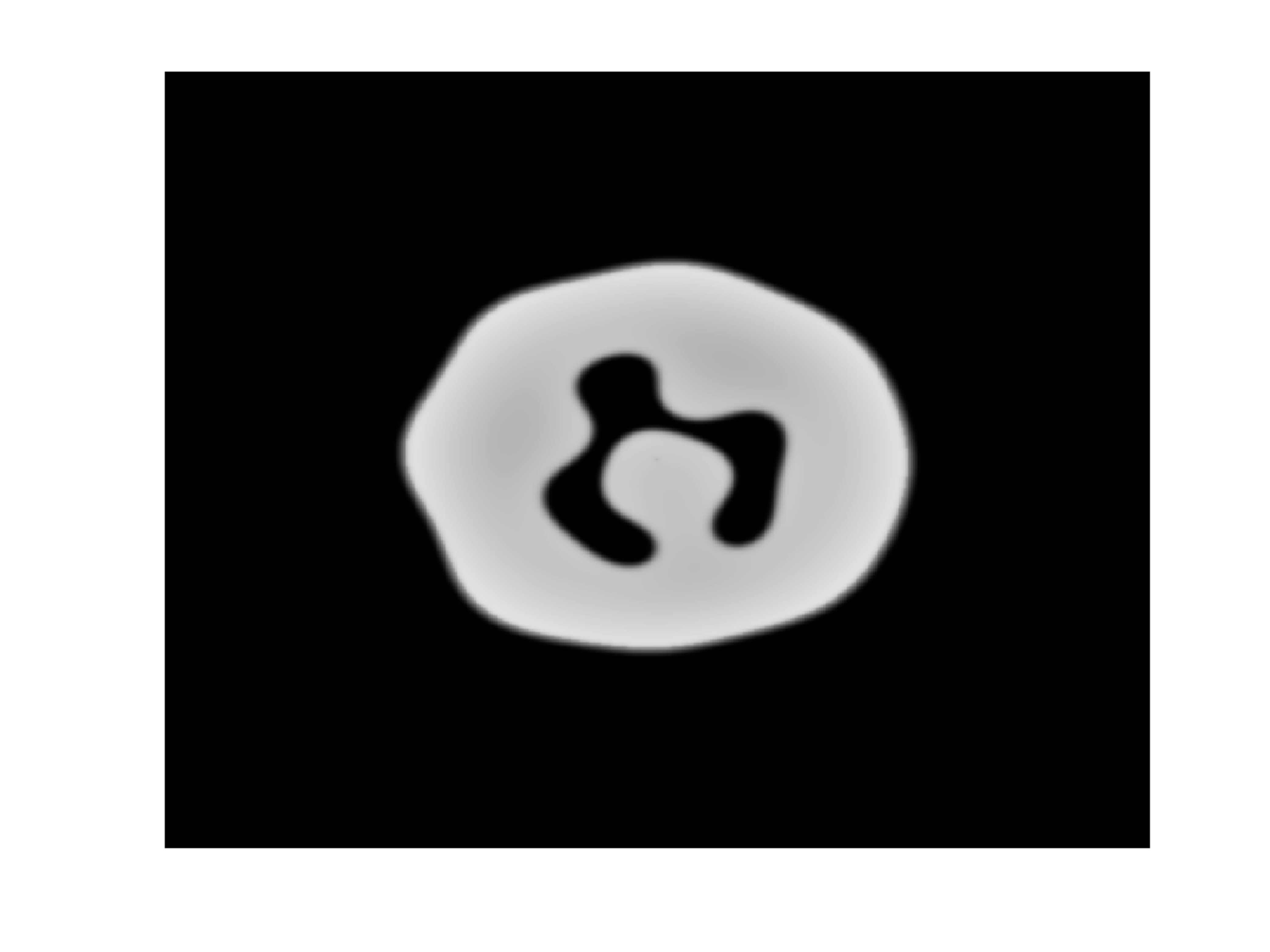} 
\caption{Two-dimensional simulation with heterogeneous domain quality ($K=1$ in the center, $K=0.6$ at the edges): $t=2000$, $t=4000$. Host density shown in shades of grey from $N=0$ (black) to $N=1$ (white).}
\label{hetero K 06 in}
\end{center}
\end{figure}

\begin{figure}[htbp]
\begin{center}
\includegraphics[width=.4\textwidth,height=.4\textwidth]{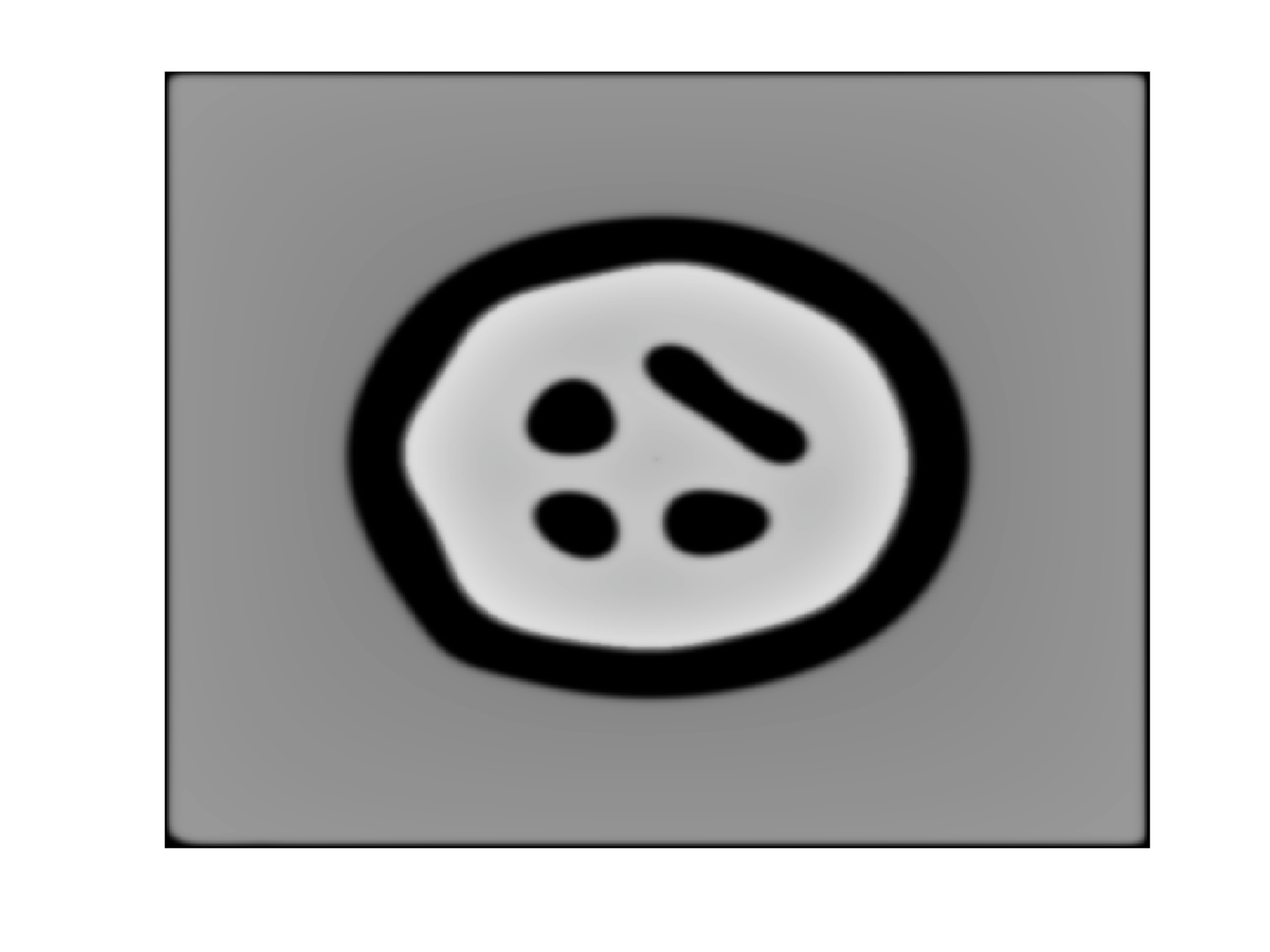} 
\caption{Two-dimensional simulation with heterogeneous domain quality ($K=1$ in the center, $K=0.6$ at the edges) at $t=6000$. Host density shown in shades of grey from $N=0$ (black) to $N=1$ (white).}
\label{hetero K 06 out}
\end{center}
\end{figure}

To simulate heterogeneity, we place a roughly circular region with $K=1$ in the center of the domain, and lower $K$ outside this area. In Figure~\ref{hetero K 09 out}, $K=0.9$ in the lower-quality region. Note that this value of $K$ is high enough for patchiness in a homogeneous environment. We place an initial outbreak in the upper-right of the domain, completely outside the region with higher $K$. The outbreak begins to form extinction areas, and easily spreads into the higher-quality habitat. Local extinctions begin to form around the edges of that area, and then inside it. Finally, the domain is filled and local extinctions largely group around the edge of the region with higher $K$, though unmistakably on its exterior (the region is visible as a rough circle of greater host density).

\begin{figure}[p]
\begin{center}
\includegraphics[width=.4\textwidth,height=.4\textwidth]{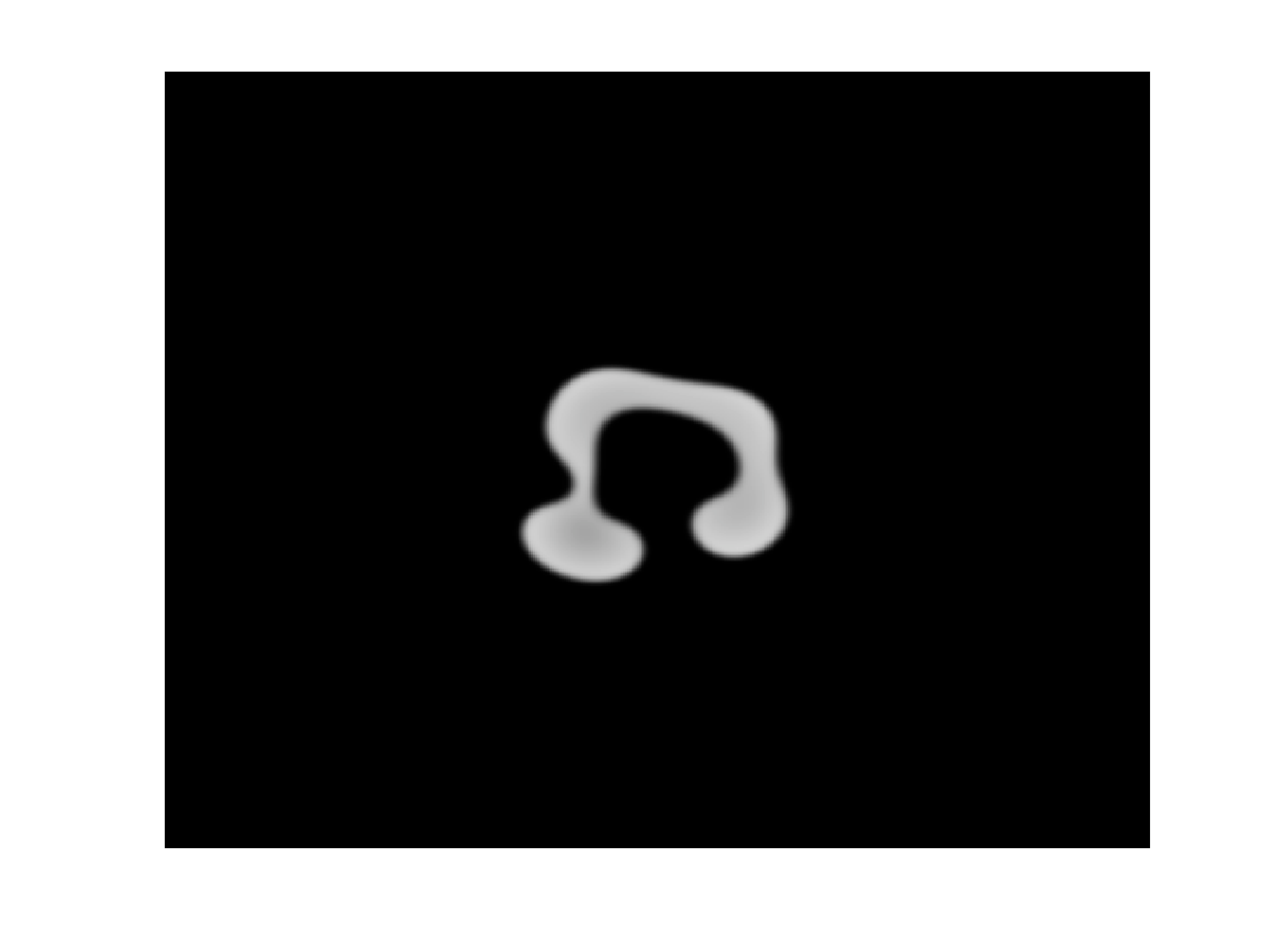} 
\includegraphics[width=.4\textwidth,height=.4\textwidth]{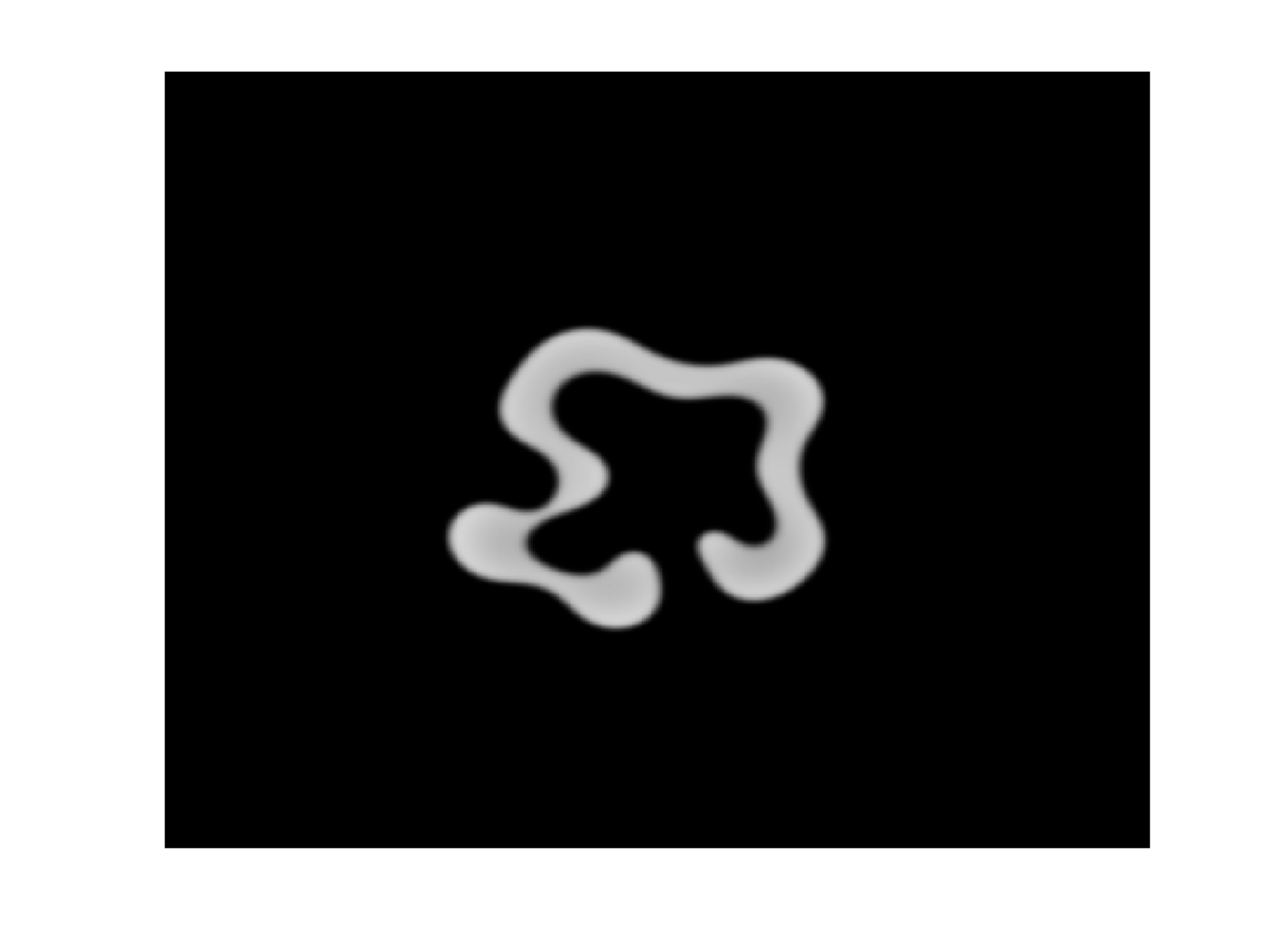} 
\includegraphics[width=.4\textwidth,height=.4\textwidth]{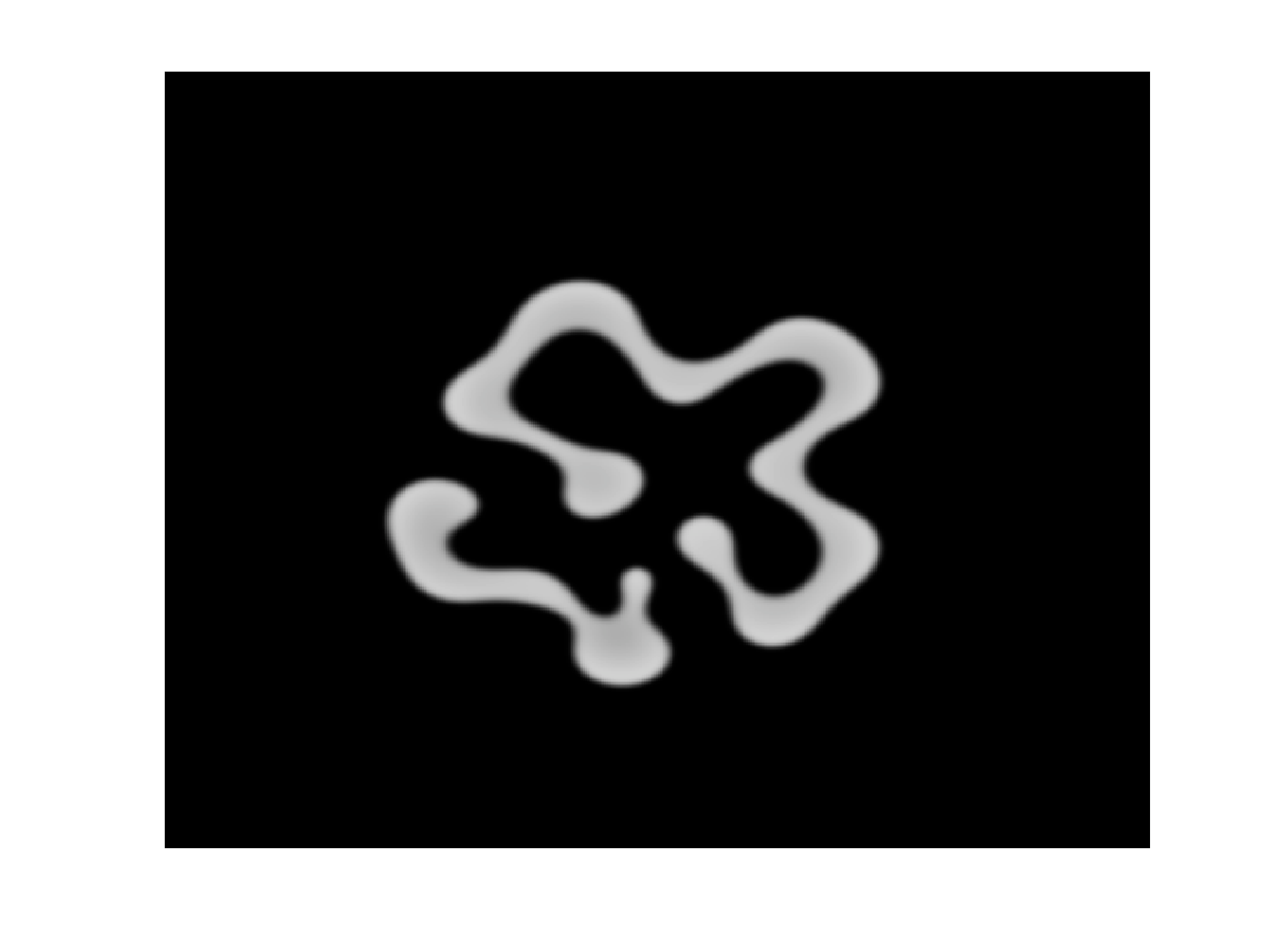} 
\includegraphics[width=.4\textwidth,height=.4\textwidth]{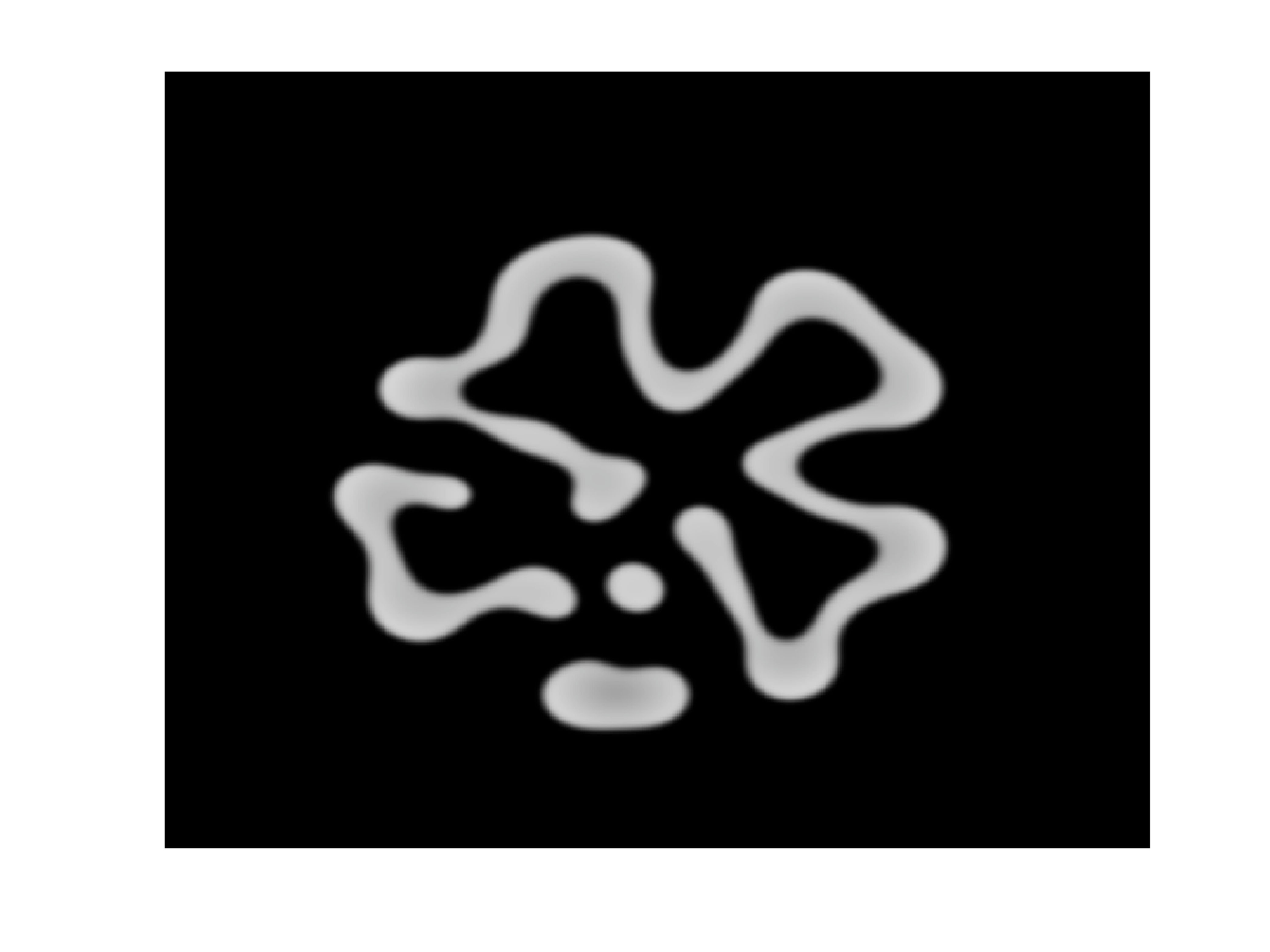} 
\includegraphics[width=.4\textwidth,height=.4\textwidth]{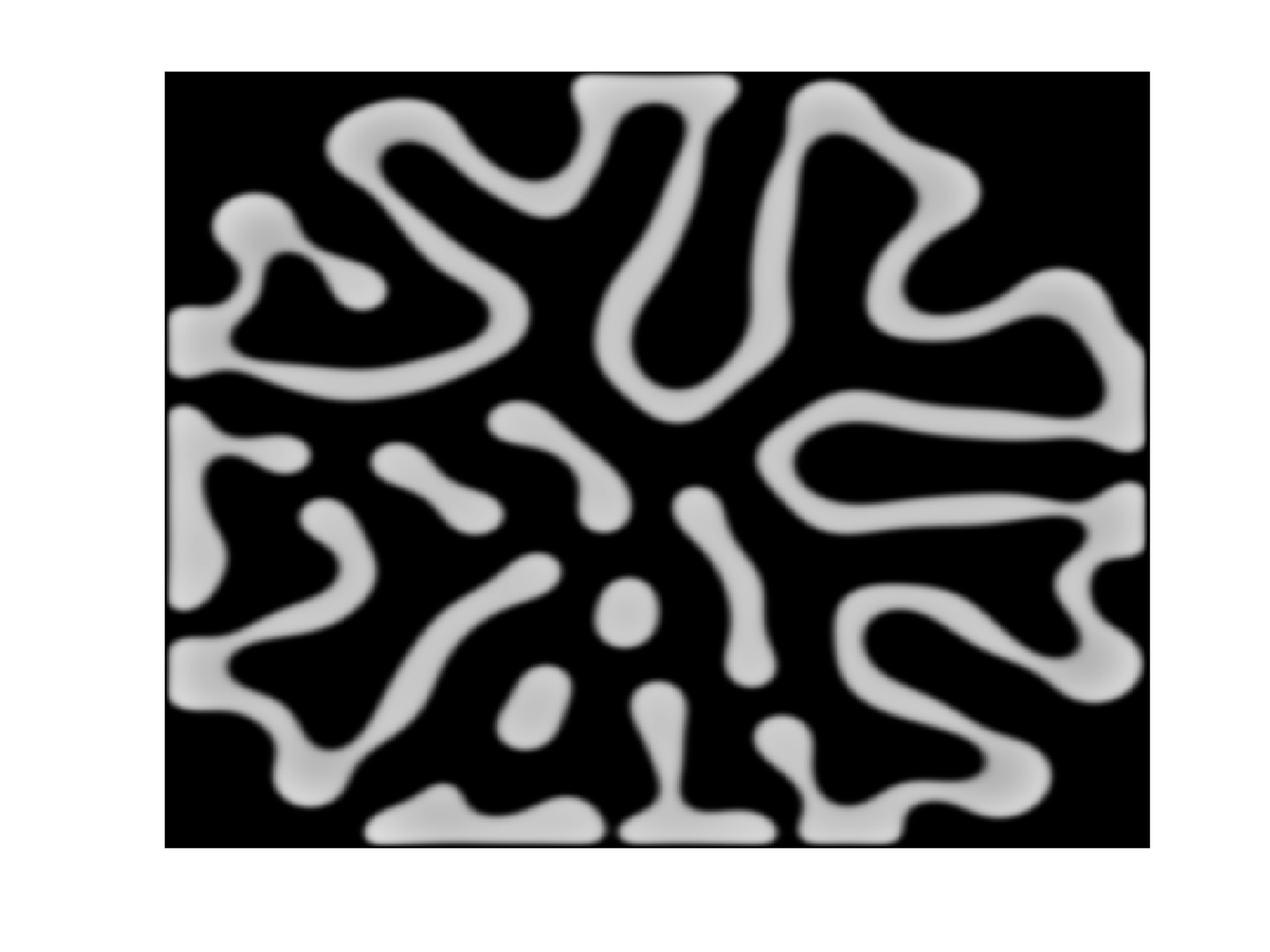} 
\includegraphics[width=.4\textwidth,height=.4\textwidth]{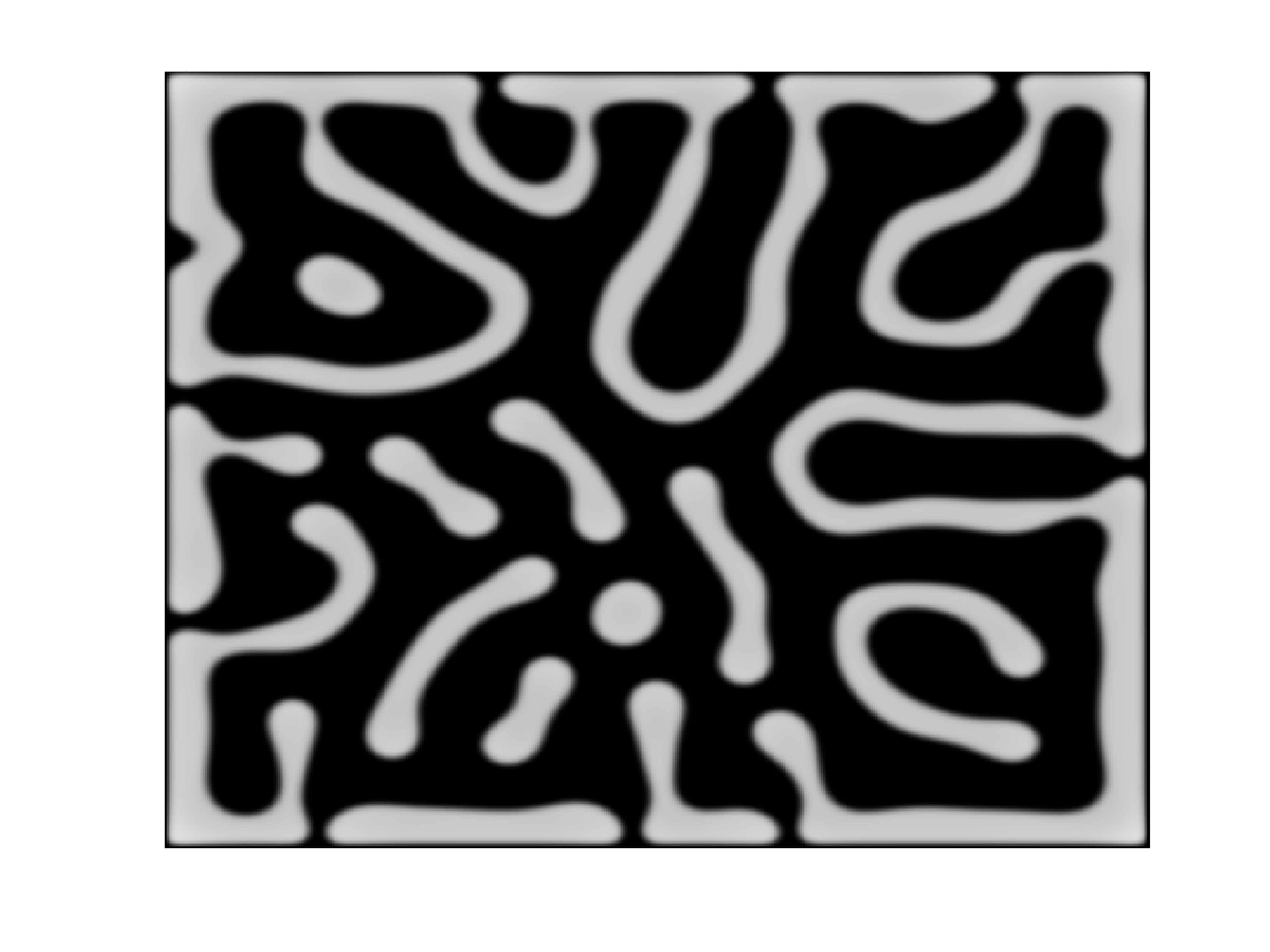} 
\caption{Two-dimensional simulation with $K=1$ and $a=4$: $t=500,1000,1500,2000,4000,6000$. Host density shown in shades of grey from $N=0$ (black) to $N=1$ (white).}
\label{homo a 4}
\end{center}
\end{figure}

If the initial outbreak lies inside the region of higher carrying capacity, it easily spreads outside and the long-term state is qualitatively identical to that in Figure~\ref{hetero K 09 out}. If we lower $K$ outside the high-quality region to $K=0.8$, which as we saw above produces no patchiness if the habitat is homogeneous, the outcome is essentially the same. Hosts can spread across the heterogeneity in either direction, and in the end there is a solid ring of local extinction around the high-quality region, and now with no local extinctions away from it. However, if we further lower carrying capacity outside to $K=0.6$, an outbreak beginning inside the region of high carrying capacity cannot escape it. This is shown in Figure~\ref{hetero K 06 in}. The high-capacity region is soon filled, but spread is stalled at its edge and the interior extinctions simply begin to rearrange. On the other hand, as shown in Figure~\ref{hetero K 06 out}, if the outbreak starts from the low-quality region, it can spread throughout the domain. 

Returning briefly to the homogeneous case with $K=1$ throughout the domain, if we take $a$ to be quite large, the shape of the host's spread becomes interesting and similar to that observed for a partial differential equation model with an Allee effect in the victim. Figure~\ref{homo a 4} shows the resulting simulation with the same initial condition as in Figure~\ref{homo K 1}. The host's spatial behavior -- coalescing into thin ribbons and breaking into patches while spreading -- is reminiscent of the spread of the prey in~\cite{Petrovskii}, where it is termed ``patchy invasion".

\section{Discussion}
We began this chapter with an analysis of an Allee effect induced by saturating predation. In many victim-exploiter models, the exploiter nullcline is a vertical line. One prominent example, for a continuous-time model, is found in~\cite{Rosenzweig}. In such cases, the specifics of the host nullcline -- other than its humped shape -- may not be important. However, with the nonspatial model (\ref{no dispersal N})-(\ref{no dispersal P}), in which the parasitoid nullcline has finite slope, our implicit assumptions require that the host nullcline have relatively small slope to the left of its maximum, as it does with the growth function we used in Chapter~\ref{Spontaneous Patchiness}. 

While the quadratic function we have used to describe growth is not realistic for many instances of the Allee effect in nature, we saw that it might closely match the properties of a growth function derived from mechanistic principles based on the reality of the system motivating our model. Specifically, an Allee effect induced by saturating predation is likely to generate a growth function of very similar shape and behavior to our quadratic approximation. 

We saw numerically how increased predation on the pupal stage of the host strengthens the Allee effect. Importantly, and in contrast to Allee growth not induced by a predator, the host nullcline is roughly quadratic even for a relatively weak Allee effect (Figure~\ref{fig} for low $m$). We found the parameter condition under which the effect is strong -- i.e., such that the host is bistable in the absence of parasitoids. Evidence from the field suggests that the effect is indeed strong there~\cite{Harrison2}. Finally, we saw how the induced growth function, and therefore the host nullcline, changes with a change in carrying capacity $K$. 

The movement of the growth function upon a change in $K$ has interesting implications for the effects of varying habitat quality. This is largely due to the paradox of enrichment~\cite{Rosenzweig}, but its destabilizing effects are mitigated by dispersal, with spatial pattern formation striking a balance between stability and extinction. Also, in the cases we considered numerically, increasing carrying capacity within any parameter range that produces a fixed number of patches seems to lead to increased overall host abundance, though the abundance falls slightly upon patch division.

Although spatial spread is not stopped in a homogeneous environment for our continuous-space model, as it may be for metapopulation models with homogeneous sites~\cite{Hadjiavgousti,Keitt}, the heterogeneities required to stop spread are fairly mild. We have also seen how adjacent regions with differing habitat quality affect one another. Similar situations, for continuous-time models, are outlined in~\cite{Fagan}. The effects of varying the quality of only a portion of the habitat can be counterintuitive. If the carrying capacity in one area is made sufficiently large, the overall abundance of the host may actually decrease. This result is conceptually similar to theoretical studies of marine systems which have predicted that, in many cases, establishment of a reserve could lower the overall abundance of the protected species in the presence of predation~\cite{Micheli} or infectious disease~\cite{McCallum}.

%
%
%
%

    \newchapter{Persistence of Nematodes with an Alternate Host} 
{Persistence of Nematodes Augmented by an Alternate Host} 
{Persistence of Parasitic Nematodes Augmented by a Scarce Alternate Host} 
\label{Nematode Chapter}

\section{Introduction}
As explained in Chapter~\ref{The Lupine Habitat}, the roots of the bush lupine provide shelter and sustenance to ghost moth larvae, which are in turn parasitized by entomopathogenic nematodes. In this way, the nematodes act to promote the health of the shrubs in a trophic cascade. However, a single infected ghost moth larva can produce hundreds of thousands of nematodes, leading to pronounced year-to-year population cycles, which can result in the local extinction of the nematode when stochasticity at low numbers is taken into account \cite{Dugaw}. 

The ability of nematodes to disperse between roots and form a metapopulation is questionable, which calls into question the global persistence seen in nematode-ghost moth systems in the field \cite{Dugaw2}. The aim of this chapter is to explore a new possibility to explain the observed persistence of nematodes -- the influence of an alternate host.

Cases in which the presence of a second host is pivotal to the survival of a parasitic population, even if that host is inferior to the first, have been observed both theoretically~\cite{Holt} and in the field~\cite{VanAlphen}. To model a second host for the nematode, we will build upon the deterministic model from \cite{Dugaw}, in which wet season dynamics (from $t=0$ to $t=T$) are given by 
\begin{equation}
 H'(t) =  - k_H H(t) - \beta H(t)N(t),    \;    H(0)=H_0
\label{dugaw model 1}
\end{equation}
\begin{equation}
 N'(t) =  - k_N N(t) - \beta H(t)N(t) + \beta \Lambda (t - \tau )H(t - \tau )N(t - \tau ), 
\label{dugaw model 2}
\end{equation}
with the next year's initial nematode numbers, $N(0)$, determined by
\begin{equation}
\lambda _o N(T) + \lambda _i \beta \int\limits_{T - \tau }^T {\Lambda (t)H(t)N(t)dt} .
\label{dugaw model 3}
\end{equation}
First we will examine the dynamics of this original model in slightly greater detail than previously done. We will then formulate a new model accounting for a second host. Finally, through numerical simulation and reasoning from the original model's dynamics, we will see how the addition of another host, even in very small numbers, may positively influence the persistence of the parasitoid.

\section{Details from the Original Model}
\label{Details from the Original Model}
We will see shortly how certain properties of the model given in \cite{Dugaw} may contribute to the ability of an alternate host to enhance the persistence of the parasitoids. It will be fruitful to examine these before moving on to the formulation of the new model.

\subsection{In-Year Dynamics}
The behavior of the model during an overexploited wet season is vitally important for our purposes since this behavior leads to the dangerously low densities of the subsequent season. In truth, the danger caused by overexploitation of the host is not the removal of the host from the system, since it is replenished every wet season. The danger lies in the premature removal of the host, such that no infections occur late enough in the wet season for the resulting nematodes to be protected from the high mortality of the dry season.

\begin{figure}[htbp]
\begin{center}
\includegraphics[width=.46\textwidth]{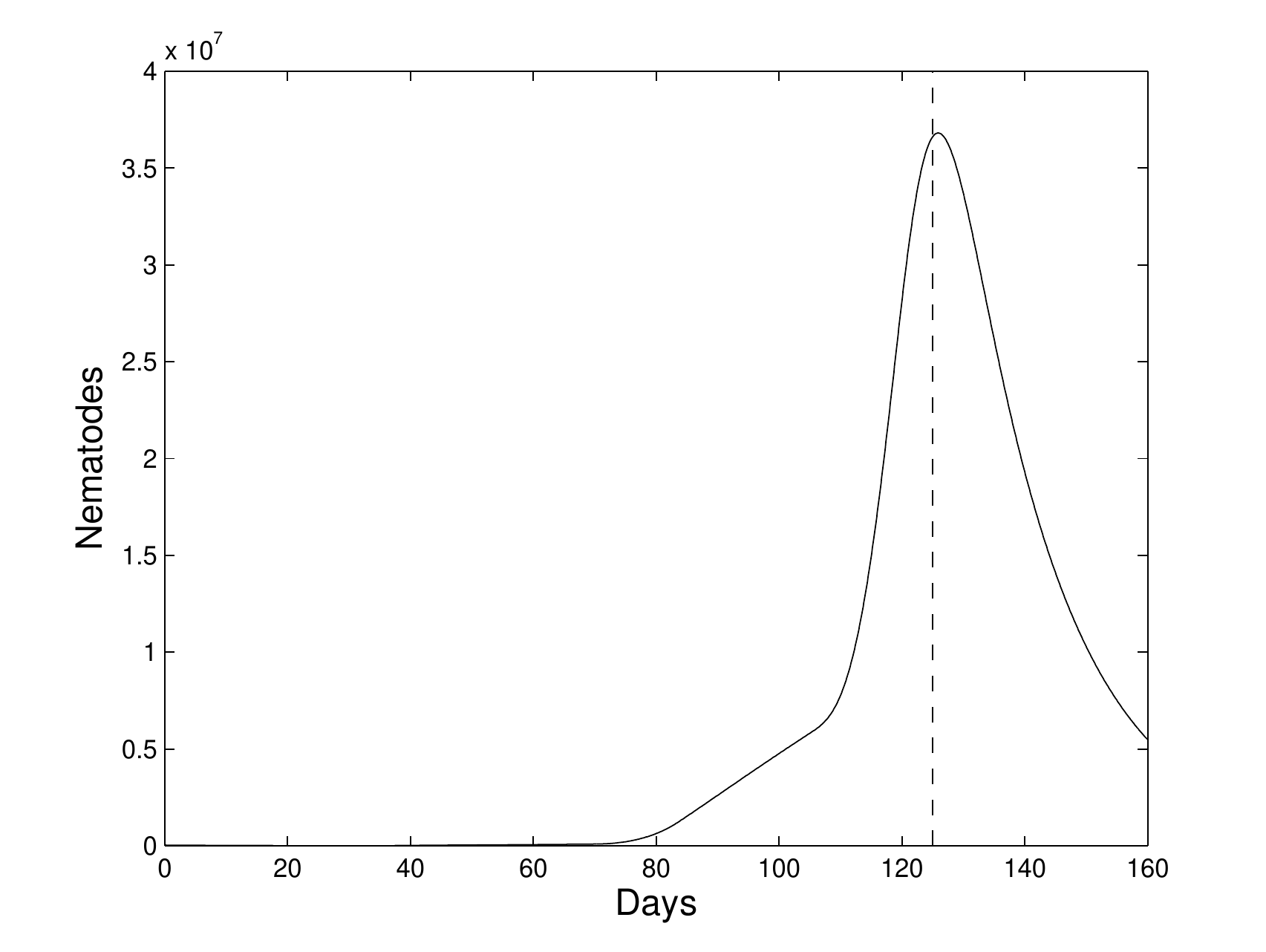}
\includegraphics[width=.46\textwidth]{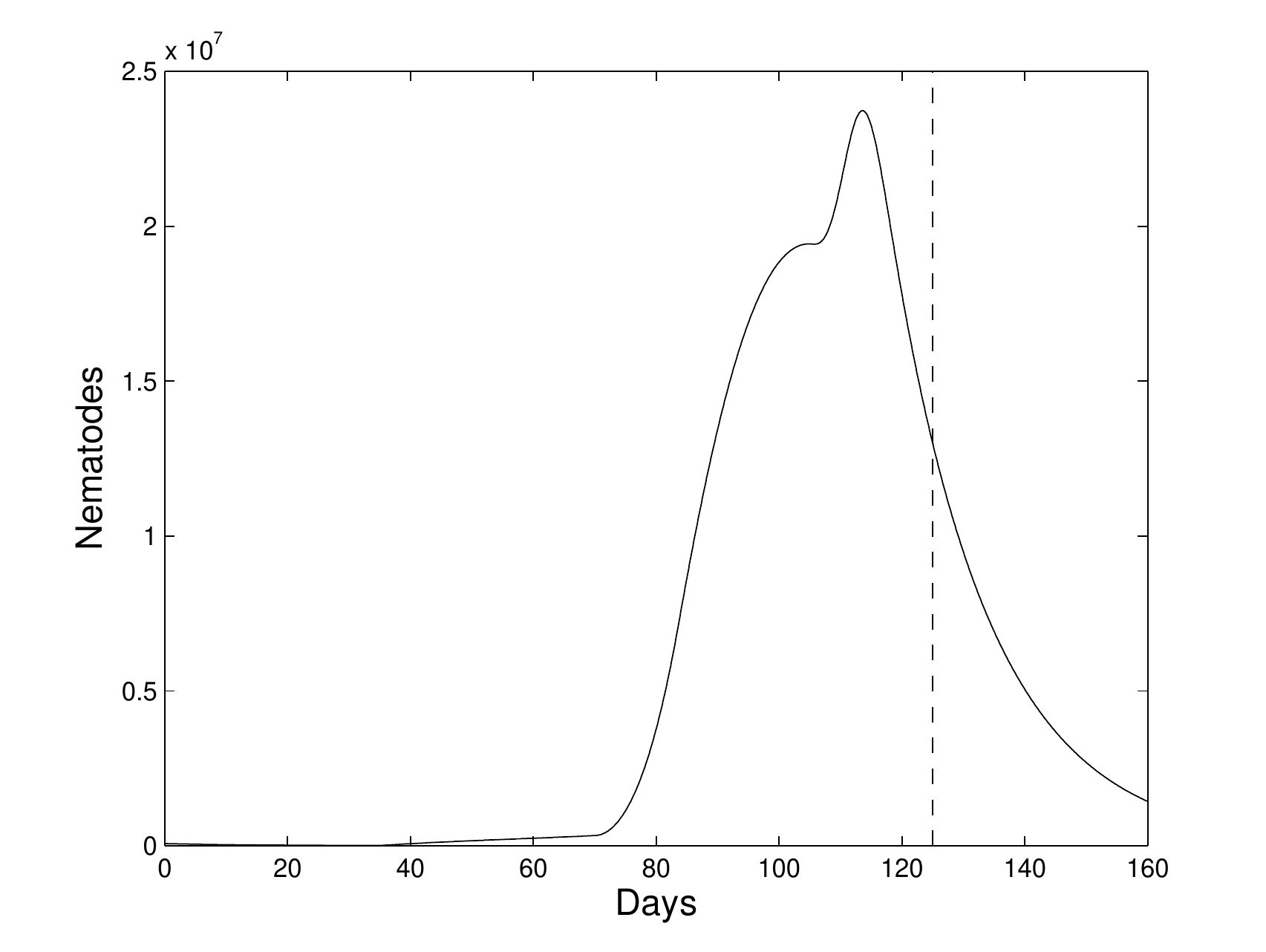}
\caption{Wet season dynamics during the overexploited year of the two cycle with $\beta=10^{-7}$ (left) and $\beta=2\times 10^{-7}$ (right). The threshold time $T-\tau=125$ is marked with a vertical line. Other parameters: $k_H=0.0001$, $k_N=0.063$, $\lambda_o = 10^{-6}$, $\lambda_i=10^{-3}$, $\Lambda(t)=\min\left\{   10,000 e^{0.09 t}  , \; 800,000    \right\}$, $H_0 = 100$.}
\label{overexploit}
\end{center}
\end{figure}

Recall that infections occurring after the time $T-\tau$ produce nematodes with reduced mortality during the dry season \cite{Dugaw}. For parameter values producing violent two-cycles in the yearly map, the last infections in the overexploited year take place before this crucial window. However, it is important to note that these infections may occur late enough that the nematode density during the window is quite high. Figure~\ref{overexploit} demonstrates these dynamics for two values of the infectivity $\beta$. For $\beta = 10^{-7}$, the nematodes reach their greatest numbers at the beginning of the window. For lower $\beta$ this peak occurs even later. Even with $\beta = 2\times 10^{-7}$, nematode numbers are significantly higher in the crucial window than in much of the rest of the season.

\subsection{Bifurcation Diagram of the Map}
\begin{figure}[htbp]
\begin{center}
\includegraphics[width=.7\textwidth]{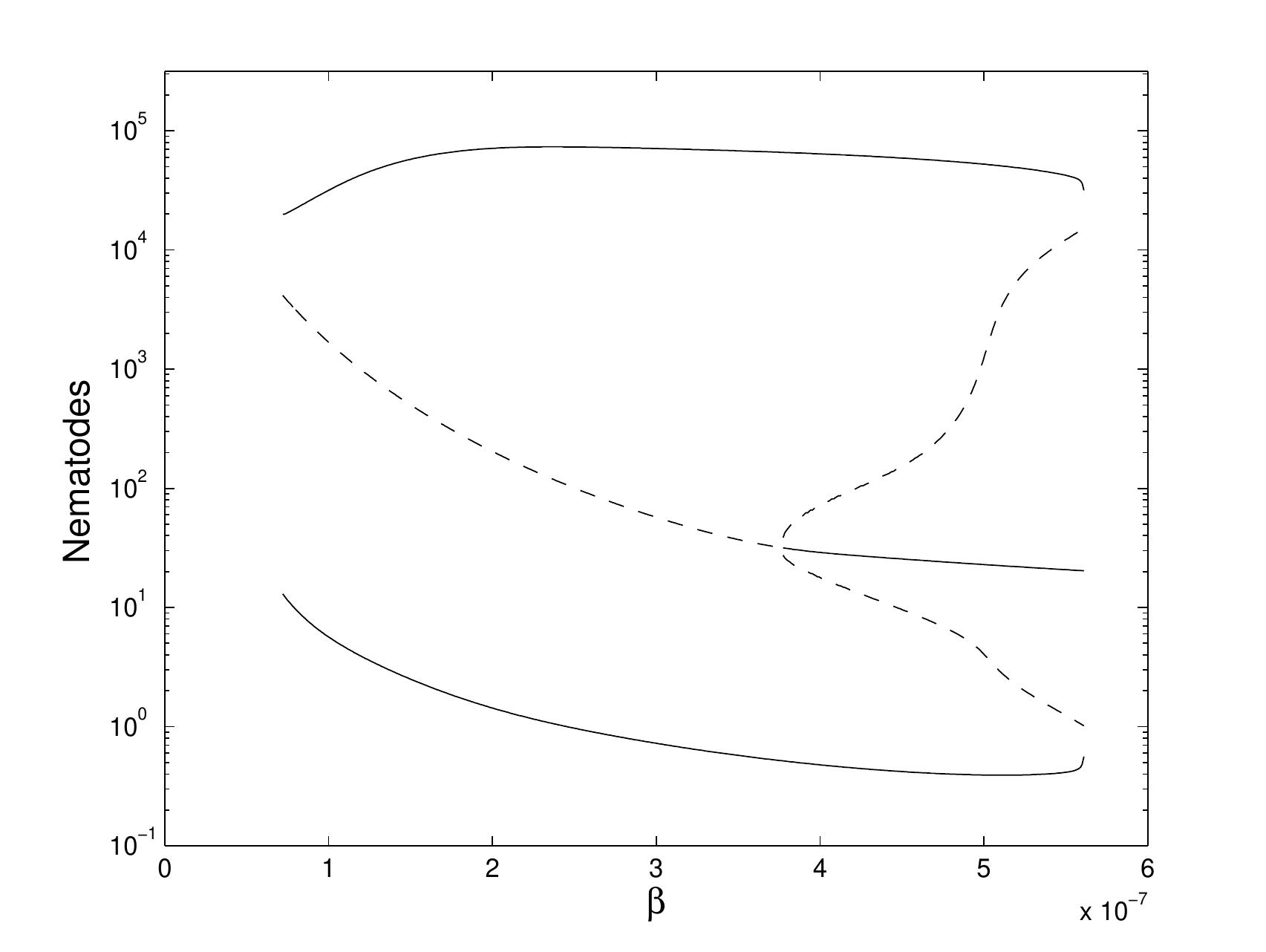}
\caption{Bifurcation diagram for values of $\beta$ producing a two-cycle. Solid curves indicate stable fixed points or cycles, dashed curves unstable. Parameters are as given in Figure~\ref{overexploit}.}
\label{bifdiag}
\end{center}
\end{figure}

Let us take note of the location and nature of the fixed point and two-cycles of the yearly map. Figure~\ref{bifdiag} gives a bifurcation diagram on the parameter range producing the stable two-cycle that motivates this work. For much of this range, the fixed point is at a fairly small density. Notice also that as the infectivity $\beta$ increases, the fixed point becomes stable through a flip bifurcation before the two-cycle vanishes in a fold bifurcation. The basin of attraction of the fixed point becomes quite large, and extends downward almost to $N=1$.

\section{The New Model}
\subsection{Unsuitability of Deterministic Modeling}
During the wet season, we consider the infection of the alternate host $H_\text{alt}$ with a per-host infection rate proportional to parasitoid density $N$. Each infection produces $\Lambda_\text{alt}$ nematodes after some incubation period, which we will take to be the same as in the primary host, $\tau$. Over the course of the dry season, all soil-dwelling nematodes experience high mortality $\lambda_o$. Nematodes remaining in alternate hosts at the beginning of the dry season, like those in primary hosts, experience lower mortality, which we also take to be $\lambda_i$. 

Written in the deterministic (mean field) form of (\ref{dugaw model 1})-(\ref{dugaw model 3}), our model would be \[
 H'(t) =  - k_H H(t) - \beta H(t)N(t),    \;   H(0)=H_0\]\[
  H_\text{alt}'(t) =  - \beta_\text{alt} N(t) H_\text{alt}(t),     \;   H_\text{alt}(0)=H_1\]\[
 N'(t) =  - k_N N(t) - \beta H(t)N(t)  + \beta \Lambda (t - \tau )H(t - \tau )N(t - \tau )   \]\[
  - \beta_\text{alt} N(t) H_\text{alt}(t)  + \beta_\text{alt} \Lambda_\text{alt}H_\text{alt}(t - \tau )N(t - \tau ), 
\] with the next year's initial nematode numbers determined by \[
\lambda _o N(T) + \lambda _i \left(  \beta \int\limits_{T - \tau }^T {\Lambda (t)H(t)N(t)dt}
+ \beta_\text{alt}\Lambda_\text{alt} \int\limits_{T - \tau }^T {H_\text{alt}(t)N(t)dt} \right) .
\]

However, if an alternate host for the nematode exists, it is likely not nearly as abundant as the primary host. As such, we wish to examine the repercussions of very few alternate host individuals per year on nematode persistence, and so modeling this secondary host in a deterministic fashion is out of the question.

\subsection{The Model and Its Simulation}
The alternate host and its interaction with the nematode population will be modeled as a continuous-time stochastic process, with each individual host subject to an infection rate of $\beta_\text{alt} N$. In the event of an infection, three transitions occur: the alternate host population is reduced by 1, the nematode population is reduced by 1, and $\tau$ days after the infection, $\Lambda_\text{alt}$ new nematodes are produced -- unless the dry season has begun. 

The stochastic process described above occurs simultaneously with the deterministic wet season dynamics. This is numerically simulated by calculating, at each time step of the differential equation solver, the approximate probability that a given alternate host will be infected during that time step: $\beta_\text{alt} N(t) \Delta t$, where $\Delta t$ is the length of the time step. Events are then carried out according to simulation of the corresponding random variable for each alternate host. The next year's initial nematode density is given by (\ref{dugaw model 3}), with the addition of the term  $\lambda_i\Lambda_\text{alt}$ multiplied by the number of infections that occurred after the cutoff time $T-\tau$.

In all simulations we will use a fourth-order Runge-Kutta routine with step size $\Delta t = 0.1$, and the parameters from Figure~\ref{overexploit}, unless otherwise indicated. Limited tests with varying $\Delta t$ indicate that this method is stable and accurate.

\section{Means of Persistence}
With the model formulated, we will now see how dynamics such as those described in Section~\ref{Details from the Original Model} contribute to enhanced persistence in the presence of a scarce alternate host. 

\subsection{Mitigated Crashing}
The obvious means by which the addition of an alternate host to the highly cyclic victim-exploiter interaction can enhance nematode persistence is by increasing nematode densities in low years.

The number of infective juveniles present in the soil at the end of an overexploited season, in the crucial window during which an infection may produce new nematodes sheltered from the high mortality of the dry season, can be orders of magnitude greater than most of the rest of the season, as seen in Figure~\ref{overexploit}. So, given a proportional relationship between nematode density and infection probability, rare alternate host infections may be most likely to occur during this window.  

\begin{figure}[htbp]
\begin{center}
\includegraphics[width=.7\textwidth]{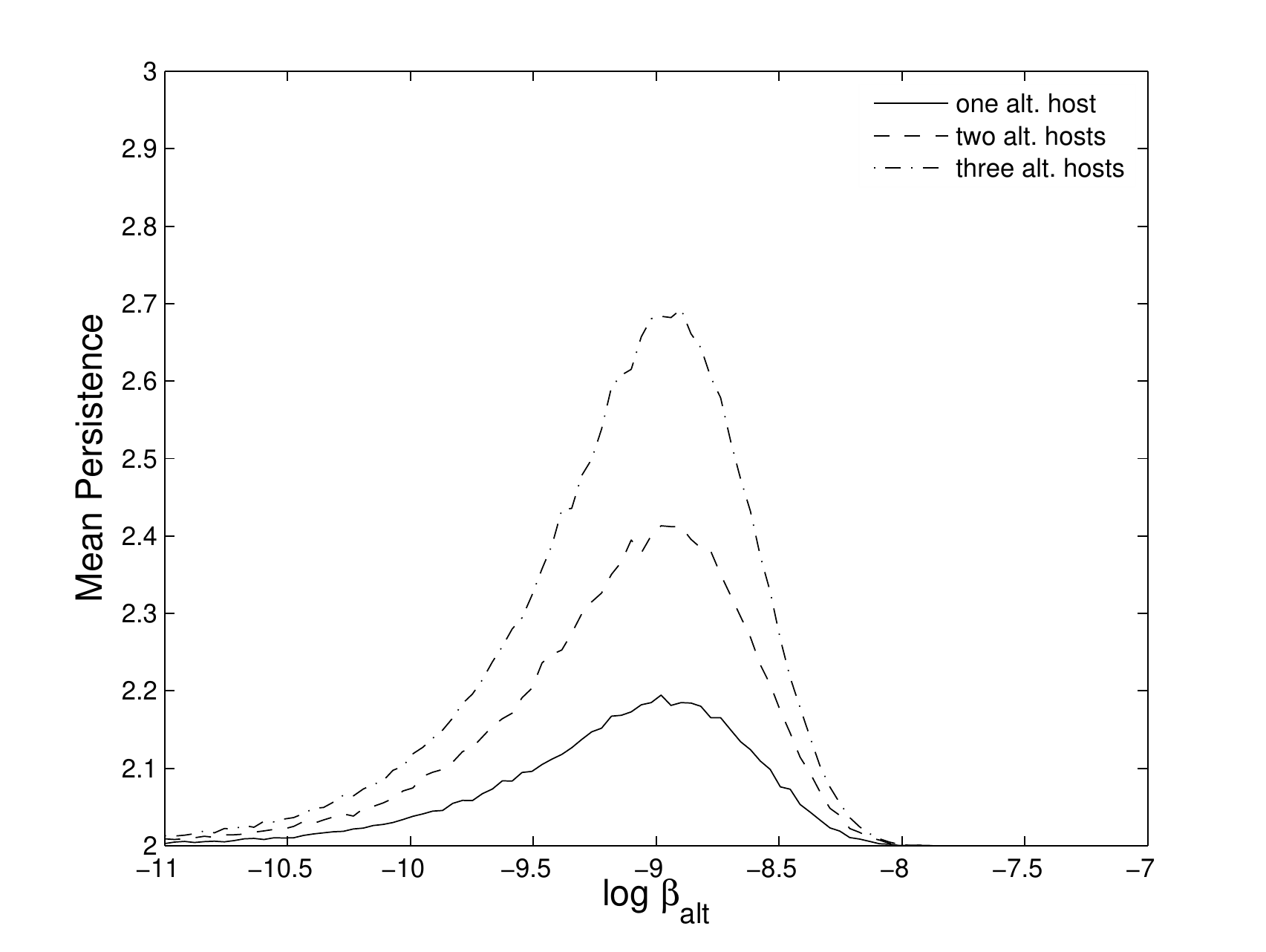}
\caption{Dependence of persistence on alternate host infectivity and number of individual alternate hosts per year.}
\label{persistence Ha}
\end{center}
\end{figure}

Figure~\ref{persistence Ha} demonstrates the dependence of persistence on $\beta_\text{alt}$, as well as the results of increasing the number of alternate hosts available in each wet season. We have used $\beta=2\times10^{-7}$ and $\Lambda_\text{alt}=25,000$, averaging on 20,000 randomized trials at 100 points along the $\beta_\text{alt}$ axis. In each simulation we start with $N=5$ and count persistence until $N$ falls below 5. Mean persistence time seems to increase proportionally with the number of alternate hosts available each year. Also note, most strikingly, that there is effectively a $\beta_\text{alt}$ threshold above which the alternate host has no effect on persistence.

\subsection{Transient Survival}
\begin{figure}[htbp]
\begin{center}
\includegraphics[width=.7\textwidth]{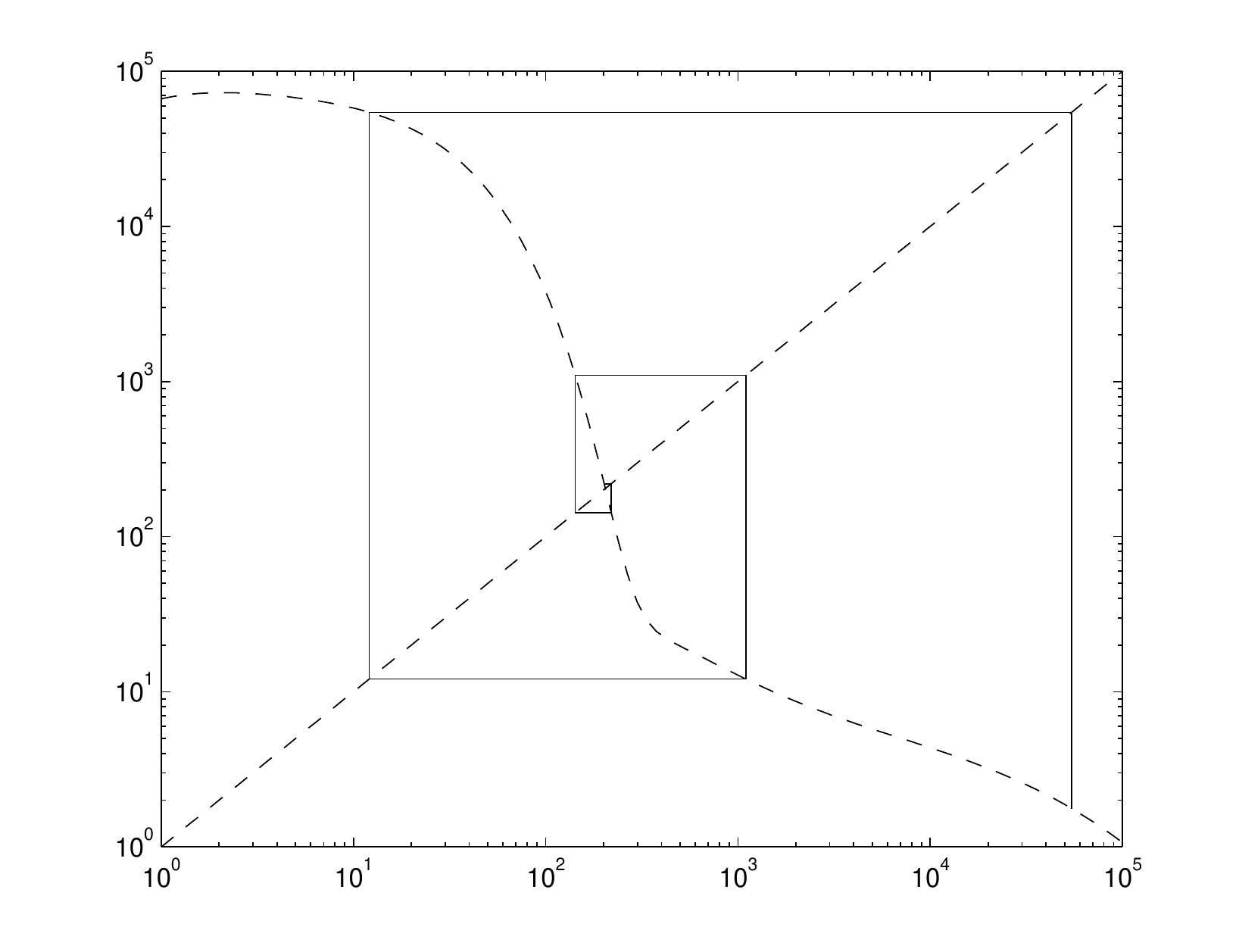}
\caption{Cobweb diagram of transient dynamics after a fortuitous infection at the end of an overexploited wet season.}
\label{transient}
\end{center}
\end{figure}

An alternate host in our cyclic interaction can enhance nematode persistence by raising nematode density to an intermediate level near the fixed point of the year-to-year dynamics, perhaps producing a lengthy coexistence transient. As discussed previously, the fixed point, even when unstable, often occurs at a relatively low density. For the parameters used in Figure~\ref{overexploit}, with $\beta = 2\times 10^{-7}$, the fixed point is around $N=200$. So if an infection late in the overexploited wet season were to produce around 200,000 sheltered infective juveniles, by our model parameters there would remain just slightly more than 200 nematodes at the beginning of the next wet season. The resulting cobweb diagram is shown in Figure~\ref{transient}.  In this particular case, nematodes do not return to low numbers (less than 10) until five years after the late-season infection.

\begin{figure}[htbp]
\begin{center}
\includegraphics[width=.7\textwidth]{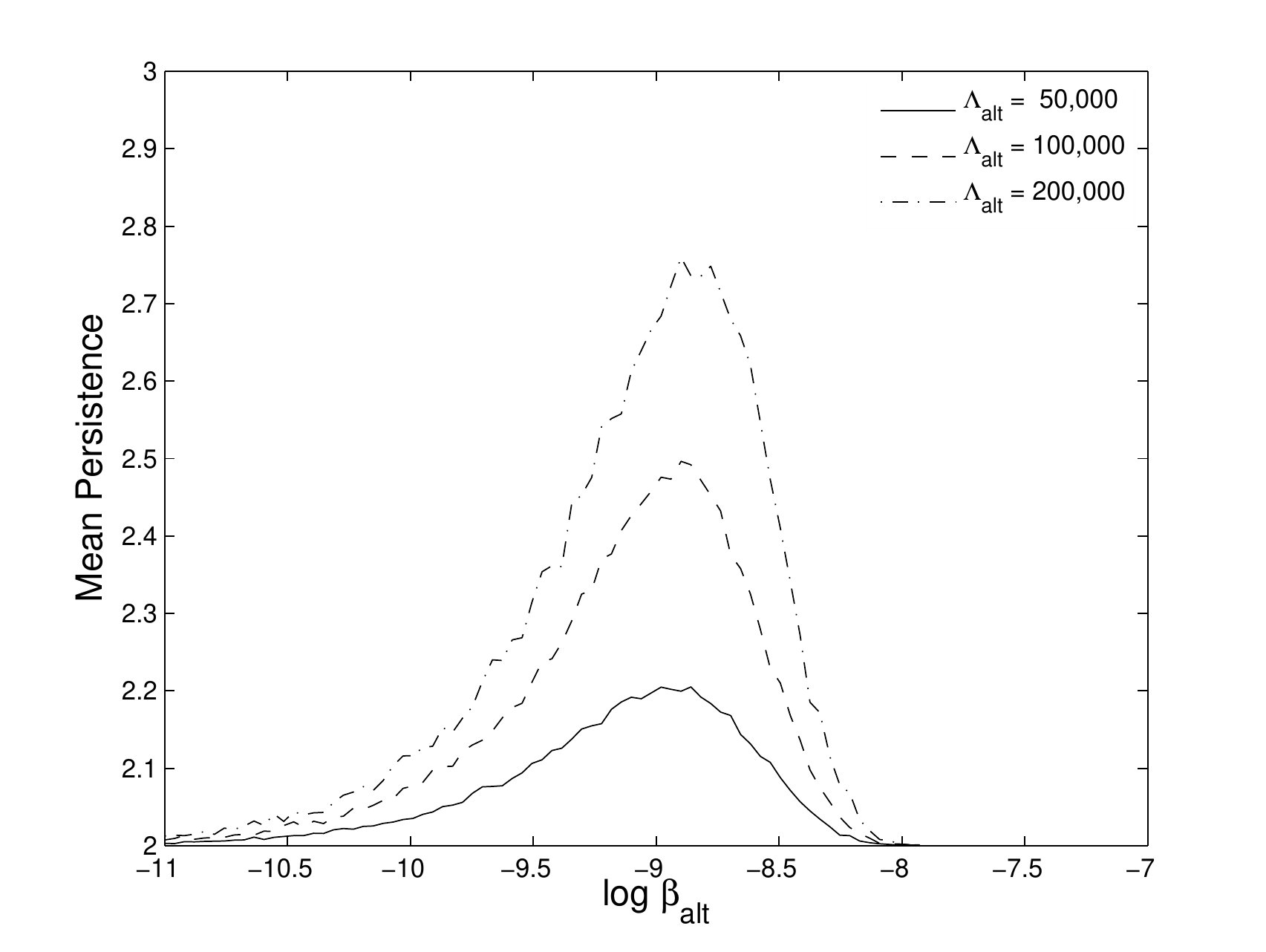}
\caption{Dependence of persistence on alternate host productivity.}
\label{persistence La}
\end{center}
\end{figure}

Figure~\ref{persistence La} shows the results of increasing the productivity of alternate host infections. Note that doubling $\Lambda_\text{alt}$ to 50,000 has little to no effect on persistence, but as $\Lambda_\text{alt}  \lambda_{i}$ approaches the fixed point, persistence is increased. However, infection output must be increased eightfold to have roughly the same effect as tripling the number of individual alternate hosts. This disparity is due to the fact that only a single alternate host infection in the crucial window is necessary to lengthen persistence by two years. The probability of such an infection increases roughly proportionally with additional hosts; increasing productivity has no bearing on that probability, and the transient effects of a given alternate infection are only lengthened significantly by choosing unreasonably specific nematode densities.

\subsection{Deterministic Stability}
As seen in Figure~\ref{bifdiag}, for larger values of the infectivity $\beta$, the yearly map exhibits not only a fixed point at low density, but stability of that fixed point with a fairly large basin of attraction. For such values, a single alternate infection before an otherwise low-density year could move the dynamics into asymptotically stable coexistence.

\begin{figure}[htbp]
\begin{center}
\includegraphics[width=.7\textwidth]{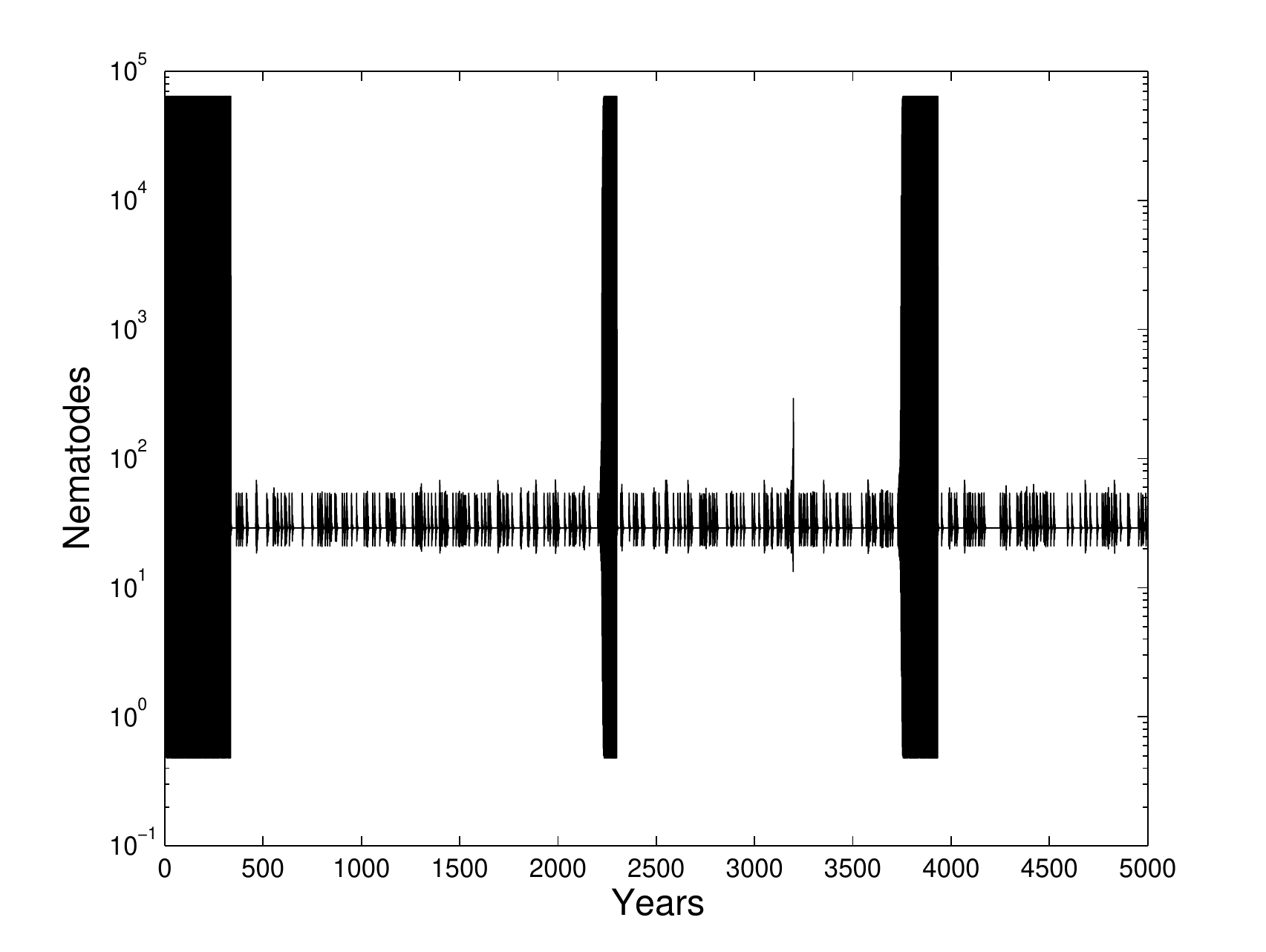}
\caption{Time series of dynamics with two attractors.}
\label{ts5000}
\end{center}
\end{figure}

Figure~\ref{ts5000} shows some of the dramatic results possible when infectivity of the primary host is chosen such that the system has a stable fixed point in addition to its limit cycle. In this case, $\beta = 4.0\times 10^{-7}$, a value close to the flip bifurcation (thus the fixed point's basin of attraction is relatively small). We have also set $\beta_\text{alt} = 10^{-10}$ and $\Lambda_\text{alt} = 25,000$, with one alternate host individual available per year. The simulation is started on its limit cycle.

Meaningful alternate host infections during overexploited years of the two-cycle may be highly unlikely, since peak in-season nematode densities occur well before the crucial window for such a high value of $\beta$ (see Figure~\ref{overexploit}); however, once an infection occurs, dynamics stay near the fixed point -- and away from nematode extinction -- for extended periods. Note that the dynamics approach the fixed point and are perturbed away by alternate infections frequently. Meaningful infections take place much more often in this range of densities than in the overexploited years of the limit cycle, because here nematode density peaks later in the wet season. In fact, these infections can take place often enough to move the system out of the basin of attraction, allowing it to approach the dangerous limit cycle. For example, this happens at 2,210 and at 3,730 years in Figure~\ref{ts5000}. Note, however, that at 3,200 years the system moves out of the basin of attraction, only to be immediately returned to it by an alternate host infection. 

Clearly, the length of time represented by Figure~\ref{ts5000} is not meant to be realistic for a time series of the dynamics in a natural rhizosphere; it simply serves to demonstrate the possible persistence of each of the stable states of the original model under the influence of an alternate host.

\section{Discussion}
The efficacy of our alternate host in lengthening the persistence of the nematode is very dependent on the value of $\beta$, the infectivity of the primary host. However, the means by which persistence may be extended are such that, for much of the range of values of $\beta$ that produce violent two-cycles, the presence of the alternate host can have dramatic effects. For low $\beta$, meaningful alternate host infections -- those which produce new nematodes sheltered from the high mortality of the dry season -- become quite likely during overexploited years. For high $\beta$, meaningful infections may become less likely, but only a single infection is needed to approach stable persistence. 

It is crucial to the survival of the parasitoid population that some host organisms be infected at the end of the wet season. We have seen how the nature of the wet season dynamics allows for this if a host exists that is significantly less susceptible to the nematode. This host, even in small numbers, then maintains the parasitoid during a crucial period when the primary host is unavailable. This is reminiscent of an alternate prey of predatory mites, studied in~\cite{Walde}, which is present earlier in the season than the prey which these mites are meant to control, thus allowing the predator population to grow larger -- and therefore more effective -- before the targeted species' numbers rise significantly.

Note that our conclusions would hold even if the only difference between the primary and alternate hosts were their infectivities, $\beta$ and $\beta_\text{alt}$, respectively. That is, the model and results would  apply to a single host population in which a few individuals are significantly less susceptible to parasitism, through genetic disposition, an effective refuge, or some other means. In such a situation we can see that, paradoxically, the existence of individuals that are much less susceptible than the rest of the population should have a very positive effect on nematode persistence.

    %
    %
    
\end{document}